\documentclass[prd,aps,twocolumn,preprintnumbers, showpacs, nofootinbib,superscriptaddress,notitlepage]{revtex4-1}
\usepackage{mathrsfs}
\usepackage{amsfonts}
\usepackage{amsmath}
\usepackage{slashed}
\usepackage{array}
\usepackage{verbatim}
\usepackage{epsfig}
\usepackage{graphicx}
\usepackage{color}
\usepackage[dvipsnames]{xcolor}
\usepackage[colorlinks,linkcolor=red,anchorcolor=green,citecolor=blue,CJKbookmarks=True]{hyperref}
\definecolor{red}{rgb}{1,0,0}

\usepackage{multirow}

\newcommand{\beq}{\begin{eqnarray}}
\newcommand{\eeq}{\end{eqnarray}}

\def\be{\begin{equation}}
\def\ee{\end{equation}}
\def\bea{\begin{eqnarray}}
\def\eea{\end{eqnarray}}
\def\bal#1\eal{\begin{align}#1\end{align}}

\begin{document}

\title{Charmed meson masses and decay constants in the continuum from the tadpole improved clover ensembles}

\collaboration{\bf{CLQCD Collaboration}}

\author{
\includegraphics[scale=0.30]{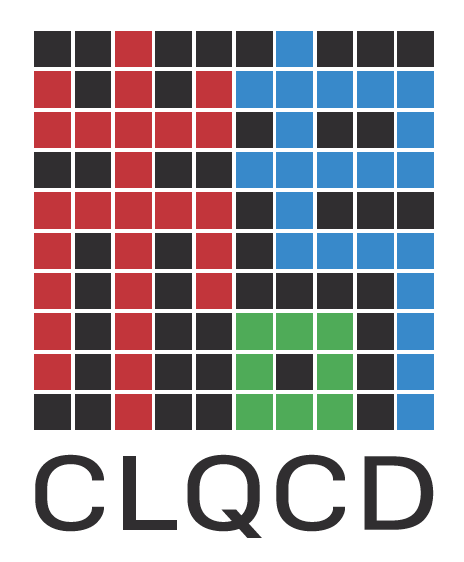}\\
Hai-Yang Du}
\affiliation{University of Chinese Academy of Sciences, School of Physical Sciences, Beijing 100049, China}
\affiliation{CAS Key Laboratory of Theoretical Physics, Institute of Theoretical Physics, Chinese Academy of Sciences, Beijing 100190, China}

\author{
Bolun Hu}
\affiliation{CAS Key Laboratory of Theoretical Physics, Institute of Theoretical Physics, Chinese Academy of Sciences, Beijing 100190, China}

\author{
Mengchu Cai}
\affiliation{CAS Key Laboratory of Theoretical Physics, Institute of Theoretical Physics, Chinese Academy of Sciences, Beijing 100190, China}

\author{Ying Chen}
\affiliation{University of Chinese Academy of Sciences, School of Physical Sciences, Beijing 100049, China}
\affiliation{Institute of High Energy Physics, Chinese Academy of Sciences, Beijing 100049, China}

\author{Heng-Tong Ding}
\affiliation{Key Laboratory of Quark \& Lepton Physics (MOE) and Institute of Particle Physics, Central China Normal University, Wuhan 430079, China}

\author{
Xiangyu Jiang}
\affiliation{CAS Key Laboratory of Theoretical Physics, Institute of Theoretical Physics, Chinese Academy of Sciences, Beijing 100190, China}

\author{Chuan Liu}
\affiliation{School of Physics, Peking University, Beijing 100871, China}
\affiliation{Center for High Energy Physics, Peking University, Beijing 100871, China}
\affiliation{Collaborative Innovation Center of Quantum Matter, Beijing 100871, China}

\author{Liuming Liu}
\affiliation{Institute of Modern Physics, Chinese Academy of Sciences, Lanzhou, 730000, China}
\affiliation{University of Chinese Academy of Sciences, Beijing 100049, China. }

\author{
Yu Meng}
\affiliation{School of Physics, Zhengzhou University, Zhengzhou, Henan 450001, China}

\author{Peng Sun}
\email[Corresponding author: ]{pengsun@impcas.ac.cn}
\affiliation{Institute of Modern Physics, Chinese Academy of Sciences, Lanzhou, 730000, China}

\author{
Ji-Hao Wang}
\affiliation{University of Chinese Academy of Sciences, School of Physical Sciences, Beijing 100049, China}
\affiliation{CAS Key Laboratory of Theoretical Physics, Institute of Theoretical Physics, Chinese Academy of Sciences, Beijing 100190, China}

\author{Yi-Bo Yang}
\email[Corresponding author: ]{ybyang@itp.ac.cn}
\affiliation{University of Chinese Academy of Sciences, School of Physical Sciences, Beijing 100049, China}
\affiliation{CAS Key Laboratory of Theoretical Physics, Institute of Theoretical Physics, Chinese Academy of Sciences, Beijing 100190, China}
\affiliation{School of Fundamental Physics and Mathematical Sciences, Hangzhou Institute for Advanced Study, UCAS, Hangzhou 310024, China}
\affiliation{International Centre for Theoretical Physics Asia-Pacific, Beijing/Hangzhou, China}

\author{
Dian-Jun Zhao}
\affiliation{University of Chinese Academy of Sciences, School of Physical Sciences, Beijing 100049, China}
\affiliation{CAS Key Laboratory of Theoretical Physics, Institute of 
Theoretical Physics, Chinese Academy of Sciences, Beijing 100190, China}

\date{\today}

\begin{abstract}
We present the determination of the charm quark mass, the masses, and decay constants of charmed mesons using thirteen 2+1 flavor gauge ensembles at five different lattice spacings $a\in[0.05,0.11]$ fm, 
8 pion masses $m_{\pi}\in(130,360)$ MeV, and several values of the strange quark mass, which facilitate us to do the chiral and continuum extrapolation. These ensembles are generated through the stout smeared clover fermion action and Symanzik gauge actions with the tadpole improvement.
{By absorbing the discretization errors into the masses and field normalization of the charm quark, we manage to suppress the discretization error of the charmed meson mass and all the S-wave open charmed meson decay constants to a few percent or even less at lattice spacing \( a \sim 0.1 \) fm. Moreover, discretization errors for other quantities are also significantly reduced. The continuum extrapolated charm quark mass, $m_c(m_c)=1.2933(72)(95)$ GeV in $\overline{\textrm{MS}}$ scheme, is determined using QED-subtracted $D_s$ meson mass and non-perturbative renormalization. Predictions of the open and close charm mesons using this charm quark mass agree with the experimental values at 0.1-0.5\% level uncertainty. We obtained $D_{(s)}$ decay constants and also by far the most precise $D_{(s)}^*$ decay constants $f_{D^*}=0.2292(26)(17)$ GeV and $f_{D^*_s}=0.2691(30)(03)$ GeV. }
\end{abstract}

\maketitle

\section{Introduction}\label{sec:intro}

Lattice QCD has proven to be highly effective in providing first-principle predictions for the non-perturbative characteristics of light quarks with $m_{q}\ll \Lambda_{\rm QCD}$. On the other hand, the lattice QCD prediction of the hadron spectrum and structure involving heavy quarks with $m_Q\gg \Lambda_{\rm QCD}$ is also highly demanded. This is necessary for constraining the CKM matrix elements through the weak decay constant of hadrons, as well as for determining Higgs coupling to heavy quarks. Furthermore, lattice QCD calculations are crucial for revealing the nature of extraordinary hadron states such as tetraquarks~\cite{BESIII:2013ris,Belle:2013yex} and pentaquarks~\cite{Wu:2010jy,LHCb:2015yax,LHCb:2019kea}.

However, the lattice QCD calculations for heavy quarks with $m_Q\gg \Lambda_{\rm QCD}$ are significantly more challenging. 
{As an effective theory of QCD with an intrinsic cutoff of the order of the inverse lattice spacing $1/a$, lattice QCD is subject to power corrections of the form \( \left[\frac{\Lambda_{\rm QCD}}{1/a}\right]^2 = \Lambda_{\rm QCD}^2 a^2 \). These corrections, which can also be viewed as discretization errors, can be eliminated by the continuum extrapolation \( a \rightarrow 0 \) (or equivalently \( 1/a \rightarrow \infty \)). In the context of heavy quark physics, additional power corrections arise at \( {\cal O}(m_Q^2 a^2) \) for most of the discretized fermion actions. Even} the $m_Q^4a^4$ corrections can become significant when $a\sim 0.1$ fm and $m_Q\sim 1$ GeV, {based on a naive power counting. Since performing lattice calculations with finer spacing and physical light quark masses demands substantially more computational resources than calculations using either heavier light quark masses or coarser lattice spacing, the precision calculation of the heavy flavor physics can be much more costly compared to the light flavor physics.}

{ 
Similar to other effective theories, the convergence of the \( {\cal O}(m_Q^2 a^2) \) corrections can be improved by redefining the expansion parameter \( m_Q \) at a given cutoff of \( 1/a \), analogous to replacing \( F_{\pi}(0) \) with \( F_{\pi}(m_{\pi}) \) in chiral perturbation theory. Moreover, most of the $m_Q a$ dependence to the charmed hadron matrix elements can be absorbed through the redefinition of the charm quark field at finite \( 1/a \), as we will demonstrate in this work. These improvements enable us to explore heavy quark physics, including its dependence on light quark mass, at relatively small values of \( 1/a \) with controllable power correction. Additionally, one can estimate residual power corrections by using ensembles at different lattice spacings but heavier light quark masses.}

{In this work, we conduct a systematic study of how charm physics observables, including the masses of open and hidden charm mesons as well as their decay constants, depend on the lattice spacing $a$ after implementing the improvements mentioned above.  At \( a \sim 0.1 \) fm, the charmed meson mass and all the S-wave open charmed meson decay constants deviate from their continuum extrapolated values by only a few percent or even less, and discretization errors for other quantities are also significantly reduced. Although the redefined charm quark mass is subject to substantial \({\cal O}(a^4)\) errors, its continuum extrapolated value remains consistent with current lattice averages~\cite{FlavourLatticeAveragingGroupFLAG:2021npn}, exhibiting an 1\% uncertainty.}

The organization of this work is outlined as follows. Section~\ref{sec:setup} presents the {theoretical framework, which includes the numerical setup, redefinition of the quark mass, and also renormalization of the matrix elements with the redefined charm quark field.} Section~\ref{sec:result} shows the continuum extrapolation of the renormalized charm mass and hadron properties, and also a comparison with the experiment and previous calculations. Finally, Section~\ref{sec:summary} offers a concise summary and discussion.

\section{Theoretical framework}\label{sec:setup}

\begin{table*}[ht!]                   
\caption{Gauge coupling $\hat{\beta}=10/(g^2u_0^4)$, lattice spacing $a$ with the second error from that of the scale parameter $w_0$, dimensionless bare quark mass parameters $\tilde{m}^{\rm b}_{l,s}$, Lattice size $\tilde{L}^3\times \tilde{T}$, corresponding pion mass $m_{\pi}$ and $\eta_s$ mass $m_{\eta_s}${, the value of $m_{\pi}L$, valence} bare strange and charm quark mass parameters $\tilde{m}^{\rm v}_{s,c}$, and the statistics information.}  
\resizebox{2.1\columnwidth}{!}{
\begin{tabular}{|c c c | c c | c c  c  c | c  c | c | } 
\hline
Symbol & $\hat{\beta}$& $a$ (fm)  & $\tilde{m}^{\rm b}_{l}$ & $\tilde{m}^{\rm b}_s$ & $\tilde{L}^3\times \tilde{T}$ & $m_{\pi}$ (MeV) & $m_{\eta_s}$ (MeV)&$m_\pi L$ & $\tilde{m}^{\rm v}_s$ & $\tilde{m}^{\rm v}_c$ & $n_{\rm cfg}\times n_{\rm src}$ \\
\hline 
C24P34 &6.200 & 0.10524(05)(62) &  $-$0.2770& $-$0.2310 & $24^3\times 64$ &340.2(1.7) &748.61(75)&4.360(22) & $-$0.2396(1) & 0.4072(07) &  $200\times 32$\\ 
C24P29 &&  &$-$0.2770 &$-$0.2400 &   $24^3\times 72$ &292.3(1.0) & 657.83(64)&3.746(13) &$-$0.2356(1) & 0.4159(07) &$760\times \ \ \!3$ \\
C32P29 && & $-$0.2770 &$-$0.2400 &   $32^3\times 64$ & 293.1(0.8) &  658.80(43)&5.008(14)&$-$0.2358(1) & 0.4150(06)&$489\times \ \ \!3$\\
C32P23 && & $-$0.2790 &$-$0.2400 &    $32^3\times 64$ &227.9(1.2) & 643.93(45)&3.894(20) &$-$0.2337(1) & 0.4190(07)&$400\times \ \ \!3$\\
C48P23 & & & $-$0.2790 &$-$0.2400 &   $48^3\times 96$ &224.1(1.2) & 644.08(62)&5.743(30) &$-$0.2338(1) & 0.4196(08)&$\ \ \!62\times \ \ \!3$\\
C48P14 & & & $-$0.2825 &$-$0.2310 &   $48^3\times 96$ &136.4(1.7) &706.55(39)&3.495(44)  &$-$0.2335(1) & 0.4205(07) &$188\times \ \ \!3$\\
\hline
E28P35  & 6.308 & 0.08973(20)(53) & $-$0.2490 & $-$0.2170& $28^3\times 64$ &351.4(1.4) & 717.94(93)&4.462(17) &$-$0.2201(3)  & 0.2823(25)&$147\times \ \ \!4$\\
\hline
F32P30  & 6.410 & 0.07753(03)(45) & $-$0.2295 &$-$0.2050 &  $32^3\times 96$ &300.4(1.2) & 675.98(97)&3.780(15) & $-$0.2038(1)& 0.1974(05) &$250\times \ \ \!3$\\
F48P30 & & & $-$0.2295 &$-$0.2050 &   $48^3\times 96$  & 302.7(0.9) &674.76(58)&5.713(16) & $-$0.2037(1)& 0.1965(04)&$\ \ \!99\times \ \ \!3$\\
F32P21 & & & $-$0.2320&$-$0.2050 &   $32^3\times 64$ &210.3(2.3)  & 658.79(94)&2.646(28)&$-$0.2023(1)& 0.1996(04)&$194\times \ \ \!3$\\
F48P21 & & &$-$0.2320 &$-$0.2050 & $48^3\times 96$ &207.5(1.1)  & 661.94(64)&3.917(22)&$-$0.2025(1)& 0.1997(04)&$\ \ \!82\times 12$\\
\hline
G36P29 & 6.498 & 0.06887(12)(41) &$-$0.2150 &$-$0.1926 & $36^3\times 108$ &297.2(0.9)  & 693.05(46)&3.731(11)&$-$0.1928(1) & 0.1433(12) &$ 270\times \ \ \!4$\\
\hline
H48P32 & 6.720 & 0.05199(08)(31)&$-$0.1850&$-$0.1700& $48^3\times 144$ &316.6(1.0)  & 691.88(65)&4.000(12)& $-$0.1701(1)& 0.0551(07) &$157\times 12$\\
\hline
\end{tabular}  
}
\label{tab:ensem}
\end{table*}

{Lattice QCD discretizes the fermion and gauge actions on the lattice using dimensionless input parameters $\tilde{m}^{\rm b}_q$ for the masses of different quark flavors, as well as the effective gauge coupling $\hat{\beta} = 10/(g^2)$. Here, we use $\tilde{O}$ to denote the dimensionless value of any quantity $O$. After the fermion fields are integrated out, their impact can be expressed as the determinants of the D-slash operator as a function of the gauge field and $\tilde{m}^{\rm b}_q$. The functional integration of the gauge field on a lattice with dimensionless volume $\tilde{L}^3 \times \tilde{T}$ is approximated by statistical samples generated through the hybrid Monte Carlo (HMC) algorithm~\cite{Duane:1987de}. The physical scale of QCD emerges when we calculate the dimensional gauge quantities, which are proportional to $\Lambda_{\text{QCD}}^n$, such as the scale parameter $w_0$=0.1736(9)~fm~\cite{FlavourLatticeAveragingGroupFLAG:2021npn, BMW:2012hcm, HotQCD:2014kol, RBC:2014ntl} for the system with $N_f=2+1$. The lattice spacing $a$ can then be obtained from the ratio of $w_0$ to its dimensionless value $\tilde{w}_0$.}

{The results in this work, are based on the 2+1 flavor (degenerate up and down quarks plus the strange quark)
ensembles~\cite{Hu:2023jet} from the CLQCD collaboration using the tadpole improved tree level Symanzik (TITLS) gauge action and the tadpole improved tree level Clover (TITLC) fermion action.} As discussed in the recent lattice average of flavor physics~\cite{FlavourLatticeAveragingGroupFLAG:2021npn}, the systematic uncertainty stemming from the absence of the charm quark as a dynamical degree of freedom in the action is expected to {vanish within the uncertainty due to the decoupling of the heavy quark with $m_Q\gg \Lambda_{\rm QCD}$. More details of the gauge ensembles and verification of the independence between statistical samples of all the ensembles can be found in Sec. A of the supplemental materials.}

\begin{figure}[thb]
\includegraphics[width=0.45\textwidth]{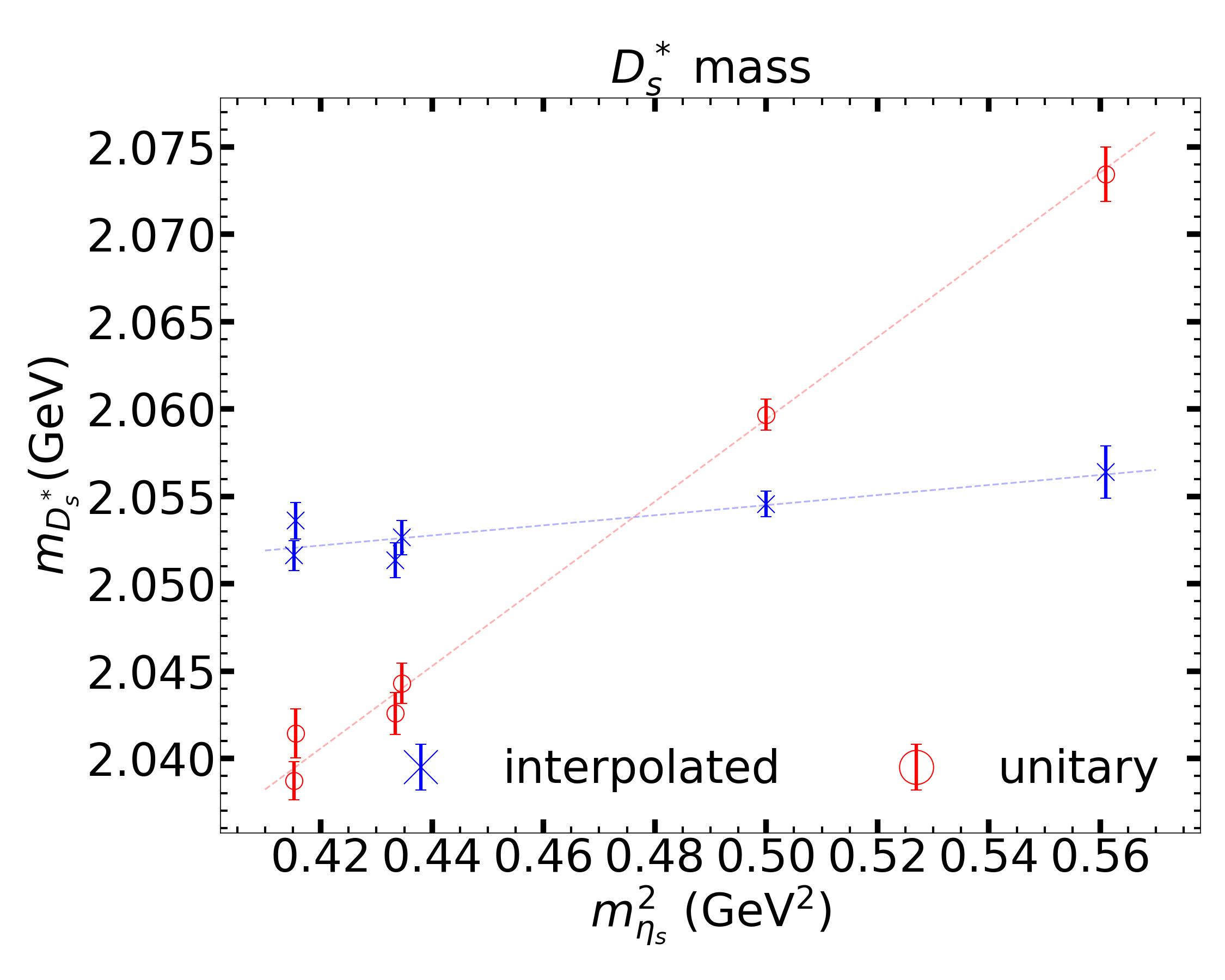}
\caption{The $m_{\eta_s}^2$ dependence of the $D_s^*$ mass at $a=0.1053$ fm, with {sea} strange quark mass parameter (red dots) and {valence} one (blue crosses).}
\label{fig:D_s^*_mass}
\end{figure}

{The parameters employed for the simulation are outlined in Table~\ref{tab:ensem}. The values of $m_{\pi}$ of all the ensembles except F32P21 are chosen to ensure $m_{\pi}L\ge$3.7, thereby efficiently suppressing the finite volume effect. The ``fictitious" meson \(\eta_s\) corresponds to the octet pseudoscalar meson with strange quark mass, which can be precisely determined on the lattice once the physical strange quark mass is derived from other physical quantities, such as \(m_K\) or \(m_{\Omega}\). For instance, we obtained \(m_{\eta_s} = 687.4(2.2) \, \text{MeV}\) using the physical strange quark mass derived from \(m_K\)~\cite{Hu:2023jet} on the CLQCD ensembles.}

{Based on the framework of partially quenched QCD, the mass of the quark associated with the hadron interpolation field—referred to as the ``valence" mass—can differ from that in the gauge configuration, which corresponds to that in the ``sea'' quark loop. Usually the impact of the valence quark mass is much stronger than that of the sea quark mass. Thus one can tune the valence strange quark mass by the most precise lattice determination $m_{\eta_s}=$ 689.89(49) MeV~\cite{Borsanyi:2020mff} on each ensemble to suppress the impact of unphysical strange quark mass on the physical observables.}

Fig.~\ref{fig:D_s^*_mass} illustrates the strange quark mass dependence (through $m_{\eta_s}^2$) of the $D_s^*$ meson mass at $a=0.1053$ fm and the {tuned valence} charm quark mass, using either the {sea} strange quark mass parameter $\tilde{m}_s$ (red circles) or the {valence one} $\tilde{m}^{\rm v}_s$ (blue crosses). The strange quark mass dependence of the latter case is only 10\% of the former case and then allows a more reliable continuum extrapolation since the contribution of the cross term likes $m_{\eta_s}^2a^2$ will be highly suppressed. The residual {deviation between the sea strange quark mass and the physical one} will be addressed through a joint fit using all the ensembles.

Determining the valence charm quark mass can present additional complexities. The physical $J/\psi$ or $\eta_c$ mass should account for both the disconnected charm quark contribution and QED effect, making it unsuitable for accurately determining the charm quark mass using pure QCD calculations with only the connected charm quark contribution. Conversely, the open-charm pseudoscalar meson masses are free of the disconnected valence charm quark contribution, and its QED effect has been computed in previous literature~\cite{Giusti:2017dmp} and updated in Ref.~\cite{DiCarlo:2019thl}. Compared to the strange quark, the valence light quark with physical mass is noisy, costly, and also susceptible to significant finite-volume effects on most of our ensembles. Thus the optimal choice to determine the interpolated valence charm quark mass parameter $\tilde{m}^{\rm v}_{c}$ on each ensemble, is requiring the $D_s$ meson mass to be the QED effect subtracted physical value~\cite{DiCarlo:2019thl} under the GRS renormalization scheme $m^{\overline{\textrm{MS}}}_{q, \rm QCD+QED}(2\mathrm{GeV})=m^{\overline{\textrm{MS}}}_{q, \rm QCD}(2\mathrm{GeV})$~\cite{Gasser:2003hk},
\begin{align}
m^{{\rm QCD}}_{D_s}&=m_{D_s}^{\rm phys}-\Delta^{\rm QED} m_{D_s}=1966.7(1.5)\ \mathrm{MeV}.
\end{align} 

{Since the mass \(m_{D_s}\) remains unchanged when we interchange the up and down quark masses in pure QCD, all terms of the form \((m_d - m_u)^{2n+1}\) will vanish. The next leading order isospin-breaking effects will thus be of the order \({\cal O}((m_d - m_u)\alpha)\) and \({\cal O}(\alpha^2)\), which are negligible in comparison to the leading order QED effects.}

The values of the {bare valence} strange and charm quark masses $\tilde{m}^{\rm v}_{s,c}$ are also presented in Table~\ref{tab:ensem}, based on the polynomial interpolation of the calculations of $m_{\eta_s,D_s}$ using three strange (charm) mass parameters. 

{We observe that \(\tilde{m}^{\rm v}_{s,c}\) on ensembles with the same lattice spacing exhibits slight variations, which can be attributed to the differences in the light and strange sea quark masses.}

Similar to the previous CLQCD study on the light and strange quarks~\cite{Hu:2023jet}, we define the PCAC charm quark mass through the PCAC relation~\cite{JLQCD:2007xff}:
\begin{align}
m_q^{\rm PC}&=\frac{m_{\rm PS}\sum_{\vec{x}}\langle A_4(\vec{x},t)P^\dag(
\vec{0},0) \rangle}{2\sum_{\vec{x}}\langle P(\vec{x},t)P^\dag(\vec{0},0) \rangle}|_{t\rightarrow \infty}\label{eq:mass_v1}.
\end{align}
where $A_{\mu}=\bar{\psi}\gamma_{5}\gamma_{\mu}\psi$, $P=\bar{\psi}\gamma_{5}\psi$, and $m_{\rm PS}$ is the corresponding pseudoscalar meson mass. The renormalized quark mass can be subsequently defined as $m^R_q=Z_A/Z_P m^{\rm PC}_q$, where $Z_{A,P}$ are the renormalization constant of corresponding currents. 

Following the previous light quark study on the same ensembles, we use the coulomb wall source propagator to suppress the statistical uncertainty and extract the meson mass and local operator matrix elements through the joint fit of the wall-to-point and wall-to-wall two-point functions~\cite{Hu:2023jet}. The number of the configuration $n_{\rm cfg}$ and sources used per configuration $n_{\rm src}$ are also collected in Table~\ref{tab:ensem}. In the case of charmed hadrons, a multi-state fit is essential to accurately determine the ground state mass. As an illustration, we present our fitting results on the H48P32 ensemble, which features the finest lattice spacing, {in Sec. B of the supplementary materials. Consistency check on two determinations of $m_s^{\rm PC}+m_c^{\rm PC}$ from quarkonium ($\eta_s$ and $\eta_c$) and also $D_s$ can be found there.}

Then the meson decay constants can be extracted from the matrix elements with proper normalization,
\begin{align}
    Z_V \langle 0|V^{\mu}|{\rm V}_i\rangle&=\epsilon^{\mu}_{i}m_{\rm V}f_{\rm V},\nonumber\\ 
    Z_T \langle 0|T^{4i}|{\rm V}_j\rangle&=\delta^{i}_{j}m_{\rm V}f^T_{\rm V},\nonumber\\ 
Z_A \langle 0|P|{\rm PS}\rangle&=\frac{m_{\rm PS}^2}{2m^{\rm PC}_q}f_{\rm PS},
\end{align}
where $V_{\mu}=\bar{\psi}\gamma_{\mu}\psi${, $T_{\mu\nu}=\bar{\psi}\gamma_{\mu}\gamma_{\nu}\psi$}, $|{\rm PS}\rangle$ ($|{\rm V}\rangle$) is the pseudoscalar (vector) meson state, and $Z_{T,V,A}$ is the tensor, vector and axial-vector current normalization constants, respectively. 
{Additionally, one can define the decay constants of the P-wave charmonium as
\begin{align}
    Z_S\langle 0|\bar{c}c|\chi_{c0}\rangle&=m_{\chi_{c0}}f_{\chi_{c0}},\nonumber\\
    Z_A\langle 0|\bar{c}\gamma_3\gamma_5 c|\chi_{c1}\rangle&=m_{\chi_{c1}}f_{\chi_{c1}},\nonumber\\Z_T\langle 0|\bar{c}\sigma_3c|h_{c}\rangle&=m_{h_c}f_{h_c},
\end{align}
using the renormalization constants $Z_{S,A,T}$.} 

All renormalization constants can be obtained by taking their ratio over $Z_V$, once $Z_V$ is determined. But
unlike its continuum counterpart, $Z_V$ under lattice regularization is subject to both $\alpha_s$ and $a^2$ corrections and can be determined from the vector current conservation condition~\cite{Martinelli:1994ty,Zhang:2020rsx}, 
\begin{align}
Z_{V}(H)\frac{\langle H|V_4|H\rangle}{\langle H|H\rangle}=1,
\end{align}
where $H$ represents an arbitrary hadronic state. 
$Z_{V}(H)$ should be independent of $H$ in the continuum due to charge conservation, while it can suffer from a significant quark mass dependence at finite lattice spacing. {As shown in Ref.~\cite{Capitani:2000xi} and verified by our practical calculation, this dependence majorly comes from the massive quark propagator on the lattice and can be absorbed into the redefinition of the massive quark field.}

{Thus we propose to use the heavy quark improved normalization factor $Z_V^c\equiv Z_V(\eta_c)$ for the $\bar{c}\gamma_{\mu}c$ current, and define that of the other charm quark bilinear current $X$ through the chiral-extrapolated ratio $\frac{Z_X}{Z_V}$ under the $\overline{\mathrm{MS}}$ scheme through the intermediate RI/MOM scheme~\cite{Martinelli:1994ty}, $Z_X^{c}\equiv Z_V^c\frac{Z_X}{Z_V}$.} The improvement can be extended to the flavor-changed current, like $Y\equiv \bar{c}\Gamma l$, by utilizing the improved renormalization constant $Z_{Y}^c\equiv \sqrt{Z_VZ_V^c}\frac{Z_Y}{Z_V}$. {Thus the improvement here can be understood by a redefinition of the charm quark field normalization at finite $a$. Based on the numerical tests on $f_{J/\psi}$, $f_{D^*_s}$ and $g_{S,{\rm ME}}(H)\equiv \frac{\langle H|S|H\rangle}{\langle H|H\rangle}$, this heavy quark improved renormalization can suppress the discretization error of heavy quark significantly, as detailed in Sec. C of the supplemental materials. The values of the renormalization constants{, and also the lattice spacing dependence of the transverse decay constant of the charmed vector mesons and those of the P-wave charmonium} can also be found there.}

\section{Results}\label{sec:result}

Similar to our previous work~\cite{Hu:2023jet}, we
perform bootstrap re-samplings on each ensemble
and conduct the correlated global fit based on these bootstrap samples. {Since we require the lattice spacing $a$ and $Z_V$ to be the same with similar $\beta$ and to correspond to physical light and strange quark masses, $a$ and $Z_V$ are obtained as the fit parameters through joint fits using kinds of quark masses and $\beta$. Then the bootstrap samples are generated for $a$ and $Z_V$ to keep the correlation between different ensembles at the same $\beta$. This treatment ensures the statistical uncertainty of $a$ and $Z_V$ contribute to that of our final predictions properly.} 

\begin{figure}[thb]
\includegraphics[width=0.45\textwidth]{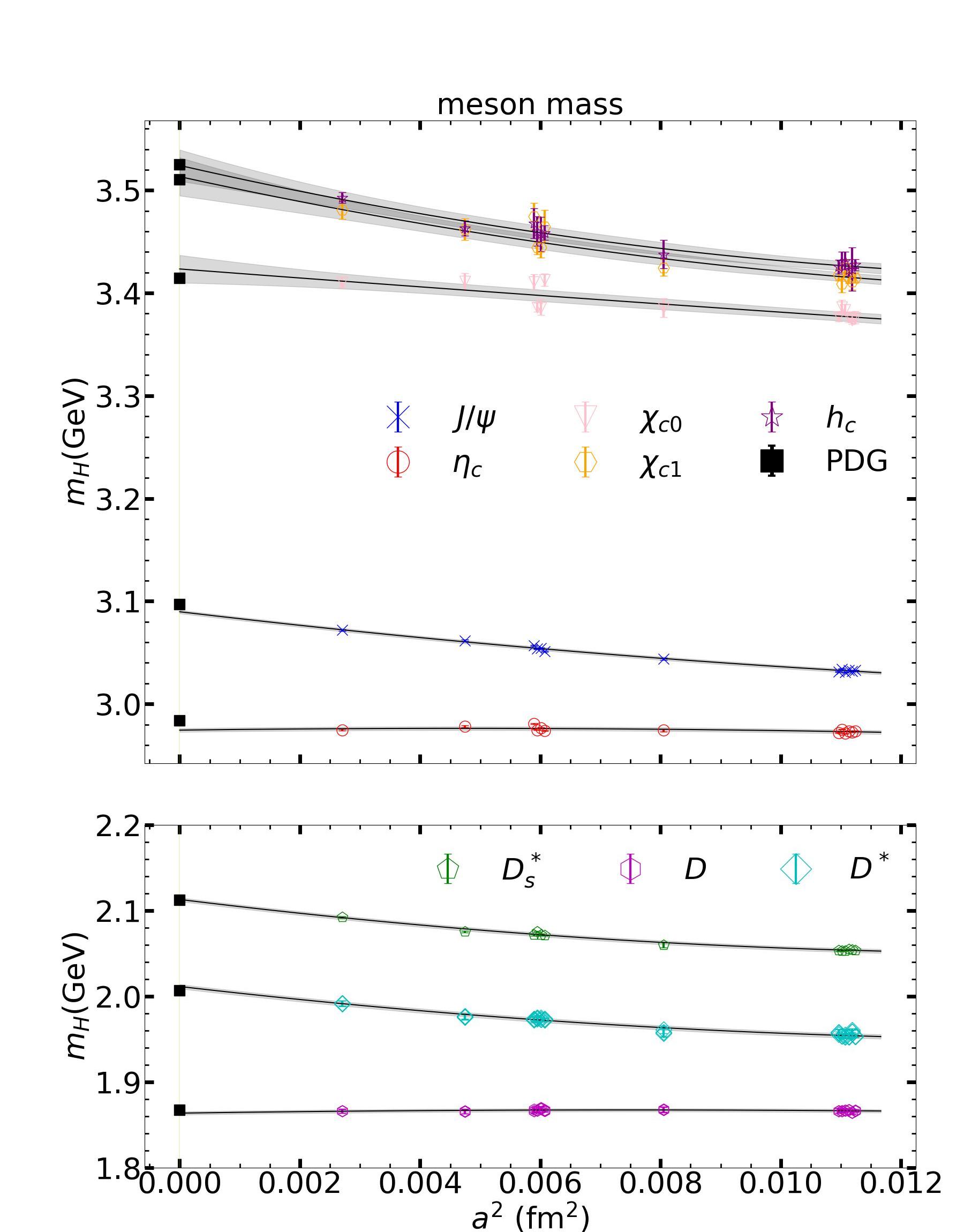}
\caption{The lattice spacing dependence of the {charmonium masses (upper panel), and pseudoscalar and vector open-charm meson masses (lower panel) except the $m_{D_s}$ case which is used to tune the charm quark mass}. The experimental values are shown as black stars for comparison. The light and strange sea quark masses have been corrected to their physical values using the parameters determined from the joint fit.}
\label{fig:meson_lat_dep}
\end{figure}

{For the dimensional quantity $X$ in this section, we employ the following joint fit, which describes the results well with $\chi^2$/d.o.f $\sim$ 1,
\begin{align}
    &X(m_{\pi},m_{\eta_s},a)=X(m_{\pi}^{\rm phys},m_{\eta_s}^{\rm phys},0)\nonumber\\
    &\quad \quad \quad +d^X_1 (m_{\pi}^2-(m_{\pi}^{\rm phys})^2)+d^X_2 (m_{\eta_s}^2-(m_{\eta_s}^{\rm phys})^2)\nonumber\\
    &\quad \quad \quad + d^X_3 a^2+ d^X_4 a^4,\label{eq:x_dep}
\end{align}
where $m_{\pi}^{\rm phys}=134.98$ MeV is the physical $\pi^0$ mass, and the squares of the pion and $\eta_s$ masses are used to account for the dependence on the light quark mass and strange sea quark masses (as the strange valence quark mass has been tuned to its physical value), respectively. The $a^2+a^4$ corrections are also introduced to describe the discretization error, and we will see that the $a^4$ term is essential for certain cases, especially the charm quark mass under $\overline{\textrm{MS}}$ 2 GeV. The $a\alpha_s$ term is not taken into account as it should be mainly removed by the non-perturbative renormalization, and our previous study~\cite{Hu:2023jet} suggests that $a^2$ extrapolation works well for light flavor physics. Based on the joint fit, one can further eliminate the impact from the unphysical light and strange quark masses by subtracting the $d^X_{1,2}$ terms from $X(m_{\pi},m_{\eta_s},a)$ and obtain $X(m_{\pi}^{\rm phys},m_{\eta_s}^{\rm phys},a)$ with physical light and strange quark masses but at a finite lattice spacing.}

While our systematic uncertainties stemming from discretization errors and unphysical light and strange quark masses have been mitigated through the joint fit outlined in Eq.~(\ref{eq:x_dep}), there remains a systematic uncertainty arising from the fit ansatz (such as the cross term $a^2m_{\pi}^2$ and/or higher-order term $m_{\pi}^4$). Thus we can also fit the data using the following modified parameterization,
\begin{align}
    &X(m_{\pi},m_{\eta_s},a)=\big[X'(m_{\pi}^{\rm phys},m_{\eta_s}^{\rm phys},0)\nonumber\\
    &\quad \quad \quad +d'^X_1 (m_{\pi}^2-(m_{\pi}^{\rm phys})^2)+d'^X_2 (m_{\eta_s}^2-(m_{\eta_s}^{\rm phys})^2)\big]\nonumber\\
    &\quad \quad \quad \big(1 + d'^X_3 a^2+ d'^X_4 a^4\big),\label{eq:x_dep2}
\end{align}
and estimate the systematic uncertainty from the fit ansatz. Such an estimate turns out to be much smaller than the other uncertainties and can be refined when ensembles with a wider range of lattice spacings and quark masses become available.

{The systematic uncertainty from the finite volume correction is estimated by including additional $e^{-m_{\pi}L}$ term in the joint fit and the impact turns out to be negligible compared to the statistical uncertainty. A similar treatment is used to estimate the systematic uncertainty from the possible ${\cal O}(a\alpha_s)$ term, which is also small.}

{Since we used FLAG average value of the scale parameter $w_0$=0.1736(9)~fm~\cite{FlavourLatticeAveragingGroupFLAG:2021npn} when determining the lattice spacing, this error of $w_0$ will enter the final result as system uncertainty. To estimate this effect, we vary the central value of $w_0$ by 1 sigma, redo the entire analysis, and consider the difference in the central value of the result as a systematic uncertainty. Since we determine the bare charm quark mass parameter on each ensemble using the physical $D_s$ mass, the impact of the $w_0$ uncertainty on the charmed hadron mass will be partially canceled and then can be smaller than 0.5\%.
}

Fig~\ref{fig:meson_lat_dep} shows the lattice spacing dependence of $m^{\rm QCD}_{D}$, $m^{\rm QCD}_{D^*}$ and $m^{\rm QCD}_{D_s^*}$, with the impact from the unphysical light and strange quark masses subtracted using the {fit ansatz defined in Eq.~(\ref{eq:x_dep}). The good agreement of the fit curve as a function of $a^2$ with our data points (with unphysical light and strange quark mass effects subtracted) also justifies our fit ansatz}. Our predictions of $m_D$ based on the joint fit are
\begin{align}
    m^{\rm QCD}_{D}&=1863.7(1.8)(1.5)~\mathrm{MeV},\nonumber\\
   m^{\rm QCD}_{D^{\pm}}-m^{\rm QCD}_{D^0}&= d^D_1\frac{(m_{\pi}^{\rm phys})^2}{m_l}(m_d-m_u)(1+{\cal O}(\frac{(m_{\pi}^{\rm phys})^2}{\Lambda^2_{\chi}}))\nonumber\\
   &=2.88(22) ~\mathrm{MeV},
\end{align}
where we use the leading order chiral perturbative theory approximation and $\frac{(m_{\pi}^{\rm phys})^2}{m_l}(m_d-m_u)$=0.0116(8) GeV$^{2}$ based on our previous work~\cite{Hu:2023jet}. They perfectly agree with the experimental values with the QED correction from literature (provided in Ref.~\cite{Giusti:2017dmp} and updated in Ref.~\cite{DiCarlo:2019thl}) subtracted, $m^{\rm QCD}_{D}$=1866.3(0.7) MeV and $m^{\rm QCD}_{D^{\pm}}-m^{\rm QCD}_{D^0}$=2.5(0.5) MeV. 

At the same time, our prediction of the $D^*$ mass (light blue) and the $D_s^*$ mass (green) are consistent with the experimental value. The unknown QED effect of the vector meson masses can be significant if the statistical uncertainty can be further suppressed.

For the charmonium mass, we can see from Fig.~\ref{fig:meson_lat_dep} that the central values of both $m_{J/\psi}$ and $m_{\eta_c}$ are a few MeV lower than the experimental values, respectively. The disconnected diagram and QED effects would not be important at the present precision.

At the same time, our prediction of the hyperfine splitting is 
\begin{align}
    \Delta^{\rm QCD, conn}_{{\rm HFS}, \bar{c}c}&=m_{J/\psi}^{\rm QCD, conn}-m_{\eta_c}^{\rm QCD, conn}\nonumber\\
    &=115.3(1.4)(1.1)~\mathrm{MeV},
\end{align}
and {slightly smaller} than the previous HPQCD study~\cite{Hatton:2020qhk} using staggered fermions,
118.6(1.1) MeV. Such a value is slightly larger than the experimental value 112.8(4)~\cite{ParticleDataGroup:2022pth}. {The difference can be attributed mostly to the quark annihilation effect (The correction from the QED effect is around 1 MeV~\cite{Hatton:2020qhk}). Direct calculations using anisotropic $N_f$=2 charm quark sea ensemble at $a_t=0.0205$ fm with $a_s/a_t=5$ indicate that the inclusion of charm quark annihilation diagram almost does not change the $J/\psi$ mass, but pushes the $\eta_c$ mass up by approximately 3-4 MeV~\cite{ZhangRenQiang:2021gnn,Zhang:2021xvl}, regardless of whether the mixing with the pseudo-scalar glueball is considered.}

{Given the approximation
\begin{align}
    m_{h_c}=\frac{1}{9}(m_{\chi_{c0}}+3m_{\chi_{c1}}+5m_{\chi_{c2}})
\end{align}
which has been verified by the experiment at 0.1 MeV level and can be explained by the short-range feature of the spin-spin interaction in the context of potential models~\cite{Kwong:1987mj,Lucha:1991vn}, we skipped the calculation of the $\chi_{c2}$ mass which requires additional complication in the interpolating field, and extract the following three splittings based on $m_{\chi_{c0},\chi_{c1},h_c}$ we obtained,
\begin{align}
    \Delta^{\rm QCD, conn}_{1P-1S, \bar{c}c}&=m_{h_c}-\frac{1}{4}(3m_{J/\psi}+m_{\eta_c})\nonumber\\
    &=462(14)(09)~\mathrm{MeV},\nonumber\\
    \Delta^{\rm QCD, conn}_{1P_{\rm spin-orbit}, \bar{c}c}&=\frac{1}{3}(3m_{h_c}-m_{\chi_{c0}}-2m_{\chi_{c1}})\nonumber\\
    &=40(18)(06)~\mathrm{MeV},\nonumber\\
    \Delta^{\rm QCD, conn}_{1P_{\rm tensor}, \bar{c}c}&=\frac{1}{5} (-m_{h_c}-m_{\chi_{c0}}+2m_{\chi_{c1}})\nonumber\\
    &=15.4(7.5)(0.2)~\mathrm{MeV}.
\end{align}
All three splittings agree with the experimental value and previous LQCD calculation using the HISQ fermion~\cite{Burch:2009az} well within the uncertainty.}

From Fig.~\ref{fig:meson_lat_dep}, we can also see that the pseudo-scalar meson masses have much smaller lattice spacing dependence than the vector meson masses since our valence strange and charm quark masses are also determined using the pseudoscalar masses. The lattice spacing dependence of the vector meson masses also suggests that the hyperfine splitting $\Delta_{\rm HFS}$ will exhibit a significant variation, which is crucial for ensuring that the continuum extrapolated values of $\Delta_{\rm HFS}$ closely match the physical values.

\begin{figure}[thb]
\includegraphics[width=0.45\textwidth]{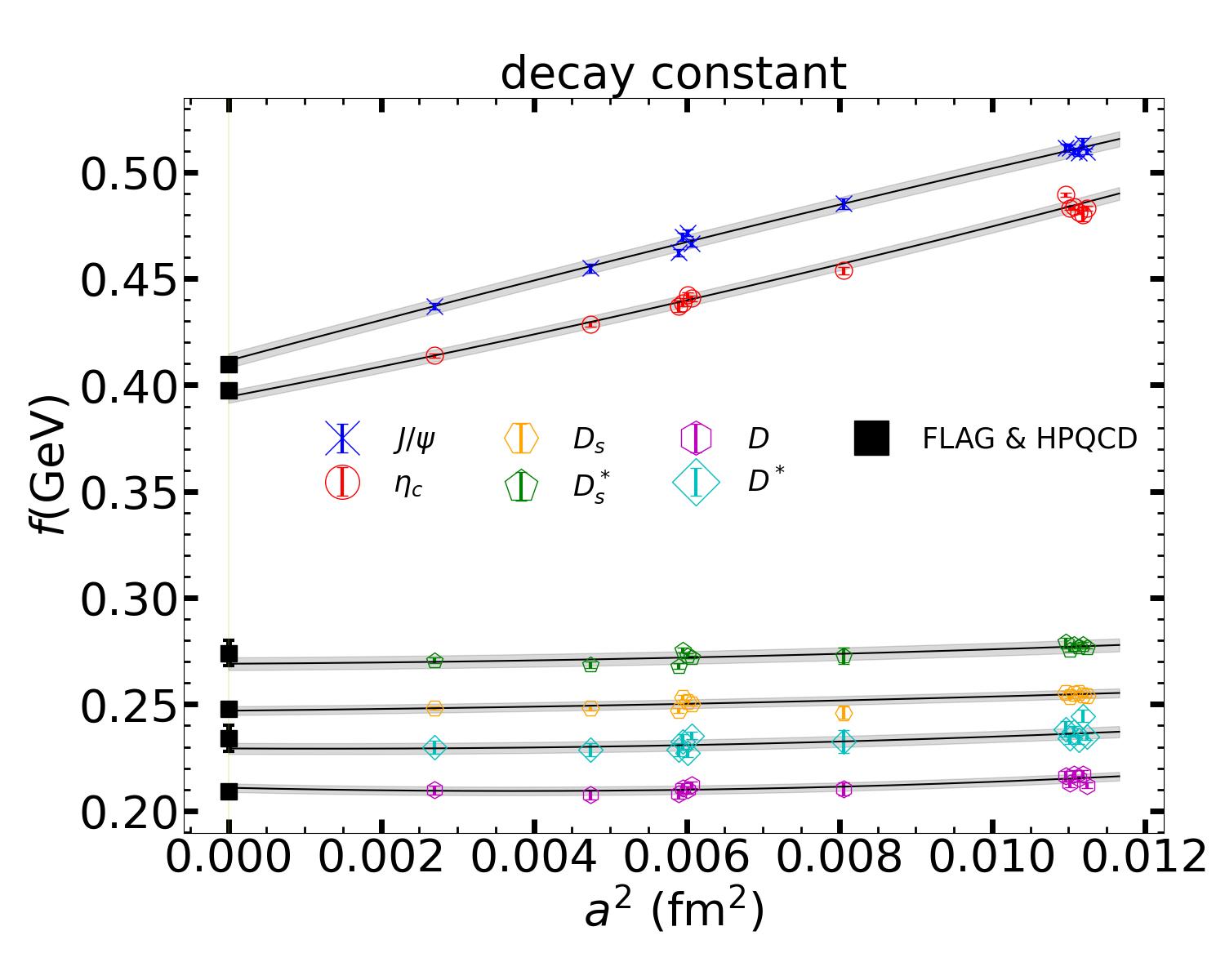}
\caption{Similar to Fig.~\ref{fig:meson_lat_dep} but for the decay constants.}
\label{fig:decay_constant_lat_dep}
\end{figure}

Using the charm quark improved normalization, the decay constants of the pseudo-scalar and vector open-charm mesons and charmonium at different lattice spacings are illustrated in Fig.~\ref{fig:decay_constant_lat_dep}. The decay constants of the open-charm mesons exhibit negligible lattice spacing dependence and are consistent with the previous HPQCD results using the HISQ action within a 1-2\% uncertainty. {For the values of $f_{J/\psi}$ and $f_{\eta_c}$, after using the charm quark improved normalization, the lattice spacing dependence is weakened and our final predictions are consistent with that from HPQCD.}

\begin{figure}[thb]
\includegraphics[width=0.45\textwidth]{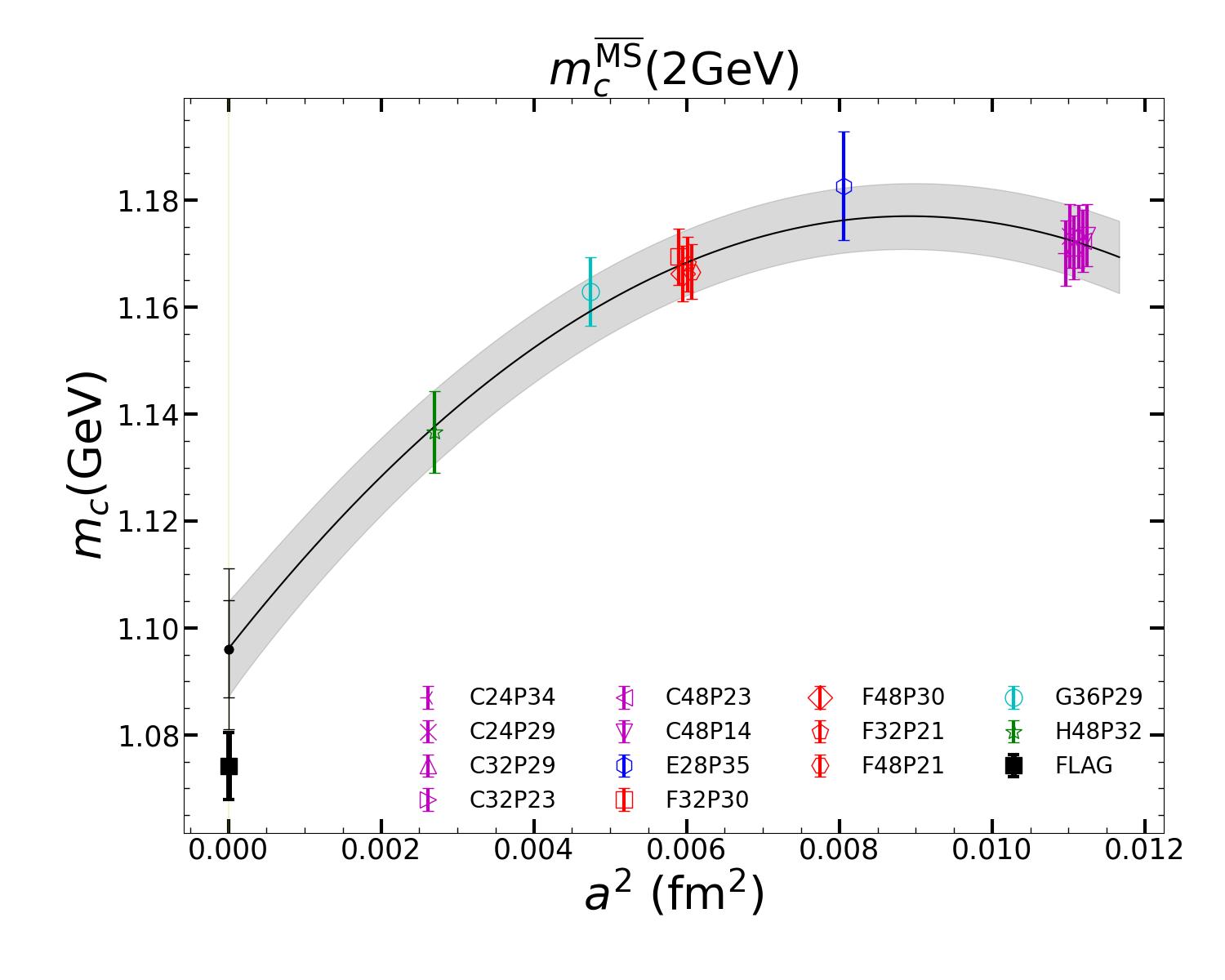}
\caption{The lattice spacing dependence of the physical charm quark mass under $\overline{\textrm{MS}}$ scheme 2 GeV on all the ensembles at different lattice spacings. The light and strange sea quark masses have been adjusted to their physical values based on the parameters derived from the joint fit.}
\label{fig:charm_quark_mass}
\end{figure}

Using a similar approach as outlined in Ref.~\cite{Hu:2023jet} for the light and strange quarks, the renormalized charm quark mass at $\overline{\textrm{MS}}$ 2 GeV with $N_f=3$ matching and running is determined for each ensemble. One can further obtain the charm quark mass in the continuum and also physical light and strange quark mass by using the joint fit of the data on different ensembles. After correcting for small effects arising from the unphysical light and strange quark masses, the renormalized charm quark masses on all the ensembles with different lattice spacings are depicted in Fig.~\ref{fig:charm_quark_mass}. It is evident that at the largest lattice spacing, $m_c$ deviates by $\sim$10\% from its continuum extrapolated value, based on a $a^2+a^4$ extrapolation. Converting the charm quark mass to $m_c(m_c)$ using perturbative running {results in a value of 1.2933(72)(95) GeV, where the two uncertainties represent the statistical and systematic errors, respectively.}

\begin{table*}[ht!]                   
\caption{Charm quark mass under the $\overline{\textrm{MS}}$ scheme, charmed meson masses, and also related decay constants. Note that our predictions and those from the literature do not include the contribution from either the annihilation diagram or QED correction. All the values are in the unit of GeV, and all the Experimental values come from PDG~\cite{ParticleDataGroup:2022pth}.}  
\begin{tabular}{c|lllllll}              
   & $m_c(m_c)$ & & $m_{D}$ & $m_{D^*}$ & $m_{D_s^*}$ \\
\hline 
This work & 1.2933(72)(95) & & 1.8637(18)(15)&2.0115(25)(28) &2.1130(22)(22) \\
FLAG/Exp. &1.276(05)~\cite{FlavourLatticeAveragingGroupFLAG:2021npn,McNeile:2010ji,Yang:2014sea,Nakayama:2016atf,Petreczky:2019ozv,Heitger:2021apz} & &1.8672(02)\ \ &2.0086(01) &2.1122(04) \\
\hline
   &  $m_{\eta_c}$ & $m_{J/\psi}$ & $m_{\chi_{c0}}$ & $m_{\chi_{c1}}$ & $m_{h_{c}}$ \\
\hline 
This work &  2.9745(18)(31) & 3.0898(16)(27)& 3.423(13)(03)& 3.511(16)(06) & 3.522(13)(11) \\
Exp.  & 2.9841(04)& 3.09690(1) &3.41471(30) & 3.51067(5)&3.52537(14) \\
\hline
      & $\Delta_{\rm{HFS}, \bar{c}l}$ & $\Delta_{\rm{HFS}, \bar{c}s}$ &  $\Delta_{\rm{HFS}, \bar{c}c}$ &  
    $\Delta_{1P_{\rm spin-orbit}, \bar{c}c}$&
    $\Delta_{1P_{\rm tensor}, \bar{c}c}$&
    $\Delta_{1P-1S, \bar{c}c}$ \\\hline
This work &0.1478(19)(18) &0.1463(22)(13) & 0.1153(14)(11)& 0.040(18)(06)& 0.0154(75)(02)&0.462(14)(09) \\
Literature  & & & 0.1186(11)~\cite{HPQCD2020} & 0.0433(66)$^{+10}_{-0}$~\cite{Burch:2009az} & 0.0150(23)$^{+3}_{-0}$~\cite{Burch:2009az} &
0.469(11)$^{+10}_{-0}$~\cite{Burch:2009az} \\
Exp. & 0.1413(01)& 0.1438(04)& 0.1128(04)&0.04669(18) & 0.016252(69)&0.45659(11) \\\hline
   & $f_D$ & $f_{D^*}$ & $f^T_{D^*}$ & $f_{D_s}$ & $f_{D^*_s}$ & $f^T_{D_s^*}$ \\
\hline    
This work & 0.2108(20)(11) & 0.2292(26)(17) & 0.2068(27)(14) & 0.2470(21)(04) & 0.2691(30)(03) &0.2447(34)(06) \\
Literature & 0.2084(15)~\cite{Kuberski:2024pms} & 0.234(6)~\cite{Chen:2020qma} &0.192(14)~\cite{Li:2024vtx}  &0.2468(13)~\cite{Kuberski:2024pms} & 0.274(6)~\cite{HPQCD2014} & 0.252(10)~\cite{Li:2024vtx} \\
\hline 
   & $f_{\eta_c}$ & $f_{J/\psi}$ &  $f^T_{J/\psi}$ & $f_{\chi_{c0}}$ & $f_{\chi_{c1}}$ & $f_{h_{c}}$ \\
\hline    
This work & 0.3944(29)(18) &0.4114(34)(22)& 0.3893(32)(24)&0.314(17)(04)& 0.2103(79)(34) & 0.1465(73)(31)\\
Literature & 0.3975(10)~\cite{HPQCD2020} & 0.4096(16)~\cite{HPQCD2020}& 0.3911(56)~\cite{Li:2024vtx} \\
\end{tabular}  
\label{tab:result}
\end{table*}

Eventually, we just collect our predictions in Table. ~\ref{tab:result} for comparison. {Besides systematic uncertainties mentioned above, those from the experimental value of $m_{D_s}$ and perturbative matching are not negligible. For the details of the systematic uncertainties estimation and error budget, see Sec. D of Supplementary Material.} 

{Using the results of $f_D$ and $f_{D_s}$, one can extract the CKM matrix elements $|V_{cd}|$ and $|V_{cs}|$, respectively. Taking into account the experimental constraints on the products of decay constants and CKM matrix elements~\cite{ParticleDataGroup:2022pth}, as given below
\begin{eqnarray}\label{eq:CKM_exp}
    f_{D^+}|V_{cd}|&=&\ \ \! 45.8(1.1)~\textrm{MeV}, \nonumber \\
    f_{D_s^+}|V_{cs}|&=&243.5(2.7)~\textrm{MeV},
\end{eqnarray}
and combining with our results summarized in Table.~\ref{tab:result}, we finally obtain 
\begin{equation}
    |V_{cd}|=0.2173(21)_{\rm lat}(52)_{\rm ex}, \quad |V_{cs}|=0.986(08)_{\rm lat}(11)_{\rm ex},
\end{equation}
where the first errors come from the lattice results and the second result from the uncertainties from the experiments in Eq.~(\ref{eq:CKM_exp}). The above results are consistent with those quoted by the PDG, but with slightly larger errors. Using the present PDG value of $|V_{cb}|=0.041(1)$~\cite{ParticleDataGroup:2022pth} which is negligible comparing to the uncertainties of $|V_{cd(s)}|$, we predict,
\begin{equation}
    |V_{cd}|^2+|V_{cs}|^2+|V_{cb}|^2=1.021(17)(22),
\end{equation}
and verify the unity of the unitarity of the CKM matrix elements involving the charm quark.}

Based on previous CLQCD work~\cite{Hu:2023jet} on similar ensembles, one can extract $f_K/f_{\pi}=1.1907(76)_{\rm stat}(17)_{\rm sys}$, where the systematic uncertainty from the renormalization is largely canceled out. By integrating this result with the experimental constraint $\frac{|V_{us}|}{|V_{ud}|}\frac{f_{K}}{f_{\pi}}=0.27683(29)_{\rm exp}(20)_{\rm th}$~\cite{ParticleDataGroup:2022pth}. and the unitarity constraint is expressed as
\begin{equation}
1=|V_{ud}|^2+|V_{us}^2|+|V_{ub}|^2=|V_{ud}|^2+|V_{us}^2|+0.0035^2,
\end{equation}
we derive the values
\begin{equation}
|V_{ud}|=0.9740(03)_{\rm lat}(01)_{\rm ph},\ |V_{us}|=0.2265(14)_{\rm lat}(03)_{\rm ph}.
\end{equation}
Consequently, the CKM relations $|V_{ud}|=|V_{cs}|$ and $|V_{us}|=|V_{cd}|$ are substantiated.

{Our predictions of the quantities $f^{(T)}_{D^*}$ and $f^{(T)}_{D_s^*}$ are by far the most precise. Compared with the previous resutls~\cite{Chen:2020qma,HPQCD2014,Li:2024vtx}, our calculation provides much better control on the systematic uncertainties on the unphysical sea quark masses and discretization error, and then also by far the most accurate. Therefore, we can provide insight into the experimental search for the purely leptonic decay of $D^*$ and $D_s^*$ particles. Most recently, the BESIII collaboration has reported the first experimental measurements on the purely leptonic decays $D^*\rightarrow l\nu_l$~\cite{BESIII:2024kxe} and $D_s^*\rightarrow l\nu_l$~\cite{BESIII:2023zjq} with $l=e,\mu$. No significant signal is observed for the channel $D^*\rightarrow l\nu_l$, and only upper limits are set as $\operatorname{Br}(D^*\rightarrow e\nu_e)< 1.1\times 10^{-5}$ and $\operatorname{Br}(D^*\rightarrow \mu\nu_{\mu})<4.3\times 10^{-6}$. For $D_s^*\rightarrow l\nu_l$, the BESIII gives the branching fraction of this decay as $(2.1^{+1.2}_{-0.9_{\textrm{stat.}}}\pm 0.2_{\textrm{syst.}})\times 10^{-5}$. Theoretically, the decay widths of these pure leptonic decays can be parameterized by the $f_{D^*}$ and $f_{D_s^*}$ via the following way
\begin{equation}
    \Gamma=\frac{G_F^2}{12\pi}|V_{cd(s)}|f_{V}^2m_V^3\left(1-\frac{m_l^2}{m_V^2}\right)^2\left(1+\frac{m_l^2}{2m_V^2}\right),
\end{equation}
where the symbol $V$ denotes the vector mesons $D^*$ and $D_s^*$, respectively, and $m_l$ is the mass of lepton $l=e,\mu$. Using the most precise CKM matrix elements $V_{cd}=0.221(4),V_{cs}=0.975(7)$~\cite{FlavourLatticeAveragingGroupFLAG:2021npn} determined from the lattice calculation of the $D_{(s)}$ meson semileptonic decays and our predictions of $f_{D^*/D_s^*}$, we then find the decay width,
\begin{eqnarray}
    \Gamma(D^*\rightarrow l\nu_l)_{l=e,\mu}&=& 3.388(98)\times 10^{-7}~\mathrm{keV}, \nonumber \\
     \Gamma(D_s^*\rightarrow l\nu_l)_{l=e,\mu}&=& 2.432(83)\times 10^{-6}~\mathrm{keV}, 
\end{eqnarray}
The branching fractions of them can then be determined by combining with the total decay width $\Gamma_{\textrm{total}}(D^*)=83.4(1.8)~\textrm{keV}$~\cite{ParticleDataGroup:2022pth} and $\Gamma_{\textrm{total}}(D_s^*)=0.0587(54)~\textrm{keV}$~\cite{Meng:2024gpd}. The results are given as $\operatorname{Br}(D^*\rightarrow l\nu_l)=4.06(09)(12)\times 10^{-9}$ and $\operatorname{Br}(D_s^*\rightarrow l\nu_l)=4.08(38)(10)\times 10^{-5}$.
}

Our major results are summarized in Table~\ref{tab:result}, where the first error encompasses the statistical {uncertainties from the correlation function, lattice spacing determination, pion and $\eta_s$ masses in the sea, and also non-perturbative renormalization}. The second error includes the systematic uncertainty stemming from the fit parameterization, the FLAG scale parameter $w_0$, {finite volume correction, missing ${\cal O}(a\alpha_s)$ effects, and also} perturbative matching of the renormalization. The values from PDG~\cite{ParticleDataGroup:2022pth}, FLAG~\cite{FlavourLatticeAveragingGroupFLAG:2021npn}, HPQCD~\cite{HPQCD2020,HPQCD2014} and ALPHA collaborations~\cite{Kuberski:2024pms} are also listed there for comparison. We use the values of $f_{D_{(s)}}$~\cite{Bussone:2023kag,Kuberski:2024pms} from the ALPHA collaboration instead of the FLAG averages~\cite{FlavourLatticeAveragingGroupFLAG:2021npn,Davies:2010ip,FermilabLattice:2011njy,Na:2012iu,Yang:2014sea,Boyle:2017jwu} since they are slightly more precise. {RBC/UKQCD also obtained high-precision values of $f_{D_s}/f_D=1.1740(51)(68)$~\cite{Boyle:2018knm} and $m_c(m_c)=1.292(12)$ GeV (using the heavy-quark improved renormalization)~\cite{DelDebbio:2024hca} which are consistent with our determination $f_{D_s}/f_D=1.1712(85)(41)$ and $m_c(m_c)=1.2933(72)(95)$ GeV, but not included in the FLAG averages yet.} 

\section{Summary}\label{sec:summary}

In this work, we {tuned} the valence strange quark mass using fictitious $\eta_s$ mass~\cite{Borsanyi:2020mff} which corresponds to the physical strange quark mass and effectively suppress the impact of the {miss-tuning effect strange quark mass of the CLQCD ensembles}. We further {tune the valence} charm quark mass based on the {pure QCD} $m_{D_s}^{\rm QCD}$ for each ensemble, effectively reducing the uncertainty in our prediction stemming from lattice spacing determination, the scale parameter $w_0$, and discretization errors in the charm quark mass.

Comparing to the overlap fermion where $Z_V=Z_A$ only depends on $m_{\rm PS}^2a^2$ with a tiny coefficient $\sim 0.02$~\cite{Wang:2020nbf,He:2022lse}, {the $Z_V$ of} the clover fermion we used here has a much stronger $m_{\rm PS}^2a^2$ dependence with a coefficient $\sim 0.5$. It makes {the heavy quark improved $Z^c_V$ normalization to be essential here to suppress the discretization error of the charmed meosn matrix elements}.

{Eventually, we have succeeded in reducing the discretization error of both the charmed hadron mass and the decay constants of open charm hadrons to a few percent level at the coarsest lattice spacing of $a=0.105$ fm. Such a lattice spacing is fairly cost-effective for investigating the dependence of charmed hadrons on light and strange quarks with high precision. As evidence, the mass difference between $D^+$ and $D^0$ is predicted to be 5.4(5) MeV, incorporating QED corrections from the literature~\cite{DiCarlo:2019thl}, which aligns well with the experimental value of 4.8(1) MeV. We anticipate that the uncertainty in our prediction can be significantly reduced through direct QED correction calculations on the CLQCD ensemble using infinite volume reconstruction techniques~\cite{Feng:2018qpx}.}

We also verified the unitarily of the CKM matrix elements involving the charm quark at 3\% level, updated the constraints on the leptonic decay width of $D^*_{(s)}$, and provided the decay constants of the P-wave charmonium decay constants in the continuum with physical quark masses. 

{However, the hyperfine splittings such as $m_{\rm V}-m_{\rm PS}$ and $\Delta_{1P_{\rm spin-orbit},\bar{c}c}$ still have significant discretiation error, when the charm quark employs the identical discretized fermion action as the light quarks. This effect can hinder the efficient investigation at the coarse lattice spacing, for physics observables which are sensitive to these splittings. Introducing an additional $1+{\cal O}(a^2)$ parameter in the heavy quark action, as proposed in~\cite{Liu:2009jc}, could further reduce the discretization error of the hyperfine splitting, as indicated in~\cite{Brown:2014ena}, and warrants further exploration.}

In summary, we would expect the previous~\cite{Zhang:2021oja,Xing:2022ijm,Liu:2023feb,Liu:2023pwr,Liu:2022gxf,Meng:2024gpd,Han:2024min,Yan:2024yuq,Meng:2024nyo} and ongoing studies using the CLQCD ensembles for the charm physics can reach higher accuracy with the {tuned valence} strange and charm quark masses and also the heavy quark improved normalization scheme presented in this work. {More specifically, hopefully, the systematic uncertainties coming from the unphysical light and strange quark masses, {finite volume}, and finite lattice spacing can be controlled using the CLQCD ensembles, in the spectrum, form factors, semileptonic decay, and behaviors at the finite temperature and magnetic field of light and charmed meson and baryon.}

\section*{Acknowledgement}
We thank the CLQCD collaborations for providing us their gauge configurations with dynamical fermions~\cite{Hu:2023jet}, which are generated on HPC Cluster of ITP-CAS, the Southern Nuclear Science Computing Center(SNSC) and the Siyuan-1 cluster supported by the Center for High Performance Computing at Shanghai Jiao Tong University. 
The calculations were performed using the Chroma software suite~\cite{Edwards:2004sx} with QUDA~\cite{Clark:2009wm,Babich:2011np,Clark:2016rdz} through HIP programming model~\cite{Bi:2020wpt}. We also thank Christine Davies, Simon Kuberski, {and Tobias Tsang} for helpful discussion and comments. The numerical calculations were carried out on the ORISE Supercomputer, HPC Cluster of ITP-CAS, and Advanced Computing East China Sub-center. This work is supported in part by NSFC grants No. 12293060, 12293062, 12293061, 12293063, 12293064, 12293065, 12435002, 12047503, 12175279, 11935017, the science and education integration young faculty project of University of Chinese Academy of Sciences, the Strategic Priority Research Program of Chinese Academy of Sciences, Grant No.\ XDB34030303, XDB34030301  and YSBR-101, and also a NSFC-DFG joint grant under Grant No.\ 12061131006 and SCHA 458/22.

\bibliography{ref}

\begin{thebibliography}{68}%
\makeatletter
\providecommand \@ifxundefined [1]{%
 \@ifx{#1\undefined}
}%
\providecommand \@ifnum [1]{%
 \ifnum #1\expandafter \@firstoftwo
 \else \expandafter \@secondoftwo
 \fi
}%
\providecommand \@ifx [1]{%
 \ifx #1\expandafter \@firstoftwo
 \else \expandafter \@secondoftwo
 \fi
}%
\providecommand \natexlab [1]{#1}%
\providecommand \enquote  [1]{``#1''}%
\providecommand \bibnamefont  [1]{#1}%
\providecommand \bibfnamefont [1]{#1}%
\providecommand \citenamefont [1]{#1}%
\providecommand \href@noop [0]{\@secondoftwo}%
\providecommand \href [0]{\begingroup \@sanitize@url \@href}%
\providecommand \@href[1]{\@@startlink{#1}\@@href}%
\providecommand \@@href[1]{\endgroup#1\@@endlink}%
\providecommand \@sanitize@url [0]{\catcode `\\12\catcode `\$12\catcode
  `\&12\catcode `\#12\catcode `\^12\catcode `\_12\catcode `\%12\relax}%
\providecommand \@@startlink[1]{}%
\providecommand \@@endlink[0]{}%
\providecommand \url  [0]{\begingroup\@sanitize@url \@url }%
\providecommand \@url [1]{\endgroup\@href {#1}{\urlprefix }}%
\providecommand \urlprefix  [0]{URL }%
\providecommand \Eprint [0]{\href }%
\providecommand \doibase [0]{http://dx.doi.org/}%
\providecommand \selectlanguage [0]{\@gobble}%
\providecommand \bibinfo  [0]{\@secondoftwo}%
\providecommand \bibfield  [0]{\@secondoftwo}%
\providecommand \translation [1]{[#1]}%
\providecommand \BibitemOpen [0]{}%
\providecommand \bibitemStop [0]{}%
\providecommand \bibitemNoStop [0]{.\EOS\space}%
\providecommand \EOS [0]{\spacefactor3000\relax}%
\providecommand \BibitemShut  [1]{\csname bibitem#1\endcsname}%
\let\auto@bib@innerbib\@empty
\bibitem [{\citenamefont {Ablikim}\ \emph {et~al.}(2013)\citenamefont {Ablikim}
  \emph {et~al.}}]{BESIII:2013ris}%
  \BibitemOpen
  \bibfield  {author} {\bibinfo {author} {\bibfnamefont {M.}~\bibnamefont
  {Ablikim}} \emph {et~al.} (\bibinfo {collaboration} {BESIII}),\ }\href
  {\doibase 10.1103/PhysRevLett.110.252001} {\bibfield  {journal} {\bibinfo
  {journal} {Phys. Rev. Lett.}\ }\textbf {\bibinfo {volume} {110}},\ \bibinfo
  {pages} {252001} (\bibinfo {year} {2013})},\ \Eprint
  {http://arxiv.org/abs/1303.5949} {arXiv:1303.5949 [hep-ex]} \BibitemShut
  {NoStop}%
\bibitem [{\citenamefont {Liu}\ \emph {et~al.}(2013)\citenamefont {Liu} \emph
  {et~al.}}]{Belle:2013yex}%
  \BibitemOpen
  \bibfield  {author} {\bibinfo {author} {\bibfnamefont {Z.~Q.}\ \bibnamefont
  {Liu}} \emph {et~al.} (\bibinfo {collaboration} {Belle}),\ }\href {\doibase
  10.1103/PhysRevLett.110.252002} {\bibfield  {journal} {\bibinfo  {journal}
  {Phys. Rev. Lett.}\ }\textbf {\bibinfo {volume} {110}},\ \bibinfo {pages}
  {252002} (\bibinfo {year} {2013})},\ \bibinfo {note} {[Erratum:
  Phys.Rev.Lett. 111, 019901 (2013)]},\ \Eprint
  {http://arxiv.org/abs/1304.0121} {arXiv:1304.0121 [hep-ex]} \BibitemShut
  {NoStop}%
\bibitem [{\citenamefont {Wu}\ \emph {et~al.}(2010)\citenamefont {Wu},
  \citenamefont {Molina}, \citenamefont {Oset},\ and\ \citenamefont
  {Zou}}]{Wu:2010jy}%
  \BibitemOpen
  \bibfield  {author} {\bibinfo {author} {\bibfnamefont {J.-J.}\ \bibnamefont
  {Wu}}, \bibinfo {author} {\bibfnamefont {R.}~\bibnamefont {Molina}}, \bibinfo
  {author} {\bibfnamefont {E.}~\bibnamefont {Oset}}, \ and\ \bibinfo {author}
  {\bibfnamefont {B.~S.}\ \bibnamefont {Zou}},\ }\href {\doibase
  10.1103/PhysRevLett.105.232001} {\bibfield  {journal} {\bibinfo  {journal}
  {Phys. Rev. Lett.}\ }\textbf {\bibinfo {volume} {105}},\ \bibinfo {pages}
  {232001} (\bibinfo {year} {2010})},\ \Eprint {http://arxiv.org/abs/1007.0573}
  {arXiv:1007.0573 [nucl-th]} \BibitemShut {NoStop}%
\bibitem [{\citenamefont {Aaij}\ \emph {et~al.}(2015)\citenamefont {Aaij} \emph
  {et~al.}}]{LHCb:2015yax}%
  \BibitemOpen
  \bibfield  {author} {\bibinfo {author} {\bibfnamefont {R.}~\bibnamefont
  {Aaij}} \emph {et~al.} (\bibinfo {collaboration} {LHCb}),\ }\href {\doibase
  10.1103/PhysRevLett.115.072001} {\bibfield  {journal} {\bibinfo  {journal}
  {Phys. Rev. Lett.}\ }\textbf {\bibinfo {volume} {115}},\ \bibinfo {pages}
  {072001} (\bibinfo {year} {2015})},\ \Eprint
  {http://arxiv.org/abs/1507.03414} {arXiv:1507.03414 [hep-ex]} \BibitemShut
  {NoStop}%
\bibitem [{\citenamefont {Aaij}\ \emph {et~al.}(2019)\citenamefont {Aaij} \emph
  {et~al.}}]{LHCb:2019kea}%
  \BibitemOpen
  \bibfield  {author} {\bibinfo {author} {\bibfnamefont {R.}~\bibnamefont
  {Aaij}} \emph {et~al.} (\bibinfo {collaboration} {LHCb}),\ }\href {\doibase
  10.1103/PhysRevLett.122.222001} {\bibfield  {journal} {\bibinfo  {journal}
  {Phys. Rev. Lett.}\ }\textbf {\bibinfo {volume} {122}},\ \bibinfo {pages}
  {222001} (\bibinfo {year} {2019})},\ \Eprint
  {http://arxiv.org/abs/1904.03947} {arXiv:1904.03947 [hep-ex]} \BibitemShut
  {NoStop}%
\bibitem [{\citenamefont {Aoki}\ \emph {et~al.}(2022)\citenamefont {Aoki} \emph
  {et~al.}}]{FlavourLatticeAveragingGroupFLAG:2021npn}%
  \BibitemOpen
  \bibfield  {author} {\bibinfo {author} {\bibfnamefont {Y.}~\bibnamefont
  {Aoki}} \emph {et~al.} (\bibinfo {collaboration} {Flavour Lattice Averaging
  Group (FLAG)}),\ }\href {\doibase 10.1140/epjc/s10052-022-10536-1} {\bibfield
   {journal} {\bibinfo  {journal} {Eur. Phys. J. C}\ }\textbf {\bibinfo
  {volume} {82}},\ \bibinfo {pages} {869} (\bibinfo {year} {2022})},\ \Eprint
  {http://arxiv.org/abs/2111.09849} {arXiv:2111.09849 [hep-lat]} \BibitemShut
  {NoStop}%
\bibitem [{\citenamefont {Duane}\ \emph {et~al.}(1987)\citenamefont {Duane},
  \citenamefont {Kennedy}, \citenamefont {Pendleton},\ and\ \citenamefont
  {Roweth}}]{Duane:1987de}%
  \BibitemOpen
  \bibfield  {author} {\bibinfo {author} {\bibfnamefont {S.}~\bibnamefont
  {Duane}}, \bibinfo {author} {\bibfnamefont {A.~D.}\ \bibnamefont {Kennedy}},
  \bibinfo {author} {\bibfnamefont {B.~J.}\ \bibnamefont {Pendleton}}, \ and\
  \bibinfo {author} {\bibfnamefont {D.}~\bibnamefont {Roweth}},\ }\href
  {\doibase 10.1016/0370-2693(87)91197-X} {\bibfield  {journal} {\bibinfo
  {journal} {Phys. Lett. B}\ }\textbf {\bibinfo {volume} {195}},\ \bibinfo
  {pages} {216} (\bibinfo {year} {1987})}\BibitemShut {NoStop}%
\bibitem [{\citenamefont {Borsanyi}\ \emph {et~al.}(2012)\citenamefont
  {Borsanyi} \emph {et~al.}}]{BMW:2012hcm}%
  \BibitemOpen
  \bibfield  {author} {\bibinfo {author} {\bibfnamefont {S.}~\bibnamefont
  {Borsanyi}} \emph {et~al.} (\bibinfo {collaboration} {BMW}),\ }\href
  {\doibase 10.1007/JHEP09(2012)010} {\bibfield  {journal} {\bibinfo  {journal}
  {JHEP}\ }\textbf {\bibinfo {volume} {09}},\ \bibinfo {pages} {010} (\bibinfo
  {year} {2012})},\ \Eprint {http://arxiv.org/abs/1203.4469} {arXiv:1203.4469
  [hep-lat]} \BibitemShut {NoStop}%
\bibitem [{\citenamefont {Bazavov}\ \emph {et~al.}(2014)\citenamefont {Bazavov}
  \emph {et~al.}}]{HotQCD:2014kol}%
  \BibitemOpen
  \bibfield  {author} {\bibinfo {author} {\bibfnamefont {A.}~\bibnamefont
  {Bazavov}} \emph {et~al.} (\bibinfo {collaboration} {HotQCD}),\ }\href
  {\doibase 10.1103/PhysRevD.90.094503} {\bibfield  {journal} {\bibinfo
  {journal} {Phys. Rev. D}\ }\textbf {\bibinfo {volume} {90}},\ \bibinfo
  {pages} {094503} (\bibinfo {year} {2014})},\ \Eprint
  {http://arxiv.org/abs/1407.6387} {arXiv:1407.6387 [hep-lat]} \BibitemShut
  {NoStop}%
\bibitem [{\citenamefont {Blum}\ \emph {et~al.}(2016)\citenamefont {Blum} \emph
  {et~al.}}]{RBC:2014ntl}%
  \BibitemOpen
  \bibfield  {author} {\bibinfo {author} {\bibfnamefont {T.}~\bibnamefont
  {Blum}} \emph {et~al.} (\bibinfo {collaboration} {RBC, UKQCD}),\ }\href
  {\doibase 10.1103/PhysRevD.93.074505} {\bibfield  {journal} {\bibinfo
  {journal} {Phys. Rev. D}\ }\textbf {\bibinfo {volume} {93}},\ \bibinfo
  {pages} {074505} (\bibinfo {year} {2016})},\ \Eprint
  {http://arxiv.org/abs/1411.7017} {arXiv:1411.7017 [hep-lat]} \BibitemShut
  {NoStop}%
\bibitem [{\citenamefont {Hu}\ \emph {et~al.}(2024)\citenamefont {Hu} \emph
  {et~al.}}]{Hu:2023jet}%
  \BibitemOpen
  \bibfield  {author} {\bibinfo {author} {\bibfnamefont {Z.-C.}\ \bibnamefont
  {Hu}} \emph {et~al.} (\bibinfo {collaboration} {CLQCD}),\ }\href {\doibase
  10.1103/PhysRevD.109.054507} {\bibfield  {journal} {\bibinfo  {journal}
  {Phys. Rev. D}\ }\textbf {\bibinfo {volume} {109}},\ \bibinfo {pages}
  {054507} (\bibinfo {year} {2024})},\ \Eprint
  {http://arxiv.org/abs/2310.00814} {arXiv:2310.00814 [hep-lat]} \BibitemShut
  {NoStop}%
\bibitem [{\citenamefont {Borsanyi}\ \emph {et~al.}(2021)\citenamefont
  {Borsanyi} \emph {et~al.}}]{Borsanyi:2020mff}%
  \BibitemOpen
  \bibfield  {author} {\bibinfo {author} {\bibfnamefont {S.}~\bibnamefont
  {Borsanyi}} \emph {et~al.},\ }\href {\doibase 10.1038/s41586-021-03418-1}
  {\bibfield  {journal} {\bibinfo  {journal} {Nature}\ }\textbf {\bibinfo
  {volume} {593}},\ \bibinfo {pages} {51} (\bibinfo {year} {2021})},\ \Eprint
  {http://arxiv.org/abs/2002.12347} {arXiv:2002.12347 [hep-lat]} \BibitemShut
  {NoStop}%
\bibitem [{\citenamefont {Giusti}\ \emph {et~al.}(2017)\citenamefont {Giusti},
  \citenamefont {Lubicz}, \citenamefont {Tarantino}, \citenamefont
  {Martinelli}, \citenamefont {Sanfilippo}, \citenamefont {Simula},\ and\
  \citenamefont {Tantalo}}]{Giusti:2017dmp}%
  \BibitemOpen
  \bibfield  {author} {\bibinfo {author} {\bibfnamefont {D.}~\bibnamefont
  {Giusti}}, \bibinfo {author} {\bibfnamefont {V.}~\bibnamefont {Lubicz}},
  \bibinfo {author} {\bibfnamefont {C.}~\bibnamefont {Tarantino}}, \bibinfo
  {author} {\bibfnamefont {G.}~\bibnamefont {Martinelli}}, \bibinfo {author}
  {\bibfnamefont {F.}~\bibnamefont {Sanfilippo}}, \bibinfo {author}
  {\bibfnamefont {S.}~\bibnamefont {Simula}}, \ and\ \bibinfo {author}
  {\bibfnamefont {N.}~\bibnamefont {Tantalo}},\ }\href {\doibase
  10.1103/PhysRevD.95.114504} {\bibfield  {journal} {\bibinfo  {journal} {Phys.
  Rev. D}\ }\textbf {\bibinfo {volume} {95}},\ \bibinfo {pages} {114504}
  (\bibinfo {year} {2017})},\ \Eprint {http://arxiv.org/abs/1704.06561}
  {arXiv:1704.06561 [hep-lat]} \BibitemShut {NoStop}%
\bibitem [{\citenamefont {Di~Carlo}\ \emph {et~al.}(2019)\citenamefont
  {Di~Carlo}, \citenamefont {Giusti}, \citenamefont {Lubicz}, \citenamefont
  {Martinelli}, \citenamefont {Sachrajda}, \citenamefont {Sanfilippo},
  \citenamefont {Simula},\ and\ \citenamefont {Tantalo}}]{DiCarlo:2019thl}%
  \BibitemOpen
  \bibfield  {author} {\bibinfo {author} {\bibfnamefont {M.}~\bibnamefont
  {Di~Carlo}}, \bibinfo {author} {\bibfnamefont {D.}~\bibnamefont {Giusti}},
  \bibinfo {author} {\bibfnamefont {V.}~\bibnamefont {Lubicz}}, \bibinfo
  {author} {\bibfnamefont {G.}~\bibnamefont {Martinelli}}, \bibinfo {author}
  {\bibfnamefont {C.~T.}\ \bibnamefont {Sachrajda}}, \bibinfo {author}
  {\bibfnamefont {F.}~\bibnamefont {Sanfilippo}}, \bibinfo {author}
  {\bibfnamefont {S.}~\bibnamefont {Simula}}, \ and\ \bibinfo {author}
  {\bibfnamefont {N.}~\bibnamefont {Tantalo}},\ }\href {\doibase
  10.1103/PhysRevD.100.034514} {\bibfield  {journal} {\bibinfo  {journal}
  {Phys. Rev. D}\ }\textbf {\bibinfo {volume} {100}},\ \bibinfo {pages}
  {034514} (\bibinfo {year} {2019})},\ \Eprint
  {http://arxiv.org/abs/1904.08731} {arXiv:1904.08731 [hep-lat]} \BibitemShut
  {NoStop}%
\bibitem [{\citenamefont {Gasser}\ \emph {et~al.}(2003)\citenamefont {Gasser},
  \citenamefont {Rusetsky},\ and\ \citenamefont {Scimemi}}]{Gasser:2003hk}%
  \BibitemOpen
  \bibfield  {author} {\bibinfo {author} {\bibfnamefont {J.}~\bibnamefont
  {Gasser}}, \bibinfo {author} {\bibfnamefont {A.}~\bibnamefont {Rusetsky}}, \
  and\ \bibinfo {author} {\bibfnamefont {I.}~\bibnamefont {Scimemi}},\ }\href
  {\doibase 10.1140/epjc/s2003-01383-1} {\bibfield  {journal} {\bibinfo
  {journal} {Eur. Phys. J. C}\ }\textbf {\bibinfo {volume} {32}},\ \bibinfo
  {pages} {97} (\bibinfo {year} {2003})},\ \Eprint
  {http://arxiv.org/abs/hep-ph/0305260} {arXiv:hep-ph/0305260} \BibitemShut
  {NoStop}%
\bibitem [{\citenamefont {Ishikawa}\ \emph {et~al.}(2008)\citenamefont
  {Ishikawa} \emph {et~al.}}]{JLQCD:2007xff}%
  \BibitemOpen
  \bibfield  {author} {\bibinfo {author} {\bibfnamefont {T.}~\bibnamefont
  {Ishikawa}} \emph {et~al.} (\bibinfo {collaboration} {JLQCD}),\ }\href
  {\doibase 10.1103/PhysRevD.78.011502} {\bibfield  {journal} {\bibinfo
  {journal} {Phys. Rev. D}\ }\textbf {\bibinfo {volume} {78}},\ \bibinfo
  {pages} {011502} (\bibinfo {year} {2008})},\ \Eprint
  {http://arxiv.org/abs/0704.1937} {arXiv:0704.1937 [hep-lat]} \BibitemShut
  {NoStop}%
\bibitem [{\citenamefont {Martinelli}\ \emph {et~al.}(1995)\citenamefont
  {Martinelli}, \citenamefont {Pittori}, \citenamefont {Sachrajda},
  \citenamefont {Testa},\ and\ \citenamefont {Vladikas}}]{Martinelli:1994ty}%
  \BibitemOpen
  \bibfield  {author} {\bibinfo {author} {\bibfnamefont {G.}~\bibnamefont
  {Martinelli}}, \bibinfo {author} {\bibfnamefont {C.}~\bibnamefont {Pittori}},
  \bibinfo {author} {\bibfnamefont {C.~T.}\ \bibnamefont {Sachrajda}}, \bibinfo
  {author} {\bibfnamefont {M.}~\bibnamefont {Testa}}, \ and\ \bibinfo {author}
  {\bibfnamefont {A.}~\bibnamefont {Vladikas}},\ }\href {\doibase
  10.1016/0550-3213(95)00126-D} {\bibfield  {journal} {\bibinfo  {journal}
  {Nucl. Phys.}\ }\textbf {\bibinfo {volume} {B445}},\ \bibinfo {pages} {81}
  (\bibinfo {year} {1995})},\ \Eprint {http://arxiv.org/abs/hep-lat/9411010}
  {arXiv:hep-lat/9411010 [hep-lat]} \BibitemShut {NoStop}%
\bibitem [{\citenamefont {Zhang}\ \emph {et~al.}(2021)\citenamefont {Zhang},
  \citenamefont {Li}, \citenamefont {Huo}, \citenamefont {Sch\"afer},
  \citenamefont {Sun},\ and\ \citenamefont {Yang}}]{Zhang:2020rsx}%
  \BibitemOpen
  \bibfield  {author} {\bibinfo {author} {\bibfnamefont {K.}~\bibnamefont
  {Zhang}}, \bibinfo {author} {\bibfnamefont {Y.-Y.}\ \bibnamefont {Li}},
  \bibinfo {author} {\bibfnamefont {Y.-K.}\ \bibnamefont {Huo}}, \bibinfo
  {author} {\bibfnamefont {A.}~\bibnamefont {Sch\"afer}}, \bibinfo {author}
  {\bibfnamefont {P.}~\bibnamefont {Sun}}, \ and\ \bibinfo {author}
  {\bibfnamefont {Y.-B.}\ \bibnamefont {Yang}} (\bibinfo {collaboration}
  {\ensuremath{\chi}QCD}),\ }\href {\doibase 10.1103/PhysRevD.104.074501}
  {\bibfield  {journal} {\bibinfo  {journal} {Phys. Rev. D}\ }\textbf {\bibinfo
  {volume} {104}},\ \bibinfo {pages} {074501} (\bibinfo {year} {2021})},\
  \Eprint {http://arxiv.org/abs/2012.05448} {arXiv:2012.05448 [hep-lat]}
  \BibitemShut {NoStop}%
\bibitem [{\citenamefont {Capitani}\ \emph {et~al.}(2001)\citenamefont
  {Capitani}, \citenamefont {Gockeler}, \citenamefont {Horsley}, \citenamefont
  {Perlt}, \citenamefont {Rakow}, \citenamefont {Schierholz},\ and\
  \citenamefont {Schiller}}]{Capitani:2000xi}%
  \BibitemOpen
  \bibfield  {author} {\bibinfo {author} {\bibfnamefont {S.}~\bibnamefont
  {Capitani}}, \bibinfo {author} {\bibfnamefont {M.}~\bibnamefont {Gockeler}},
  \bibinfo {author} {\bibfnamefont {R.}~\bibnamefont {Horsley}}, \bibinfo
  {author} {\bibfnamefont {H.}~\bibnamefont {Perlt}}, \bibinfo {author}
  {\bibfnamefont {P.~E.~L.}\ \bibnamefont {Rakow}}, \bibinfo {author}
  {\bibfnamefont {G.}~\bibnamefont {Schierholz}}, \ and\ \bibinfo {author}
  {\bibfnamefont {A.}~\bibnamefont {Schiller}},\ }\href {\doibase
  10.1016/S0550-3213(00)00590-3} {\bibfield  {journal} {\bibinfo  {journal}
  {Nucl. Phys. B}\ }\textbf {\bibinfo {volume} {593}},\ \bibinfo {pages} {183}
  (\bibinfo {year} {2001})},\ \Eprint {http://arxiv.org/abs/hep-lat/0007004}
  {arXiv:hep-lat/0007004} \BibitemShut {NoStop}%
\bibitem [{\citenamefont {Hatton}\ \emph
  {et~al.}(2020{\natexlab{a}})\citenamefont {Hatton}, \citenamefont {Davies},
  \citenamefont {Galloway}, \citenamefont {Koponen}, \citenamefont {Lepage},\
  and\ \citenamefont {Lytle}}]{Hatton:2020qhk}%
  \BibitemOpen
  \bibfield  {author} {\bibinfo {author} {\bibfnamefont {D.}~\bibnamefont
  {Hatton}}, \bibinfo {author} {\bibfnamefont {C.~T.~H.}\ \bibnamefont
  {Davies}}, \bibinfo {author} {\bibfnamefont {B.}~\bibnamefont {Galloway}},
  \bibinfo {author} {\bibfnamefont {J.}~\bibnamefont {Koponen}}, \bibinfo
  {author} {\bibfnamefont {G.~P.}\ \bibnamefont {Lepage}}, \ and\ \bibinfo
  {author} {\bibfnamefont {A.~T.}\ \bibnamefont {Lytle}} (\bibinfo
  {collaboration} {HPQCD}),\ }\href {\doibase 10.1103/PhysRevD.102.054511}
  {\bibfield  {journal} {\bibinfo  {journal} {Phys. Rev. D}\ }\textbf {\bibinfo
  {volume} {102}},\ \bibinfo {pages} {054511} (\bibinfo {year}
  {2020}{\natexlab{a}})},\ \Eprint {http://arxiv.org/abs/2005.01845}
  {arXiv:2005.01845 [hep-lat]} \BibitemShut {NoStop}%
\bibitem [{\citenamefont {Workman}\ \emph {et~al.}(2022)\citenamefont {Workman}
  \emph {et~al.}}]{ParticleDataGroup:2022pth}%
  \BibitemOpen
  \bibfield  {author} {\bibinfo {author} {\bibfnamefont {R.~L.}\ \bibnamefont
  {Workman}} \emph {et~al.} (\bibinfo {collaboration} {Particle Data Group}),\
  }\href {\doibase 10.1093/ptep/ptac097} {\bibfield  {journal} {\bibinfo
  {journal} {PTEP}\ }\textbf {\bibinfo {volume} {2022}},\ \bibinfo {pages}
  {083C01} (\bibinfo {year} {2022})}\BibitemShut {NoStop}%
\bibitem [{\citenamefont {Zhang}\ \emph
  {et~al.}(2022{\natexlab{a}})\citenamefont {Zhang}, \citenamefont {Sun},
  \citenamefont {Chen}, \citenamefont {Ying Chen~and}, \citenamefont {Jiang},\
  and\ \citenamefont {Liu}}]{ZhangRenQiang:2021gnn}%
  \BibitemOpen
  \bibfield  {author} {\bibinfo {author} {\bibfnamefont {R.}~\bibnamefont
  {Zhang}}, \bibinfo {author} {\bibfnamefont {W.}~\bibnamefont {Sun}}, \bibinfo
  {author} {\bibfnamefont {F.}~\bibnamefont {Chen}}, \bibinfo {author}
  {\bibfnamefont {M.~G.}\ \bibnamefont {Ying Chen~and}}, \bibinfo {author}
  {\bibfnamefont {X.}~\bibnamefont {Jiang}}, \ and\ \bibinfo {author}
  {\bibfnamefont {Z.}~\bibnamefont {Liu}},\ }\href {\doibase
  10.1088/1674-1137/ac3d8c} {\bibfield  {journal} {\bibinfo  {journal} {Chin.
  Phys. C}\ }\textbf {\bibinfo {volume} {46}},\ \bibinfo {pages} {043102}
  (\bibinfo {year} {2022}{\natexlab{a}})},\ \Eprint
  {http://arxiv.org/abs/2110.01755} {arXiv:2110.01755 [hep-lat]} \BibitemShut
  {NoStop}%
\bibitem [{\citenamefont {Zhang}\ \emph
  {et~al.}(2022{\natexlab{b}})\citenamefont {Zhang}, \citenamefont {Sun},
  \citenamefont {Chen}, \citenamefont {Gong}, \citenamefont {Gui},\ and\
  \citenamefont {Liu}}]{Zhang:2021xvl}%
  \BibitemOpen
  \bibfield  {author} {\bibinfo {author} {\bibfnamefont {R.}~\bibnamefont
  {Zhang}}, \bibinfo {author} {\bibfnamefont {W.}~\bibnamefont {Sun}}, \bibinfo
  {author} {\bibfnamefont {Y.}~\bibnamefont {Chen}}, \bibinfo {author}
  {\bibfnamefont {M.}~\bibnamefont {Gong}}, \bibinfo {author} {\bibfnamefont
  {L.-C.}\ \bibnamefont {Gui}}, \ and\ \bibinfo {author} {\bibfnamefont
  {Z.}~\bibnamefont {Liu}},\ }\href {\doibase 10.1016/j.physletb.2022.136960}
  {\bibfield  {journal} {\bibinfo  {journal} {Phys. Lett. B}\ }\textbf
  {\bibinfo {volume} {827}},\ \bibinfo {pages} {136960} (\bibinfo {year}
  {2022}{\natexlab{b}})},\ \Eprint {http://arxiv.org/abs/2107.12749}
  {arXiv:2107.12749 [hep-lat]} \BibitemShut {NoStop}%
\bibitem [{\citenamefont {Kwong}\ \emph {et~al.}(1987)\citenamefont {Kwong},
  \citenamefont {Rosner},\ and\ \citenamefont {Quigg}}]{Kwong:1987mj}%
  \BibitemOpen
  \bibfield  {author} {\bibinfo {author} {\bibfnamefont {W.}~\bibnamefont
  {Kwong}}, \bibinfo {author} {\bibfnamefont {J.~L.}\ \bibnamefont {Rosner}}, \
  and\ \bibinfo {author} {\bibfnamefont {C.}~\bibnamefont {Quigg}},\ }\href
  {\doibase 10.1146/annurev.ns.37.120187.001545} {\bibfield  {journal}
  {\bibinfo  {journal} {Ann. Rev. Nucl. Part. Sci.}\ }\textbf {\bibinfo
  {volume} {37}},\ \bibinfo {pages} {325} (\bibinfo {year} {1987})}\BibitemShut
  {NoStop}%
\bibitem [{\citenamefont {Lucha}\ \emph {et~al.}(1991)\citenamefont {Lucha},
  \citenamefont {Schoberl},\ and\ \citenamefont {Gromes}}]{Lucha:1991vn}%
  \BibitemOpen
  \bibfield  {author} {\bibinfo {author} {\bibfnamefont {W.}~\bibnamefont
  {Lucha}}, \bibinfo {author} {\bibfnamefont {F.~F.}\ \bibnamefont {Schoberl}},
  \ and\ \bibinfo {author} {\bibfnamefont {D.}~\bibnamefont {Gromes}},\ }\href
  {\doibase 10.1016/0370-1573(91)90001-3} {\bibfield  {journal} {\bibinfo
  {journal} {Phys. Rept.}\ }\textbf {\bibinfo {volume} {200}},\ \bibinfo
  {pages} {127} (\bibinfo {year} {1991})}\BibitemShut {NoStop}%
\bibitem [{\citenamefont {Burch}\ \emph {et~al.}(2010)\citenamefont {Burch},
  \citenamefont {DeTar}, \citenamefont {Di~Pierro}, \citenamefont {El-Khadra},
  \citenamefont {Freeland}, \citenamefont {Gottlieb}, \citenamefont {Kronfeld},
  \citenamefont {Levkova}, \citenamefont {Mackenzie},\ and\ \citenamefont
  {Simone}}]{Burch:2009az}%
  \BibitemOpen
  \bibfield  {author} {\bibinfo {author} {\bibfnamefont {T.}~\bibnamefont
  {Burch}}, \bibinfo {author} {\bibfnamefont {C.}~\bibnamefont {DeTar}},
  \bibinfo {author} {\bibfnamefont {M.}~\bibnamefont {Di~Pierro}}, \bibinfo
  {author} {\bibfnamefont {A.~X.}\ \bibnamefont {El-Khadra}}, \bibinfo {author}
  {\bibfnamefont {E.~D.}\ \bibnamefont {Freeland}}, \bibinfo {author}
  {\bibfnamefont {S.}~\bibnamefont {Gottlieb}}, \bibinfo {author}
  {\bibfnamefont {A.~S.}\ \bibnamefont {Kronfeld}}, \bibinfo {author}
  {\bibfnamefont {L.}~\bibnamefont {Levkova}}, \bibinfo {author} {\bibfnamefont
  {P.~B.}\ \bibnamefont {Mackenzie}}, \ and\ \bibinfo {author} {\bibfnamefont
  {J.~N.}\ \bibnamefont {Simone}},\ }\href {\doibase
  10.1103/PhysRevD.81.034508} {\bibfield  {journal} {\bibinfo  {journal} {Phys.
  Rev. D}\ }\textbf {\bibinfo {volume} {81}},\ \bibinfo {pages} {034508}
  (\bibinfo {year} {2010})},\ \Eprint {http://arxiv.org/abs/0912.2701}
  {arXiv:0912.2701 [hep-lat]} \BibitemShut {NoStop}%
\bibitem [{\citenamefont {McNeile}\ \emph {et~al.}(2010)\citenamefont
  {McNeile}, \citenamefont {Davies}, \citenamefont {Follana}, \citenamefont
  {Hornbostel},\ and\ \citenamefont {Lepage}}]{McNeile:2010ji}%
  \BibitemOpen
  \bibfield  {author} {\bibinfo {author} {\bibfnamefont {C.}~\bibnamefont
  {McNeile}}, \bibinfo {author} {\bibfnamefont {C.~T.~H.}\ \bibnamefont
  {Davies}}, \bibinfo {author} {\bibfnamefont {E.}~\bibnamefont {Follana}},
  \bibinfo {author} {\bibfnamefont {K.}~\bibnamefont {Hornbostel}}, \ and\
  \bibinfo {author} {\bibfnamefont {G.~P.}\ \bibnamefont {Lepage}},\ }\href
  {\doibase 10.1103/PhysRevD.82.034512} {\bibfield  {journal} {\bibinfo
  {journal} {Phys. Rev. D}\ }\textbf {\bibinfo {volume} {82}},\ \bibinfo
  {pages} {034512} (\bibinfo {year} {2010})},\ \Eprint
  {http://arxiv.org/abs/1004.4285} {arXiv:1004.4285 [hep-lat]} \BibitemShut
  {NoStop}%
\bibitem [{\citenamefont {Yang}\ \emph {et~al.}(2015)\citenamefont {Yang} \emph
  {et~al.}}]{Yang:2014sea}%
  \BibitemOpen
  \bibfield  {author} {\bibinfo {author} {\bibfnamefont {Y.-B.}\ \bibnamefont
  {Yang}} \emph {et~al.},\ }\href {\doibase 10.1103/PhysRevD.92.034517}
  {\bibfield  {journal} {\bibinfo  {journal} {Phys. Rev. D}\ }\textbf {\bibinfo
  {volume} {92}},\ \bibinfo {pages} {034517} (\bibinfo {year} {2015})},\
  \Eprint {http://arxiv.org/abs/1410.3343} {arXiv:1410.3343 [hep-lat]}
  \BibitemShut {NoStop}%
\bibitem [{\citenamefont {Nakayama}\ \emph {et~al.}(2016)\citenamefont
  {Nakayama}, \citenamefont {Fahy},\ and\ \citenamefont
  {Hashimoto}}]{Nakayama:2016atf}%
  \BibitemOpen
  \bibfield  {author} {\bibinfo {author} {\bibfnamefont {K.}~\bibnamefont
  {Nakayama}}, \bibinfo {author} {\bibfnamefont {B.}~\bibnamefont {Fahy}}, \
  and\ \bibinfo {author} {\bibfnamefont {S.}~\bibnamefont {Hashimoto}},\ }\href
  {\doibase 10.1103/PhysRevD.94.054507} {\bibfield  {journal} {\bibinfo
  {journal} {Phys. Rev. D}\ }\textbf {\bibinfo {volume} {94}},\ \bibinfo
  {pages} {054507} (\bibinfo {year} {2016})},\ \Eprint
  {http://arxiv.org/abs/1606.01002} {arXiv:1606.01002 [hep-lat]} \BibitemShut
  {NoStop}%
\bibitem [{\citenamefont {Petreczky}\ and\ \citenamefont
  {Weber}(2019)}]{Petreczky:2019ozv}%
  \BibitemOpen
  \bibfield  {author} {\bibinfo {author} {\bibfnamefont {P.}~\bibnamefont
  {Petreczky}}\ and\ \bibinfo {author} {\bibfnamefont {J.~H.}\ \bibnamefont
  {Weber}},\ }\href {\doibase 10.1103/PhysRevD.100.034519} {\bibfield
  {journal} {\bibinfo  {journal} {Phys. Rev. D}\ }\textbf {\bibinfo {volume}
  {100}},\ \bibinfo {pages} {034519} (\bibinfo {year} {2019})},\ \Eprint
  {http://arxiv.org/abs/1901.06424} {arXiv:1901.06424 [hep-lat]} \BibitemShut
  {NoStop}%
\bibitem [{\citenamefont {Heitger}\ \emph {et~al.}(2021)\citenamefont
  {Heitger}, \citenamefont {Joswig},\ and\ \citenamefont
  {Kuberski}}]{Heitger:2021apz}%
  \BibitemOpen
  \bibfield  {author} {\bibinfo {author} {\bibfnamefont {J.}~\bibnamefont
  {Heitger}}, \bibinfo {author} {\bibfnamefont {F.}~\bibnamefont {Joswig}}, \
  and\ \bibinfo {author} {\bibfnamefont {S.}~\bibnamefont {Kuberski}} (\bibinfo
  {collaboration} {ALPHA}),\ }\href {\doibase 10.1007/JHEP05(2021)288}
  {\bibfield  {journal} {\bibinfo  {journal} {JHEP}\ }\textbf {\bibinfo
  {volume} {05}},\ \bibinfo {pages} {288} (\bibinfo {year} {2021})},\ \Eprint
  {http://arxiv.org/abs/2101.02694} {arXiv:2101.02694 [hep-lat]} \BibitemShut
  {NoStop}%
\bibitem [{\citenamefont {Hatton}\ \emph
  {et~al.}(2020{\natexlab{b}})\citenamefont {Hatton}, \citenamefont {Davies},
  \citenamefont {Galloway}, \citenamefont {Koponen}, \citenamefont {Lepage},\
  and\ \citenamefont {Lytle}}]{HPQCD2020}%
  \BibitemOpen
  \bibfield  {author} {\bibinfo {author} {\bibfnamefont {D.}~\bibnamefont
  {Hatton}}, \bibinfo {author} {\bibfnamefont {C.~T.~H.}\ \bibnamefont
  {Davies}}, \bibinfo {author} {\bibfnamefont {B.}~\bibnamefont {Galloway}},
  \bibinfo {author} {\bibfnamefont {J.}~\bibnamefont {Koponen}}, \bibinfo
  {author} {\bibfnamefont {G.~P.}\ \bibnamefont {Lepage}}, \ and\ \bibinfo
  {author} {\bibfnamefont {A.~T.}\ \bibnamefont {Lytle}} (\bibinfo
  {collaboration} {HPQCD}),\ }\href {\doibase 10.1103/PhysRevD.102.054511}
  {\bibfield  {journal} {\bibinfo  {journal} {Phys. Rev. D}\ }\textbf {\bibinfo
  {volume} {102}},\ \bibinfo {pages} {054511} (\bibinfo {year}
  {2020}{\natexlab{b}})},\ \Eprint {http://arxiv.org/abs/2005.01845}
  {arXiv:2005.01845 [hep-lat]} \BibitemShut {NoStop}%
\bibitem [{\citenamefont {Kuberski}\ \emph {et~al.}(2024)\citenamefont
  {Kuberski}, \citenamefont {Joswig}, \citenamefont {Collins}, \citenamefont
  {Heitger},\ and\ \citenamefont {S\"oldner}}]{Kuberski:2024pms}%
  \BibitemOpen
  \bibfield  {author} {\bibinfo {author} {\bibfnamefont {S.}~\bibnamefont
  {Kuberski}}, \bibinfo {author} {\bibfnamefont {F.}~\bibnamefont {Joswig}},
  \bibinfo {author} {\bibfnamefont {S.}~\bibnamefont {Collins}}, \bibinfo
  {author} {\bibfnamefont {J.}~\bibnamefont {Heitger}}, \ and\ \bibinfo
  {author} {\bibfnamefont {W.}~\bibnamefont {S\"oldner}} (\bibinfo
  {collaboration} {RQCD, ALPHA}),\ }\href {\doibase 10.1007/JHEP07(2024)090}
  {\bibfield  {journal} {\bibinfo  {journal} {JHEP}\ }\textbf {\bibinfo
  {volume} {07}},\ \bibinfo {pages} {090} (\bibinfo {year} {2024})},\ \Eprint
  {http://arxiv.org/abs/2405.04506} {arXiv:2405.04506 [hep-lat]} \BibitemShut
  {NoStop}%
\bibitem [{\citenamefont {Chen}\ \emph {et~al.}(2021)\citenamefont {Chen},
  \citenamefont {Wei-Feng}, \citenamefont {Gong}, \citenamefont {Liu},\ and\
  \citenamefont {Ma}}]{Chen:2020qma}%
  \BibitemOpen
  \bibfield  {author} {\bibinfo {author} {\bibfnamefont {Y.}~\bibnamefont
  {Chen}}, \bibinfo {author} {\bibfnamefont {C.}~\bibnamefont {Wei-Feng}},
  \bibinfo {author} {\bibfnamefont {M.}~\bibnamefont {Gong}}, \bibinfo {author}
  {\bibfnamefont {Z.}~\bibnamefont {Liu}}, \ and\ \bibinfo {author}
  {\bibfnamefont {Y.}~\bibnamefont {Ma}},\ }\href {\doibase
  10.1088/1674-1137/abcd8f} {\bibfield  {journal} {\bibinfo  {journal} {Chin.
  Phys. C}\ }\textbf {\bibinfo {volume} {45}},\ \bibinfo {pages} {023109}
  (\bibinfo {year} {2021})},\ \Eprint {http://arxiv.org/abs/2008.05208}
  {arXiv:2008.05208 [hep-lat]} \BibitemShut {NoStop}%
\bibitem [{\citenamefont {Li}\ \emph {et~al.}(2024)\citenamefont {Li},
  \citenamefont {Chen}, \citenamefont {Gong}, \citenamefont {Liu},
  \citenamefont {Liu},\ and\ \citenamefont {Wang}}]{Li:2024vtx}%
  \BibitemOpen
  \bibfield  {author} {\bibinfo {author} {\bibfnamefont {D.}~\bibnamefont
  {Li}}, \bibinfo {author} {\bibfnamefont {Y.}~\bibnamefont {Chen}}, \bibinfo
  {author} {\bibfnamefont {M.}~\bibnamefont {Gong}}, \bibinfo {author}
  {\bibfnamefont {K.-F.}\ \bibnamefont {Liu}}, \bibinfo {author} {\bibfnamefont
  {Z.}~\bibnamefont {Liu}}, \ and\ \bibinfo {author} {\bibfnamefont
  {T.}~\bibnamefont {Wang}},\ }\href@noop {} {\  (\bibinfo {year} {2024})},\
  \Eprint {http://arxiv.org/abs/2407.03697} {arXiv:2407.03697 [hep-lat]}
  \BibitemShut {NoStop}%
\bibitem [{\citenamefont {Donald}\ \emph {et~al.}(2014)\citenamefont {Donald},
  \citenamefont {Davies}, \citenamefont {Koponen},\ and\ \citenamefont
  {Lepage}}]{HPQCD2014}%
  \BibitemOpen
  \bibfield  {author} {\bibinfo {author} {\bibfnamefont {G.~C.}\ \bibnamefont
  {Donald}}, \bibinfo {author} {\bibfnamefont {C.~T.~H.}\ \bibnamefont
  {Davies}}, \bibinfo {author} {\bibfnamefont {J.}~\bibnamefont {Koponen}}, \
  and\ \bibinfo {author} {\bibfnamefont {G.~P.}\ \bibnamefont {Lepage}}
  (\bibinfo {collaboration} {HPQCD}),\ }\href {\doibase
  10.1103/PhysRevLett.112.212002} {\bibfield  {journal} {\bibinfo  {journal}
  {Phys. Rev. Lett.}\ }\textbf {\bibinfo {volume} {112}},\ \bibinfo {pages}
  {212002} (\bibinfo {year} {2014})},\ \Eprint {http://arxiv.org/abs/1312.5264}
  {arXiv:1312.5264 [hep-lat]} \BibitemShut {NoStop}%
\bibitem [{\citenamefont {Ablikim}\ \emph {et~al.}(2024)\citenamefont {Ablikim}
  \emph {et~al.}}]{BESIII:2024kxe}%
  \BibitemOpen
  \bibfield  {author} {\bibinfo {author} {\bibfnamefont {M.}~\bibnamefont
  {Ablikim}} \emph {et~al.} (\bibinfo {collaboration} {BESIII}),\ }\href
  {\doibase 10.1103/PhysRevD.110.012003} {\bibfield  {journal} {\bibinfo
  {journal} {Phys. Rev. D}\ }\textbf {\bibinfo {volume} {110}},\ \bibinfo
  {pages} {012003} (\bibinfo {year} {2024})},\ \Eprint
  {http://arxiv.org/abs/2405.09066} {arXiv:2405.09066 [hep-ex]} \BibitemShut
  {NoStop}%
\bibitem [{\citenamefont {Ablikim}\ \emph {et~al.}(2023)\citenamefont {Ablikim}
  \emph {et~al.}}]{BESIII:2023zjq}%
  \BibitemOpen
  \bibfield  {author} {\bibinfo {author} {\bibfnamefont {M.}~\bibnamefont
  {Ablikim}} \emph {et~al.} (\bibinfo {collaboration} {BESIII}),\ }\href
  {\doibase 10.1103/PhysRevLett.131.141802} {\bibfield  {journal} {\bibinfo
  {journal} {Phys. Rev. Lett.}\ }\textbf {\bibinfo {volume} {131}},\ \bibinfo
  {pages} {141802} (\bibinfo {year} {2023})},\ \Eprint
  {http://arxiv.org/abs/2304.12159} {arXiv:2304.12159 [hep-ex]} \BibitemShut
  {NoStop}%
\bibitem [{\citenamefont {Meng}\ \emph
  {et~al.}(2024{\natexlab{a}})\citenamefont {Meng}, \citenamefont {Dang},
  \citenamefont {Liu}, \citenamefont {Liu}, \citenamefont {Shen}, \citenamefont
  {Yan},\ and\ \citenamefont {Zhang}}]{Meng:2024gpd}%
  \BibitemOpen
  \bibfield  {author} {\bibinfo {author} {\bibfnamefont {Y.}~\bibnamefont
  {Meng}}, \bibinfo {author} {\bibfnamefont {J.-L.}\ \bibnamefont {Dang}},
  \bibinfo {author} {\bibfnamefont {C.}~\bibnamefont {Liu}}, \bibinfo {author}
  {\bibfnamefont {Z.}~\bibnamefont {Liu}}, \bibinfo {author} {\bibfnamefont
  {T.}~\bibnamefont {Shen}}, \bibinfo {author} {\bibfnamefont {H.}~\bibnamefont
  {Yan}}, \ and\ \bibinfo {author} {\bibfnamefont {K.-L.}\ \bibnamefont
  {Zhang}},\ }\href {\doibase 10.1103/PhysRevD.109.074511} {\bibfield
  {journal} {\bibinfo  {journal} {Phys. Rev. D}\ }\textbf {\bibinfo {volume}
  {109}},\ \bibinfo {pages} {074511} (\bibinfo {year} {2024}{\natexlab{a}})},\
  \Eprint {http://arxiv.org/abs/2401.13475} {arXiv:2401.13475 [hep-lat]}
  \BibitemShut {NoStop}%
\bibitem [{\citenamefont {Bussone}\ \emph {et~al.}(2024)\citenamefont
  {Bussone}, \citenamefont {Conigli}, \citenamefont {Frison}, \citenamefont
  {Herdo\'\i{}za}, \citenamefont {Pena}, \citenamefont {Preti}, \citenamefont
  {S\'aez},\ and\ \citenamefont {Ugarrio}}]{Bussone:2023kag}%
  \BibitemOpen
  \bibfield  {author} {\bibinfo {author} {\bibfnamefont {A.}~\bibnamefont
  {Bussone}}, \bibinfo {author} {\bibfnamefont {A.}~\bibnamefont {Conigli}},
  \bibinfo {author} {\bibfnamefont {J.}~\bibnamefont {Frison}}, \bibinfo
  {author} {\bibfnamefont {G.}~\bibnamefont {Herdo\'\i{}za}}, \bibinfo {author}
  {\bibfnamefont {C.}~\bibnamefont {Pena}}, \bibinfo {author} {\bibfnamefont
  {D.}~\bibnamefont {Preti}}, \bibinfo {author} {\bibfnamefont
  {A.}~\bibnamefont {S\'aez}}, \ and\ \bibinfo {author} {\bibfnamefont
  {J.}~\bibnamefont {Ugarrio}} (\bibinfo {collaboration} {Alpha}),\ }\href
  {\doibase 10.1140/epjc/s10052-024-12816-4} {\bibfield  {journal} {\bibinfo
  {journal} {Eur. Phys. J. C}\ }\textbf {\bibinfo {volume} {84}},\ \bibinfo
  {pages} {506} (\bibinfo {year} {2024})},\ \Eprint
  {http://arxiv.org/abs/2309.14154} {arXiv:2309.14154 [hep-lat]} \BibitemShut
  {NoStop}%
\bibitem [{\citenamefont {Davies}\ \emph {et~al.}(2010)\citenamefont {Davies},
  \citenamefont {McNeile}, \citenamefont {Follana}, \citenamefont {Lepage},
  \citenamefont {Na},\ and\ \citenamefont {Shigemitsu}}]{Davies:2010ip}%
  \BibitemOpen
  \bibfield  {author} {\bibinfo {author} {\bibfnamefont {C.~T.~H.}\
  \bibnamefont {Davies}}, \bibinfo {author} {\bibfnamefont {C.}~\bibnamefont
  {McNeile}}, \bibinfo {author} {\bibfnamefont {E.}~\bibnamefont {Follana}},
  \bibinfo {author} {\bibfnamefont {G.~P.}\ \bibnamefont {Lepage}}, \bibinfo
  {author} {\bibfnamefont {H.}~\bibnamefont {Na}}, \ and\ \bibinfo {author}
  {\bibfnamefont {J.}~\bibnamefont {Shigemitsu}},\ }\href {\doibase
  10.1103/PhysRevD.82.114504} {\bibfield  {journal} {\bibinfo  {journal} {Phys.
  Rev. D}\ }\textbf {\bibinfo {volume} {82}},\ \bibinfo {pages} {114504}
  (\bibinfo {year} {2010})},\ \Eprint {http://arxiv.org/abs/1008.4018}
  {arXiv:1008.4018 [hep-lat]} \BibitemShut {NoStop}%
\bibitem [{\citenamefont {Bazavov}\ \emph {et~al.}(2012)\citenamefont {Bazavov}
  \emph {et~al.}}]{FermilabLattice:2011njy}%
  \BibitemOpen
  \bibfield  {author} {\bibinfo {author} {\bibfnamefont {A.}~\bibnamefont
  {Bazavov}} \emph {et~al.} (\bibinfo {collaboration} {Fermilab Lattice,
  MILC}),\ }\href {\doibase 10.1103/PhysRevD.85.114506} {\bibfield  {journal}
  {\bibinfo  {journal} {Phys. Rev. D}\ }\textbf {\bibinfo {volume} {85}},\
  \bibinfo {pages} {114506} (\bibinfo {year} {2012})},\ \Eprint
  {http://arxiv.org/abs/1112.3051} {arXiv:1112.3051 [hep-lat]} \BibitemShut
  {NoStop}%
\bibitem [{\citenamefont {Na}\ \emph {et~al.}(2012)\citenamefont {Na},
  \citenamefont {Davies}, \citenamefont {Follana}, \citenamefont {Lepage},\
  and\ \citenamefont {Shigemitsu}}]{Na:2012iu}%
  \BibitemOpen
  \bibfield  {author} {\bibinfo {author} {\bibfnamefont {H.}~\bibnamefont
  {Na}}, \bibinfo {author} {\bibfnamefont {C.~T.~H.}\ \bibnamefont {Davies}},
  \bibinfo {author} {\bibfnamefont {E.}~\bibnamefont {Follana}}, \bibinfo
  {author} {\bibfnamefont {G.~P.}\ \bibnamefont {Lepage}}, \ and\ \bibinfo
  {author} {\bibfnamefont {J.}~\bibnamefont {Shigemitsu}},\ }\href {\doibase
  10.1103/PhysRevD.86.054510} {\bibfield  {journal} {\bibinfo  {journal} {Phys.
  Rev. D}\ }\textbf {\bibinfo {volume} {86}},\ \bibinfo {pages} {054510}
  (\bibinfo {year} {2012})},\ \Eprint {http://arxiv.org/abs/1206.4936}
  {arXiv:1206.4936 [hep-lat]} \BibitemShut {NoStop}%
\bibitem [{\citenamefont {Boyle}\ \emph {et~al.}(2017)\citenamefont {Boyle},
  \citenamefont {Del~Debbio}, \citenamefont {J\"uttner}, \citenamefont
  {Khamseh}, \citenamefont {Sanfilippo},\ and\ \citenamefont
  {Tsang}}]{Boyle:2017jwu}%
  \BibitemOpen
  \bibfield  {author} {\bibinfo {author} {\bibfnamefont {P.~A.}\ \bibnamefont
  {Boyle}}, \bibinfo {author} {\bibfnamefont {L.}~\bibnamefont {Del~Debbio}},
  \bibinfo {author} {\bibfnamefont {A.}~\bibnamefont {J\"uttner}}, \bibinfo
  {author} {\bibfnamefont {A.}~\bibnamefont {Khamseh}}, \bibinfo {author}
  {\bibfnamefont {F.}~\bibnamefont {Sanfilippo}}, \ and\ \bibinfo {author}
  {\bibfnamefont {J.~T.}\ \bibnamefont {Tsang}},\ }\href {\doibase
  10.1007/JHEP12(2017)008} {\bibfield  {journal} {\bibinfo  {journal} {JHEP}\
  }\textbf {\bibinfo {volume} {12}},\ \bibinfo {pages} {008} (\bibinfo {year}
  {2017})},\ \Eprint {http://arxiv.org/abs/1701.02644} {arXiv:1701.02644
  [hep-lat]} \BibitemShut {NoStop}%
\bibitem [{\citenamefont {Boyle}\ \emph {et~al.}(2018)\citenamefont {Boyle},
  \citenamefont {Del~Debbio}, \citenamefont {Garron}, \citenamefont {Juttner},
  \citenamefont {Soni}, \citenamefont {Tsang},\ and\ \citenamefont
  {Witzel}}]{Boyle:2018knm}%
  \BibitemOpen
  \bibfield  {author} {\bibinfo {author} {\bibfnamefont {P.~A.}\ \bibnamefont
  {Boyle}}, \bibinfo {author} {\bibfnamefont {L.}~\bibnamefont {Del~Debbio}},
  \bibinfo {author} {\bibfnamefont {N.}~\bibnamefont {Garron}}, \bibinfo
  {author} {\bibfnamefont {A.}~\bibnamefont {Juttner}}, \bibinfo {author}
  {\bibfnamefont {A.}~\bibnamefont {Soni}}, \bibinfo {author} {\bibfnamefont
  {J.~T.}\ \bibnamefont {Tsang}}, \ and\ \bibinfo {author} {\bibfnamefont
  {O.}~\bibnamefont {Witzel}} (\bibinfo {collaboration} {RBC/UKQCD}),\
  }\href@noop {} {\  (\bibinfo {year} {2018})},\ \Eprint
  {http://arxiv.org/abs/1812.08791} {arXiv:1812.08791 [hep-lat]} \BibitemShut
  {NoStop}%
\bibitem [{\citenamefont {Del~Debbio}\ \emph {et~al.}(2024)\citenamefont
  {Del~Debbio}, \citenamefont {Erben}, \citenamefont {Flynn}, \citenamefont
  {Mukherjee},\ and\ \citenamefont {Tsang}}]{DelDebbio:2024hca}%
  \BibitemOpen
  \bibfield  {author} {\bibinfo {author} {\bibfnamefont {L.}~\bibnamefont
  {Del~Debbio}}, \bibinfo {author} {\bibfnamefont {F.}~\bibnamefont {Erben}},
  \bibinfo {author} {\bibfnamefont {J.~M.}\ \bibnamefont {Flynn}}, \bibinfo
  {author} {\bibfnamefont {R.}~\bibnamefont {Mukherjee}}, \ and\ \bibinfo
  {author} {\bibfnamefont {J.~T.}\ \bibnamefont {Tsang}} (\bibinfo
  {collaboration} {RBC, UKQCD}),\ }\href {\doibase 10.1103/PhysRevD.110.054512}
  {\bibfield  {journal} {\bibinfo  {journal} {Phys. Rev. D}\ }\textbf {\bibinfo
  {volume} {110}},\ \bibinfo {pages} {054512} (\bibinfo {year} {2024})},\
  \Eprint {http://arxiv.org/abs/2407.18700} {arXiv:2407.18700 [hep-lat]}
  \BibitemShut {NoStop}%
\bibitem [{\citenamefont {Wang}\ \emph {et~al.}(2021)\citenamefont {Wang},
  \citenamefont {Liang}, \citenamefont {Draper}, \citenamefont {Liu},\ and\
  \citenamefont {Yang}}]{Wang:2020nbf}%
  \BibitemOpen
  \bibfield  {author} {\bibinfo {author} {\bibfnamefont {G.}~\bibnamefont
  {Wang}}, \bibinfo {author} {\bibfnamefont {J.}~\bibnamefont {Liang}},
  \bibinfo {author} {\bibfnamefont {T.}~\bibnamefont {Draper}}, \bibinfo
  {author} {\bibfnamefont {K.-F.}\ \bibnamefont {Liu}}, \ and\ \bibinfo
  {author} {\bibfnamefont {Y.-B.}\ \bibnamefont {Yang}} (\bibinfo
  {collaboration} {chiQCD}),\ }\href {\doibase 10.1103/PhysRevD.104.074502}
  {\bibfield  {journal} {\bibinfo  {journal} {Phys. Rev. D}\ }\textbf {\bibinfo
  {volume} {104}},\ \bibinfo {pages} {074502} (\bibinfo {year} {2021})},\
  \Eprint {http://arxiv.org/abs/2006.05431} {arXiv:2006.05431 [hep-ph]}
  \BibitemShut {NoStop}%
\bibitem [{\citenamefont {He}\ \emph {et~al.}(2022)\citenamefont {He},
  \citenamefont {Bi}, \citenamefont {Draper}, \citenamefont {Liu},
  \citenamefont {Liu},\ and\ \citenamefont {Yang}}]{He:2022lse}%
  \BibitemOpen
  \bibfield  {author} {\bibinfo {author} {\bibfnamefont {F.}~\bibnamefont
  {He}}, \bibinfo {author} {\bibfnamefont {Y.-J.}\ \bibnamefont {Bi}}, \bibinfo
  {author} {\bibfnamefont {T.}~\bibnamefont {Draper}}, \bibinfo {author}
  {\bibfnamefont {K.-F.}\ \bibnamefont {Liu}}, \bibinfo {author} {\bibfnamefont
  {Z.}~\bibnamefont {Liu}}, \ and\ \bibinfo {author} {\bibfnamefont {Y.-B.}\
  \bibnamefont {Yang}} (\bibinfo {collaboration} {\ensuremath{\chi}QCD}),\
  }\href {\doibase 10.1103/PhysRevD.106.114506} {\bibfield  {journal} {\bibinfo
   {journal} {Phys. Rev. D}\ }\textbf {\bibinfo {volume} {106}},\ \bibinfo
  {pages} {114506} (\bibinfo {year} {2022})},\ \Eprint
  {http://arxiv.org/abs/2204.09246} {arXiv:2204.09246 [hep-lat]} \BibitemShut
  {NoStop}%
\bibitem [{\citenamefont {Feng}\ and\ \citenamefont
  {Jin}(2019)}]{Feng:2018qpx}%
  \BibitemOpen
  \bibfield  {author} {\bibinfo {author} {\bibfnamefont {X.}~\bibnamefont
  {Feng}}\ and\ \bibinfo {author} {\bibfnamefont {L.}~\bibnamefont {Jin}},\
  }\href {\doibase 10.1103/PhysRevD.100.094509} {\bibfield  {journal} {\bibinfo
   {journal} {Phys. Rev. D}\ }\textbf {\bibinfo {volume} {100}},\ \bibinfo
  {pages} {094509} (\bibinfo {year} {2019})},\ \Eprint
  {http://arxiv.org/abs/1812.09817} {arXiv:1812.09817 [hep-lat]} \BibitemShut
  {NoStop}%
\bibitem [{\citenamefont {Liu}\ \emph {et~al.}(2010)\citenamefont {Liu},
  \citenamefont {Lin}, \citenamefont {Orginos},\ and\ \citenamefont
  {Walker-Loud}}]{Liu:2009jc}%
  \BibitemOpen
  \bibfield  {author} {\bibinfo {author} {\bibfnamefont {L.}~\bibnamefont
  {Liu}}, \bibinfo {author} {\bibfnamefont {H.-W.}\ \bibnamefont {Lin}},
  \bibinfo {author} {\bibfnamefont {K.}~\bibnamefont {Orginos}}, \ and\
  \bibinfo {author} {\bibfnamefont {A.}~\bibnamefont {Walker-Loud}},\ }\href
  {\doibase 10.1103/PhysRevD.81.094505} {\bibfield  {journal} {\bibinfo
  {journal} {Phys. Rev. D}\ }\textbf {\bibinfo {volume} {81}},\ \bibinfo
  {pages} {094505} (\bibinfo {year} {2010})},\ \Eprint
  {http://arxiv.org/abs/0909.3294} {arXiv:0909.3294 [hep-lat]} \BibitemShut
  {NoStop}%
\bibitem [{\citenamefont {Brown}\ \emph {et~al.}(2014)\citenamefont {Brown},
  \citenamefont {Detmold}, \citenamefont {Meinel},\ and\ \citenamefont
  {Orginos}}]{Brown:2014ena}%
  \BibitemOpen
  \bibfield  {author} {\bibinfo {author} {\bibfnamefont {Z.~S.}\ \bibnamefont
  {Brown}}, \bibinfo {author} {\bibfnamefont {W.}~\bibnamefont {Detmold}},
  \bibinfo {author} {\bibfnamefont {S.}~\bibnamefont {Meinel}}, \ and\ \bibinfo
  {author} {\bibfnamefont {K.}~\bibnamefont {Orginos}},\ }\href {\doibase
  10.1103/PhysRevD.90.094507} {\bibfield  {journal} {\bibinfo  {journal} {Phys.
  Rev. D}\ }\textbf {\bibinfo {volume} {90}},\ \bibinfo {pages} {094507}
  (\bibinfo {year} {2014})},\ \Eprint {http://arxiv.org/abs/1409.0497}
  {arXiv:1409.0497 [hep-lat]} \BibitemShut {NoStop}%
\bibitem [{\citenamefont {Zhang}\ \emph
  {et~al.}(2022{\natexlab{c}})\citenamefont {Zhang} \emph
  {et~al.}}]{Zhang:2021oja}%
  \BibitemOpen
  \bibfield  {author} {\bibinfo {author} {\bibfnamefont {Q.-A.}\ \bibnamefont
  {Zhang}} \emph {et~al.},\ }\href {\doibase 10.1088/1674-1137/ac2b12}
  {\bibfield  {journal} {\bibinfo  {journal} {Chin. Phys. C}\ }\textbf
  {\bibinfo {volume} {46}},\ \bibinfo {pages} {011002} (\bibinfo {year}
  {2022}{\natexlab{c}})},\ \Eprint {http://arxiv.org/abs/2103.07064}
  {arXiv:2103.07064 [hep-lat]} \BibitemShut {NoStop}%
\bibitem [{\citenamefont {Xing}\ \emph {et~al.}(2022)\citenamefont {Xing},
  \citenamefont {Liang}, \citenamefont {Liu}, \citenamefont {Sun},\ and\
  \citenamefont {Yang}}]{Xing:2022ijm}%
  \BibitemOpen
  \bibfield  {author} {\bibinfo {author} {\bibfnamefont {H.}~\bibnamefont
  {Xing}}, \bibinfo {author} {\bibfnamefont {J.}~\bibnamefont {Liang}},
  \bibinfo {author} {\bibfnamefont {L.}~\bibnamefont {Liu}}, \bibinfo {author}
  {\bibfnamefont {P.}~\bibnamefont {Sun}}, \ and\ \bibinfo {author}
  {\bibfnamefont {Y.-B.}\ \bibnamefont {Yang}},\ }\href@noop {} {\  (\bibinfo
  {year} {2022})},\ \Eprint {http://arxiv.org/abs/2210.08555} {arXiv:2210.08555
  [hep-lat]} \BibitemShut {NoStop}%
\bibitem [{\citenamefont {Liu}\ \emph {et~al.}(2023)\citenamefont {Liu},
  \citenamefont {Liu}, \citenamefont {Sun}, \citenamefont {Sun}, \citenamefont
  {Tan}, \citenamefont {Wang}, \citenamefont {Yang},\ and\ \citenamefont
  {Zhang}}]{Liu:2023feb}%
  \BibitemOpen
  \bibfield  {author} {\bibinfo {author} {\bibfnamefont {H.}~\bibnamefont
  {Liu}}, \bibinfo {author} {\bibfnamefont {L.}~\bibnamefont {Liu}}, \bibinfo
  {author} {\bibfnamefont {P.}~\bibnamefont {Sun}}, \bibinfo {author}
  {\bibfnamefont {W.}~\bibnamefont {Sun}}, \bibinfo {author} {\bibfnamefont
  {J.-X.}\ \bibnamefont {Tan}}, \bibinfo {author} {\bibfnamefont
  {W.}~\bibnamefont {Wang}}, \bibinfo {author} {\bibfnamefont {Y.-B.}\
  \bibnamefont {Yang}}, \ and\ \bibinfo {author} {\bibfnamefont {Q.-A.}\
  \bibnamefont {Zhang}},\ }\href {\doibase 10.1016/j.physletb.2023.137941}
  {\bibfield  {journal} {\bibinfo  {journal} {Phys. Lett. B}\ }\textbf
  {\bibinfo {volume} {841}},\ \bibinfo {pages} {137941} (\bibinfo {year}
  {2023})},\ \Eprint {http://arxiv.org/abs/2303.17865} {arXiv:2303.17865
  [hep-lat]} \BibitemShut {NoStop}%
\bibitem [{\citenamefont {Liu}\ \emph {et~al.}(2024{\natexlab{a}})\citenamefont
  {Liu}, \citenamefont {Wang},\ and\ \citenamefont {Zhang}}]{Liu:2023pwr}%
  \BibitemOpen
  \bibfield  {author} {\bibinfo {author} {\bibfnamefont {H.}~\bibnamefont
  {Liu}}, \bibinfo {author} {\bibfnamefont {W.}~\bibnamefont {Wang}}, \ and\
  \bibinfo {author} {\bibfnamefont {Q.-A.}\ \bibnamefont {Zhang}},\ }\href
  {\doibase 10.1103/PhysRevD.109.036037} {\bibfield  {journal} {\bibinfo
  {journal} {Phys. Rev. D}\ }\textbf {\bibinfo {volume} {109}},\ \bibinfo
  {pages} {036037} (\bibinfo {year} {2024}{\natexlab{a}})},\ \Eprint
  {http://arxiv.org/abs/2309.05432} {arXiv:2309.05432 [hep-ph]} \BibitemShut
  {NoStop}%
\bibitem [{\citenamefont {Liu}\ \emph {et~al.}(2024{\natexlab{b}})\citenamefont
  {Liu}, \citenamefont {He}, \citenamefont {Liu}, \citenamefont {Sun},
  \citenamefont {Wang}, \citenamefont {Yang},\ and\ \citenamefont
  {Zhang}}]{Liu:2022gxf}%
  \BibitemOpen
  \bibfield  {author} {\bibinfo {author} {\bibfnamefont {H.}~\bibnamefont
  {Liu}}, \bibinfo {author} {\bibfnamefont {J.}~\bibnamefont {He}}, \bibinfo
  {author} {\bibfnamefont {L.}~\bibnamefont {Liu}}, \bibinfo {author}
  {\bibfnamefont {P.}~\bibnamefont {Sun}}, \bibinfo {author} {\bibfnamefont
  {W.}~\bibnamefont {Wang}}, \bibinfo {author} {\bibfnamefont {Y.-B.}\
  \bibnamefont {Yang}}, \ and\ \bibinfo {author} {\bibfnamefont {Q.-A.}\
  \bibnamefont {Zhang}},\ }\href {\doibase 10.1007/s11433-023-2205-0}
  {\bibfield  {journal} {\bibinfo  {journal} {Sci. China Phys. Mech. Astron.}\
  }\textbf {\bibinfo {volume} {67}},\ \bibinfo {pages} {211011} (\bibinfo
  {year} {2024}{\natexlab{b}})},\ \Eprint {http://arxiv.org/abs/2207.00183}
  {arXiv:2207.00183 [hep-lat]} \BibitemShut {NoStop}%
\bibitem [{\citenamefont {Han}\ \emph {et~al.}(2024)\citenamefont {Han},
  \citenamefont {Hua}, \citenamefont {Ji}, \citenamefont {L\"u}, \citenamefont
  {Wang}, \citenamefont {Xu}, \citenamefont {Zhang},\ and\ \citenamefont
  {Zhao}}]{Han:2024min}%
  \BibitemOpen
  \bibfield  {author} {\bibinfo {author} {\bibfnamefont {X.-Y.}\ \bibnamefont
  {Han}}, \bibinfo {author} {\bibfnamefont {J.}~\bibnamefont {Hua}}, \bibinfo
  {author} {\bibfnamefont {X.}~\bibnamefont {Ji}}, \bibinfo {author}
  {\bibfnamefont {C.-D.}\ \bibnamefont {L\"u}}, \bibinfo {author}
  {\bibfnamefont {W.}~\bibnamefont {Wang}}, \bibinfo {author} {\bibfnamefont
  {J.}~\bibnamefont {Xu}}, \bibinfo {author} {\bibfnamefont {Q.-A.}\
  \bibnamefont {Zhang}}, \ and\ \bibinfo {author} {\bibfnamefont
  {S.}~\bibnamefont {Zhao}},\ }\href@noop {} {\  (\bibinfo {year} {2024})},\
  \Eprint {http://arxiv.org/abs/2403.17492} {arXiv:2403.17492 [hep-ph]}
  \BibitemShut {NoStop}%
\bibitem [{\citenamefont {Yan}\ \emph {et~al.}(2024)\citenamefont {Yan},
  \citenamefont {Liu}, \citenamefont {Liu}, \citenamefont {Meng},\ and\
  \citenamefont {Xing}}]{Yan:2024yuq}%
  \BibitemOpen
  \bibfield  {author} {\bibinfo {author} {\bibfnamefont {H.}~\bibnamefont
  {Yan}}, \bibinfo {author} {\bibfnamefont {C.}~\bibnamefont {Liu}}, \bibinfo
  {author} {\bibfnamefont {L.}~\bibnamefont {Liu}}, \bibinfo {author}
  {\bibfnamefont {Y.}~\bibnamefont {Meng}}, \ and\ \bibinfo {author}
  {\bibfnamefont {H.}~\bibnamefont {Xing}},\ }\href@noop {} {\  (\bibinfo
  {year} {2024})},\ \Eprint {http://arxiv.org/abs/2404.13479} {arXiv:2404.13479
  [hep-lat]} \BibitemShut {NoStop}%
\bibitem [{\citenamefont {Meng}\ \emph
  {et~al.}(2024{\natexlab{b}})\citenamefont {Meng}, \citenamefont {Dang},
  \citenamefont {Liu}, \citenamefont {Tuo}, \citenamefont {Yan}, \citenamefont
  {Yang},\ and\ \citenamefont {Zhang}}]{Meng:2024nyo}%
  \BibitemOpen
  \bibfield  {author} {\bibinfo {author} {\bibfnamefont {Y.}~\bibnamefont
  {Meng}}, \bibinfo {author} {\bibfnamefont {J.-L.}\ \bibnamefont {Dang}},
  \bibinfo {author} {\bibfnamefont {C.}~\bibnamefont {Liu}}, \bibinfo {author}
  {\bibfnamefont {X.-Y.}\ \bibnamefont {Tuo}}, \bibinfo {author} {\bibfnamefont
  {H.}~\bibnamefont {Yan}}, \bibinfo {author} {\bibfnamefont {Y.-B.}\
  \bibnamefont {Yang}}, \ and\ \bibinfo {author} {\bibfnamefont {K.-L.}\
  \bibnamefont {Zhang}},\ }\href {\doibase 10.1103/PhysRevD.110.074510}
  {\bibfield  {journal} {\bibinfo  {journal} {Phys. Rev. D}\ }\textbf {\bibinfo
  {volume} {110}},\ \bibinfo {pages} {074510} (\bibinfo {year}
  {2024}{\natexlab{b}})},\ \Eprint {http://arxiv.org/abs/2407.13568}
  {arXiv:2407.13568 [hep-lat]} \BibitemShut {NoStop}%
\bibitem [{\citenamefont {Edwards}\ and\ \citenamefont
  {Joo}(2005)}]{Edwards:2004sx}%
  \BibitemOpen
  \bibfield  {author} {\bibinfo {author} {\bibfnamefont {R.~G.}\ \bibnamefont
  {Edwards}}\ and\ \bibinfo {author} {\bibfnamefont {B.}~\bibnamefont {Joo}}
  (\bibinfo {collaboration} {SciDAC, LHPC, UKQCD}),\ }\bibfield  {booktitle}
  {\emph {\bibinfo {booktitle} {{Lattice field theory. Proceedings, 22nd
  International Symposium, Lattice 2004, Batavia, USA, June 21-26, 2004}}},\
  }\href {\doibase 10.1016/j.nuclphysbps.2004.11.254} {\bibfield  {journal}
  {\bibinfo  {journal} {Nucl. Phys. Proc. Suppl.}\ }\textbf {\bibinfo {volume}
  {140}},\ \bibinfo {pages} {832} (\bibinfo {year} {2005})},\ \bibinfo {note}
  {[,832(2004)]},\ \Eprint {http://arxiv.org/abs/hep-lat/0409003}
  {arXiv:hep-lat/0409003 [hep-lat]} \BibitemShut {NoStop}%
\bibitem [{\citenamefont {Clark}\ \emph {et~al.}(2010)\citenamefont {Clark},
  \citenamefont {Babich}, \citenamefont {Barros}, \citenamefont {Brower},\ and\
  \citenamefont {Rebbi}}]{Clark:2009wm}%
  \BibitemOpen
  \bibfield  {author} {\bibinfo {author} {\bibfnamefont {M.~A.}\ \bibnamefont
  {Clark}}, \bibinfo {author} {\bibfnamefont {R.}~\bibnamefont {Babich}},
  \bibinfo {author} {\bibfnamefont {K.}~\bibnamefont {Barros}}, \bibinfo
  {author} {\bibfnamefont {R.~C.}\ \bibnamefont {Brower}}, \ and\ \bibinfo
  {author} {\bibfnamefont {C.}~\bibnamefont {Rebbi}},\ }\href {\doibase
  10.1016/j.cpc.2010.05.002} {\bibfield  {journal} {\bibinfo  {journal}
  {Comput. Phys. Commun.}\ }\textbf {\bibinfo {volume} {181}},\ \bibinfo
  {pages} {1517} (\bibinfo {year} {2010})},\ \Eprint
  {http://arxiv.org/abs/0911.3191} {arXiv:0911.3191 [hep-lat]} \BibitemShut
  {NoStop}%
\bibitem [{\citenamefont {Babich}\ \emph {et~al.}(2011)\citenamefont {Babich},
  \citenamefont {Clark}, \citenamefont {Joo}, \citenamefont {Shi},
  \citenamefont {Brower},\ and\ \citenamefont {Gottlieb}}]{Babich:2011np}%
  \BibitemOpen
  \bibfield  {author} {\bibinfo {author} {\bibfnamefont {R.}~\bibnamefont
  {Babich}}, \bibinfo {author} {\bibfnamefont {M.~A.}\ \bibnamefont {Clark}},
  \bibinfo {author} {\bibfnamefont {B.}~\bibnamefont {Joo}}, \bibinfo {author}
  {\bibfnamefont {G.}~\bibnamefont {Shi}}, \bibinfo {author} {\bibfnamefont
  {R.~C.}\ \bibnamefont {Brower}}, \ and\ \bibinfo {author} {\bibfnamefont
  {S.}~\bibnamefont {Gottlieb}},\ }in\ \href {\doibase 10.1145/2063384.2063478}
  {\emph {\bibinfo {booktitle} {{SC11 International Conference for High
  Performance Computing, Networking, Storage and Analysis Seattle, Washington,
  November 12-18, 2011}}}}\ (\bibinfo {year} {2011})\ \Eprint
  {http://arxiv.org/abs/1109.2935} {arXiv:1109.2935 [hep-lat]} \BibitemShut
  {NoStop}%
\bibitem [{\citenamefont {Clark}\ \emph {et~al.}(2016)\citenamefont {Clark},
  \citenamefont {Jo}, \citenamefont {Strelchenko}, \citenamefont {Cheng},
  \citenamefont {Gambhir},\ and\ \citenamefont {Brower}}]{Clark:2016rdz}%
  \BibitemOpen
  \bibfield  {author} {\bibinfo {author} {\bibfnamefont {M.~A.}\ \bibnamefont
  {Clark}}, \bibinfo {author} {\bibfnamefont {B.}~\bibnamefont {Jo}}, \bibinfo
  {author} {\bibfnamefont {A.}~\bibnamefont {Strelchenko}}, \bibinfo {author}
  {\bibfnamefont {M.}~\bibnamefont {Cheng}}, \bibinfo {author} {\bibfnamefont
  {A.}~\bibnamefont {Gambhir}}, \ and\ \bibinfo {author} {\bibfnamefont
  {R.}~\bibnamefont {Brower}},\ }\href@noop {} {\  (\bibinfo {year} {2016})},\
  \Eprint {http://arxiv.org/abs/1612.07873} {arXiv:1612.07873 [hep-lat]}
  \BibitemShut {NoStop}%
\bibitem [{\citenamefont {Bi}\ \emph {et~al.}(2020)\citenamefont {Bi},
  \citenamefont {Xiao}, \citenamefont {Gong}, \citenamefont {Guo},
  \citenamefont {Sun}, \citenamefont {Xu},\ and\ \citenamefont
  {Yang}}]{Bi:2020wpt}%
  \BibitemOpen
  \bibfield  {author} {\bibinfo {author} {\bibfnamefont {Y.-J.}\ \bibnamefont
  {Bi}}, \bibinfo {author} {\bibfnamefont {Y.}~\bibnamefont {Xiao}}, \bibinfo
  {author} {\bibfnamefont {M.}~\bibnamefont {Gong}}, \bibinfo {author}
  {\bibfnamefont {W.-Y.}\ \bibnamefont {Guo}}, \bibinfo {author} {\bibfnamefont
  {P.}~\bibnamefont {Sun}}, \bibinfo {author} {\bibfnamefont {S.}~\bibnamefont
  {Xu}}, \ and\ \bibinfo {author} {\bibfnamefont {Y.-B.}\ \bibnamefont
  {Yang}},\ }\bibfield  {booktitle} {\emph {\bibinfo {booktitle} {{Proceedings,
  37th International Symposium on Lattice Field Theory (Lattice 2019): Wuhan,
  China, June 16-22 2019}}},\ }\href {\doibase 10.22323/1.363.0286} {\bibfield
  {journal} {\bibinfo  {journal} {PoS}\ }\textbf {\bibinfo {volume}
  {LATTICE2019}},\ \bibinfo {pages} {286} (\bibinfo {year} {2020})},\ \Eprint
  {http://arxiv.org/abs/2001.05706} {arXiv:2001.05706 [hep-lat]} \BibitemShut
  {NoStop}%
\bibitem [{\citenamefont {Horsley}\ \emph {et~al.}(2008)\citenamefont
  {Horsley}, \citenamefont {Perlt}, \citenamefont {Rakow}, \citenamefont
  {Schierholz},\ and\ \citenamefont {Schiller}}]{Horsley:2008ap}%
  \BibitemOpen
  \bibfield  {author} {\bibinfo {author} {\bibfnamefont {R.}~\bibnamefont
  {Horsley}}, \bibinfo {author} {\bibfnamefont {H.}~\bibnamefont {Perlt}},
  \bibinfo {author} {\bibfnamefont {P.~E.~L.}\ \bibnamefont {Rakow}}, \bibinfo
  {author} {\bibfnamefont {G.}~\bibnamefont {Schierholz}}, \ and\ \bibinfo
  {author} {\bibfnamefont {A.}~\bibnamefont {Schiller}},\ }\href {\doibase
  10.1103/PhysRevD.78.054504} {\bibfield  {journal} {\bibinfo  {journal} {Phys.
  Rev. D}\ }\textbf {\bibinfo {volume} {78}},\ \bibinfo {pages} {054504}
  (\bibinfo {year} {2008})},\ \Eprint {http://arxiv.org/abs/0807.0345}
  {arXiv:0807.0345 [hep-lat]} \BibitemShut {NoStop}%
\bibitem [{\citenamefont {Cundy}\ \emph {et~al.}(2009)\citenamefont {Cundy}
  \emph {et~al.}}]{Cundy:2009yy}%
  \BibitemOpen
  \bibfield  {author} {\bibinfo {author} {\bibfnamefont {N.}~\bibnamefont
  {Cundy}} \emph {et~al.},\ }\href {\doibase 10.1103/PhysRevD.79.094507}
  {\bibfield  {journal} {\bibinfo  {journal} {Phys. Rev. D}\ }\textbf {\bibinfo
  {volume} {79}},\ \bibinfo {pages} {094507} (\bibinfo {year} {2009})},\
  \Eprint {http://arxiv.org/abs/0901.3302} {arXiv:0901.3302 [hep-lat]}
  \BibitemShut {NoStop}%
\bibitem [{\citenamefont {L\"uscher}(2010)}]{Luscher:2010iy}%
  \BibitemOpen
  \bibfield  {author} {\bibinfo {author} {\bibfnamefont {M.}~\bibnamefont
  {L\"uscher}},\ }\href {\doibase 10.1007/JHEP08(2010)071} {\bibfield
  {journal} {\bibinfo  {journal} {JHEP}\ }\textbf {\bibinfo {volume} {08}},\
  \bibinfo {pages} {071} (\bibinfo {year} {2010})},\ \bibinfo {note} {[Erratum:
  JHEP 03, 092 (2014)]},\ \Eprint {http://arxiv.org/abs/1006.4518}
  {arXiv:1006.4518 [hep-lat]} \BibitemShut {NoStop}%
\bibitem [{\citenamefont {Liu}\ \emph {et~al.}(2014)\citenamefont {Liu},
  \citenamefont {Chen}, \citenamefont {Dong}, \citenamefont {Glatzmaier},
  \citenamefont {Gong}, \citenamefont {Li}, \citenamefont {Liu}, \citenamefont
  {Yang},\ and\ \citenamefont {Zhang}}]{Liu:2013yxz}%
  \BibitemOpen
  \bibfield  {author} {\bibinfo {author} {\bibfnamefont {Z.}~\bibnamefont
  {Liu}}, \bibinfo {author} {\bibfnamefont {Y.}~\bibnamefont {Chen}}, \bibinfo
  {author} {\bibfnamefont {S.-J.}\ \bibnamefont {Dong}}, \bibinfo {author}
  {\bibfnamefont {M.}~\bibnamefont {Glatzmaier}}, \bibinfo {author}
  {\bibfnamefont {M.}~\bibnamefont {Gong}}, \bibinfo {author} {\bibfnamefont
  {A.}~\bibnamefont {Li}}, \bibinfo {author} {\bibfnamefont {K.-F.}\
  \bibnamefont {Liu}}, \bibinfo {author} {\bibfnamefont {Y.-B.}\ \bibnamefont
  {Yang}}, \ and\ \bibinfo {author} {\bibfnamefont {J.-B.}\ \bibnamefont
  {Zhang}} (\bibinfo {collaboration} {chiQCD}),\ }\href {\doibase
  10.1103/PhysRevD.90.034505} {\bibfield  {journal} {\bibinfo  {journal} {Phys.
  Rev. D}\ }\textbf {\bibinfo {volume} {90}},\ \bibinfo {pages} {034505}
  (\bibinfo {year} {2014})},\ \Eprint {http://arxiv.org/abs/1312.7628}
  {arXiv:1312.7628 [hep-lat]} \BibitemShut {NoStop}%
\end{thebibliography}%

\clearpage

\begin{widetext}

\section*{Supplemental Materials}

\subsection{{Detailed information of CLQCD ensembles and the} autocorrelation}

\begin{table*}[ht!]                   
\caption{Gauge coupling $\hat{\beta}=10/(g^2u_0^4)$, lattice spacing $a$ with the second error from that of the scale parameter $w_0$, tadpole improvement factors $u_0$ and $v_0$, {critical bare quark mass parameters $\tilde{m}^{\mathrm{crti}}$ which drives the pion mass towards zero}, dimensionless bare quark mass parameters $\tilde{m}^{\rm b}_{l,s}$, Lattice size $\tilde{L}^3\times \tilde{T}$, corresponding pion mass $m_{\pi}$ and $\eta_s$ mass $m_{\eta_s}$, {valence }bare strange and charm quark mass parameters $\tilde{m}^{\rm v}_{s,c}$.}  
\resizebox{1.0\columnwidth}{!}{
\begin{tabular}{|c c c | c c c c c | c  c c  | c  c | } 
\hline
Symbol & $\hat{\beta}$& $a$ (fm)  & $u_0$ & $v_0$ & $\tilde{m}^{\mathrm{crti}}$ & $\tilde{m}^b_{l}$ & $\tilde{m}^b_s$ & $\tilde{L}^3\times \tilde{T}$ & $m_{\pi}$ (MeV) & $m_{\eta_s}$ (MeV) & $\tilde{m}^{\rm v}_s$ & $\tilde{m}^{\rm v}_c$ \\
\hline 
C24P34 &6.200 & 0.10524(05)(62) & 0.855453& 0.951479& $-$0.28854(7) &$-$0.2770& $-$0.2310 & $24^3\times 64$ &340.2(1.7) &748.61(75) & $-$0.2396(1) & 0.4072(07)\\ 
C24P29 && &  0.855453 &0.951479 & $-$0.28557(4)&$-$0.2770 &$-$0.2400 &   $24^3\times 72$ &292.3(1.0) & 657.83(64) &$-$0.2356(1) & 0.4159(07)  \\
C32P29 && &  0.855453  &0.951479 & $-$0.28570(3)&$-$0.2770 &$-$0.2400 &   $32^3\times 64$ & 293.1(0.8) &  658.80(43)&$-$0.2358(1) & 0.4150(06)\\
C32P23 && &  0.855520 & 0.951545 & $-$0.28432(4)&$-$0.2790 &$-$0.2400 &    $32^3\times 64$ &227.9(1.2) & 643.93(45) &$-$0.2337(1) & 0.4190(07)\\
C48P23 & & & 0.855520 &0.951545 &  $-$0.28422(5)&$-$0.2790 &$-$0.2400 &   $48^3\times 96$ &224.1(1.2) & 644.08(62) &$-$0.2338(1) & 0.4196(08)\\
C48P14 & & & 0.855548 & 0.951570 & $-$0.28436(2)&$-$0.2825 &$-$0.2310 &   $48^3\times 96$ &136.4(1.7) &706.55(39)  &$-$0.2335(1) & 0.4205(07) \\
\hline
E28P35  & 6.308 & 0.08973(20)(53) & 0.859646&0.954385 & $-$0.25871(5)&$-$0.2490 & $-$0.2170& $28^3\times 64$ &351.4(1.4) & 717.94(93) &$-$0.2201(3)  & 0.2823(25)\\
\hline
F32P30  & 6.410 & 0.07753(03)(45) & 0.863437& 0.956942 & $-$0.23539(3)&$-$0.2295 &$-$0.2050 &  $32^3\times 96$ &300.4(1.2) & 675.98(97) & $-$0.2038(1)& 0.1974(05) \\
F48P30 & & &  0.863473 & 0.956984 & $-$0.23547(2)&$-$0.2295 &$-$0.2050 &   $48^3\times 96$  & 302.7(0.9) &674.76(58) & $-$0.2037(1)& 0.1965(04)\\
F32P21 & & &  0.863488 &0.957017 & $-$0.23477(3)& $-$0.2320&$-$0.2050 &   $32^3\times 64$ &210.3(2.3)  & 658.79(94)& $-$0.2023(1)& 0.1996(04)\\
F48P21 & & &0.863499 & 0.957006& $-$0.23478(2)&$-$0.2320 &$-$0.2050 & $48^3\times 96$ &207.5(1.1)  & 661.94(64)& $-$0.2025(1)& 0.1997(04)\\
\hline
G36P29 & 6.498 & 0.06887(12)(41)&0.866476 &0.958910  & $-$0.21981(2)&$-$0.2150 &$-$0.1926 & $36^3\times 108$ &297.2(0.9)  & 693.05(46)&$-$0.1928(1) & 0.1433(12) \\
\hline
H48P32 & 6.720 & 0.05199(08)(31)&0.873378 & 0.963137 & $-$0.18884(4)& $-$0.1850& $-$0.1700& $48^3\times 144$ &316.6(1.0)  & 691.88(65)& $-$0.1701(1)& 0.0551(07) \\
\hline
\end{tabular}  
}
\label{tab:ensemv}
\end{table*}

The results in this work, are based on the 2+1 flavor 
ensembles from the CLQCD collaboration using the tadpole improved tree level Symanzik (TITLS) gauge action and the tadpole improved tree level Clover (TITLC) fermion action. 

{The TITLS gauge action, denoted as $S_g$, is defined in the following,
\bal\label{eq:gauge_action}
S_g=\frac{1}{N_c}\mathrm{Re}\sum_{x,\mu<\nu}\mathrm{Tr}\big[1- \hat{\beta} \big(\mathcal{P}^U_{\mu,\nu}(x)+\frac{c_1\mathcal{R}^U_{\mu,\nu}(x)}{1-8c_1^0}\big)],
\eal
where $N_c=3$, \mbox{$\mathcal{P}^{U}_{\mu,\nu}(x)= U_\mu(x)U_\nu(x+a\hat{\mu})U^{\dagger}_\mu(x+a\hat{\nu})U^{\dagger}_\nu(x)$}, $\mathcal{R}^{U}_{\mu,\nu}(x)= U_\mu(x)U_\mu(x+a\hat{\mu})U_\nu(x+2a\hat{\mu})U^{\dagger}_\mu(x+a\hat{\mu}+a\hat{\nu})U^{\dagger}_\mu(x+a\hat{\nu})U^{\dagger}_\nu(x)$, $U_{\mu}(x)=P[\mathrm{exp}({\rm i}g_0\int_{x+\hat{\mu}a}^x \mathrm{d}y A_{\mu}(y))]$, $\hat{\beta}=(1-8c_1^0)\frac{6}{g_0^2u_0^4}\equiv 10/(g_0^2u_0^4)$ with $c_1^0=-\frac{1}{12}$, $c_1=\frac{c_1^0}{ u_0^2}$, $u_0=\langle \frac{\mathrm{Re}\mathrm{Tr}\sum_{x,\mu<\nu}\mathcal{P}^{U}_{\mu\nu}(x)}{6N_c\tilde{V}} \rangle^{1/4}$ is the tadpole improvement factor, $\tilde{V}=\tilde{L}^3\times \tilde{T}$ is the dimensionless 4-D volume of the lattice, and we use $\tilde{O}$ for the dimensionless value of any quantity $O$.

The TITLC fermion action uses 1-step stout smeared link $V$ with smearing parameter $\rho=0.125$,
\bal\label{eq:quark_action}
&S_q(m)=\sum_{x,\mu=1,...,4,\eta=\pm}\bar{\psi}(x+\eta\hat{\mu}a)\frac{1-\eta\gamma_{\mu}}{2}V_{\eta\mu}(x)\psi(x)\nonumber\\ 
&\quad +\sum_x\bar{\psi}(x)\big[-(4+ma)\delta_{y,x})+c_{\rm sw}\sigma^{\mu\nu}g_0F^V_{\mu\nu}\big]\psi(x),
\eal
where $c_{\rm sw}=\frac{1}{v^3_{0}}$ with $v_0=\langle \frac{\mathrm{Re}\mathrm{Tr}\sum_{x,\mu<\nu}\mathcal{P}^{V}_{\mu\nu}(x)}{6N_c\tilde{V}} \rangle^{1/4}$, and 
\begin{eqnarray}
F^V_{\mu\nu} &=& \frac{i}{8a^2g_0} (\mathcal{P}^V_{\mu,\nu}-\mathcal{P}^V_{\nu,\mu}+\mathcal{P}^V_{\nu,-\mu}-\mathcal{P}^V_{-\mu,\nu} \nonumber \\
&& + \mathcal{P}^V_{-\mu,-\nu} -\mathcal{P}^V_{-\nu,-\mu} + \mathcal{P}^V_{-\nu,\mu} -\mathcal{P}^V_{\mu,-\nu}).
\end{eqnarray}
Note that the TITLC action here uses the smeared gauge field $V$ for the clover term $\sigma^{\mu\nu}F^V_{\mu\nu}$ to suppress the statistical fluctuation, and then different from the action defined in Ref.~\cite{Horsley:2008ap,Cundy:2009yy} which uses $U$ for the clover term.}

In addition to the ensembles employed for determining the light quark masses and low energy constants~\cite{Hu:2023jet}, we have included two additional ensembles, E28P35 and G36P29, with $m_{\pi}\sim 300$ MeV and gauge couplings $\hat{\beta}=6.308$ and 6.498 to manage discretization errors. The tuning of the action parameters at these two values of $\hat{\beta}$ is similar to that in Ref.~\cite{Hu:2023jet}, and the lattice spacings are determined by the scale parameter $w_0$~\cite{FlavourLatticeAveragingGroupFLAG:2021npn,BMW:2012hcm,HotQCD:2014kol,RBC:2014ntl} {which is obtained using the gradient flow~\cite{Luscher:2010iy} of the same gauge action}. The spatial volume of the E28P35 and G36P29 ensembles are similar to those of the other ensembles with $m_{\pi}\sim$ 300 MeV.
The parameters employed for the simulation are outlined in Table~\ref{tab:ensemv}.

For the lattice configurations generated by the hybrid Monte Carlo, the autocorrelation between the trajectories is unavoidable. The independent configuration can only be obtained per each ${\cal O}(10)$ trajectories.
To quantitatively evaluate the impact of autocorrelation, we employ a detailed statistical analysis focused on the variance of binned means:

\begin{enumerate}
    \item \textbf{Overall Mean:}
    The overall mean, $\bar{O}$, of a measurement is computed as:
    \[
    \bar{O} = \frac{1}{N} \sum_{i=1}^{N} O_i
    \]
    where $O_i$ denotes individual measurements within one trajectory.

    \item \textbf{Binned Means:}
    For bin sizes varying from 1 to $n_{\text{max}}$, the mean of observations within each bin is calculated as:
    \[
    \bar{O}_{n,k} = \frac{1}{n} \sum_{i=(k-1)n+1}^{kn} O_i, \quad k = 1, 2, \dots, \left\lfloor \frac{N}{n} \right\rfloor
    \]
    where $k$ indexes the bins.

    \item \textbf{Variance of Binned Means:}
    The variance of these binned means from the overall mean is given by:
    \[
    \sigma^2_{n}(O) = \frac{1}{M-1} \sum_{k=1}^{M} (\bar{O}_{n,k} - \bar{O})^2
    \]
    where $M = \left\lfloor \frac{N}{n} \right\rfloor$ is the number of complete bins.

\begin{figure*}[p] 
   \centering
   \begin{tabular}{cc}
       \includegraphics[width=.47\textwidth]{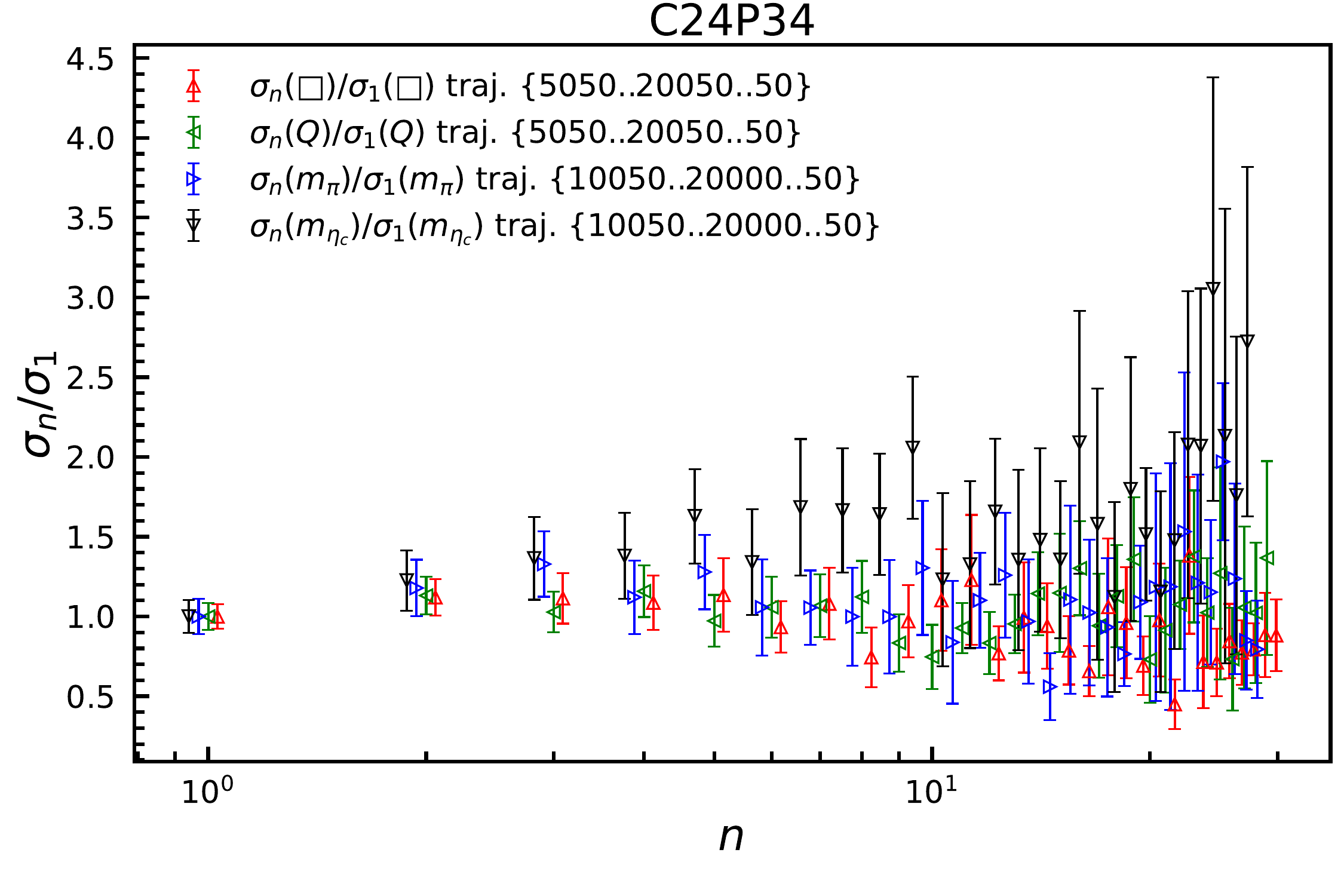} &
       \includegraphics[width=.47\textwidth]{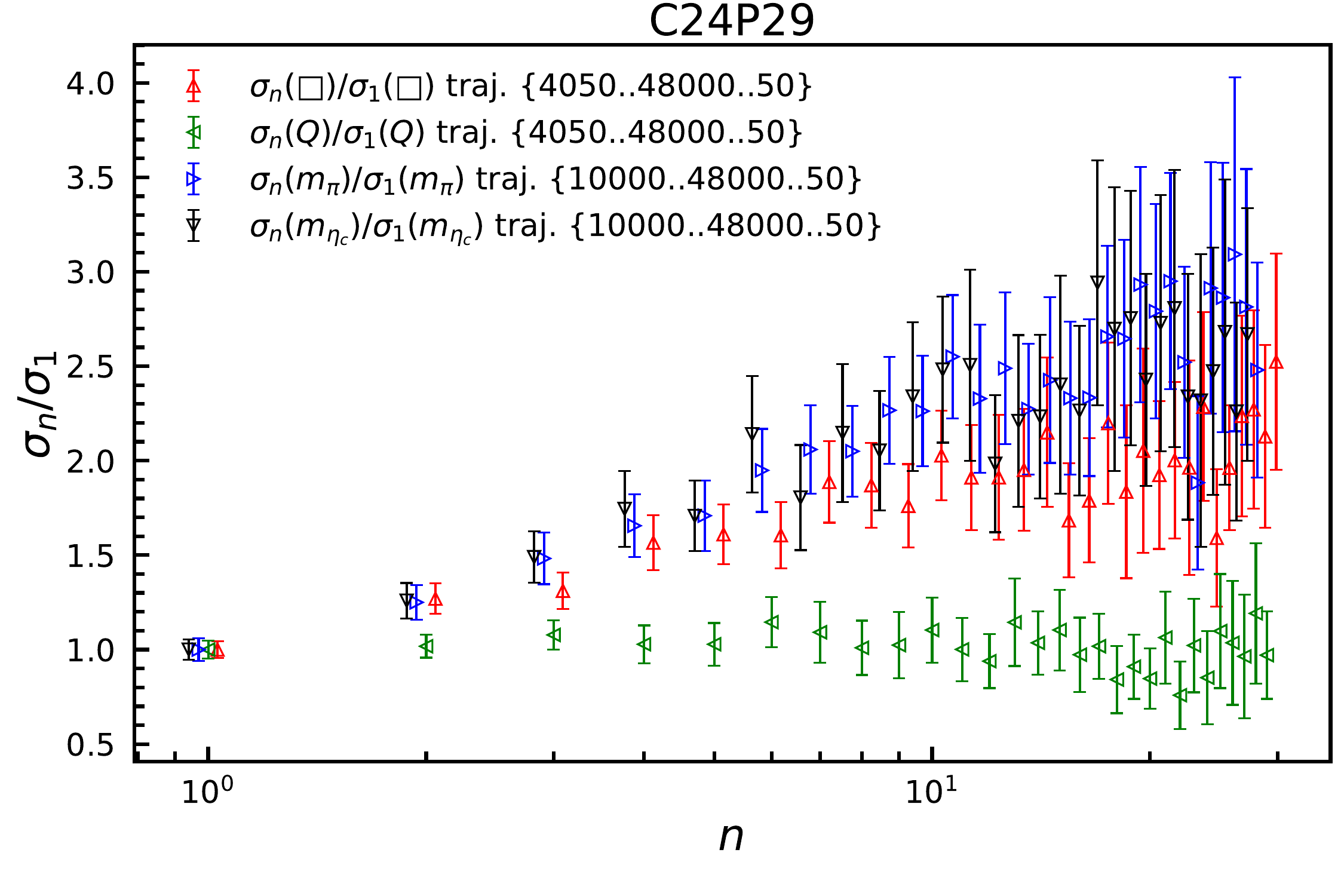} \\
       \includegraphics[width=.47\textwidth]{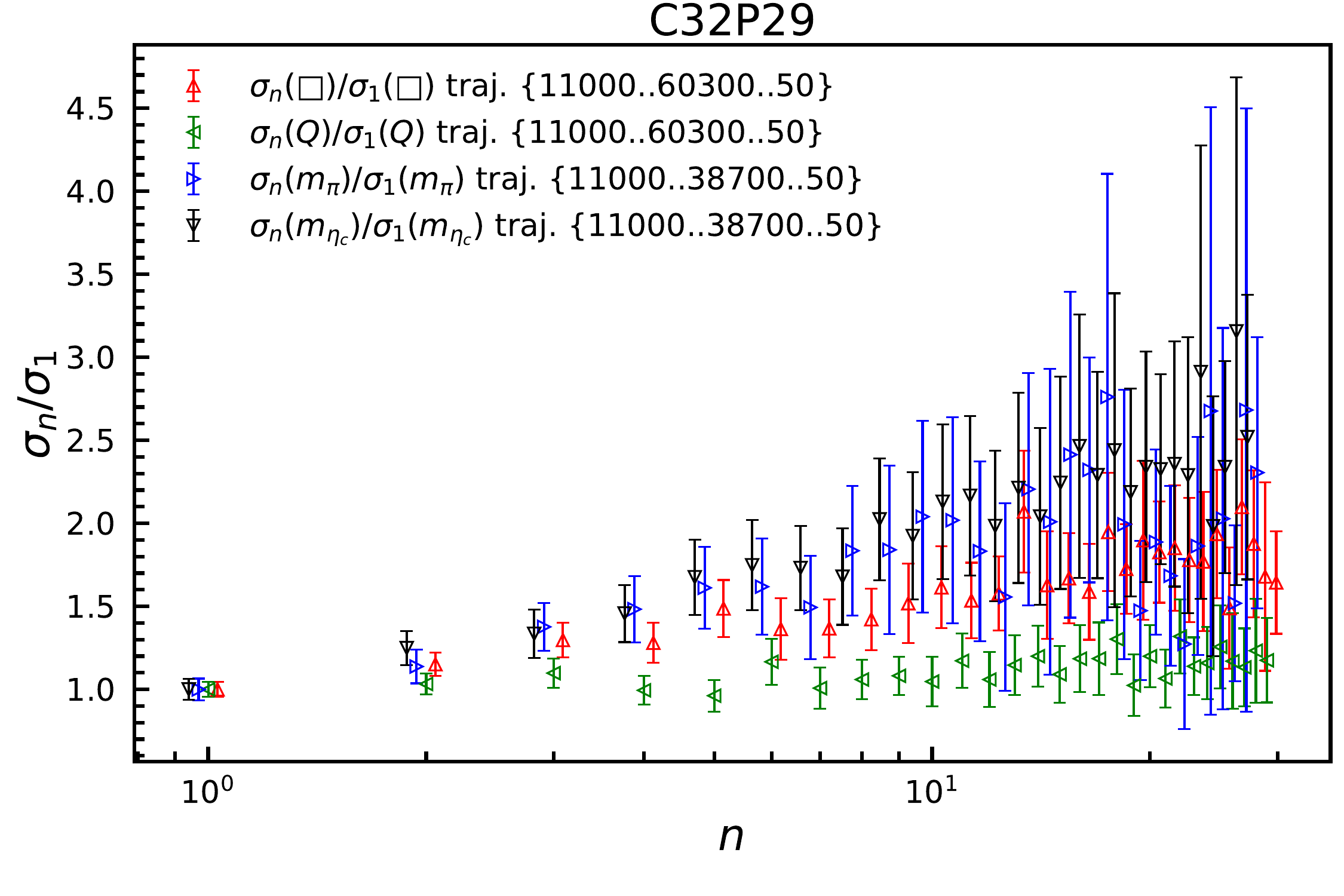} &
       \includegraphics[width=.47\textwidth]{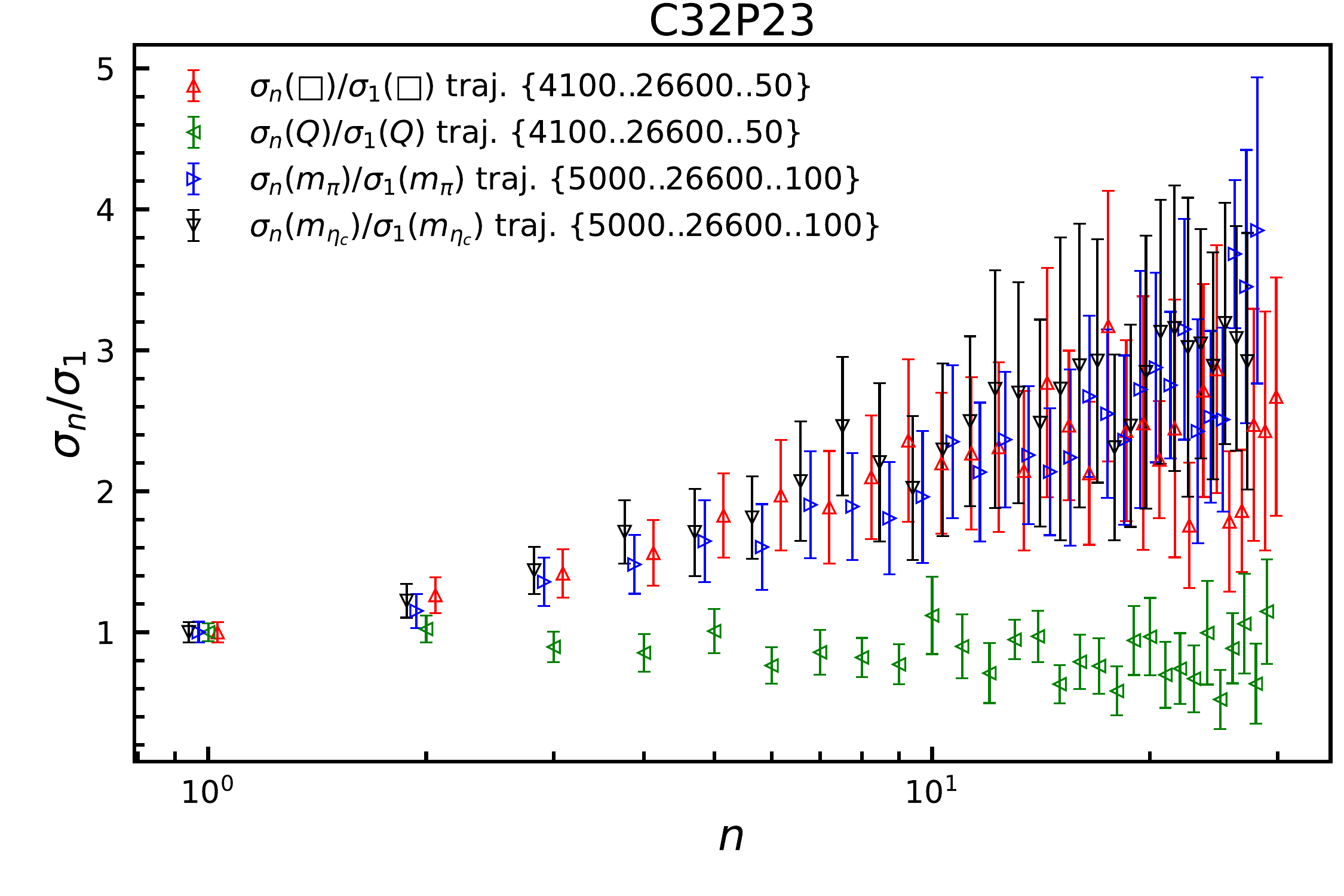} \\
       \includegraphics[width=.47\textwidth]{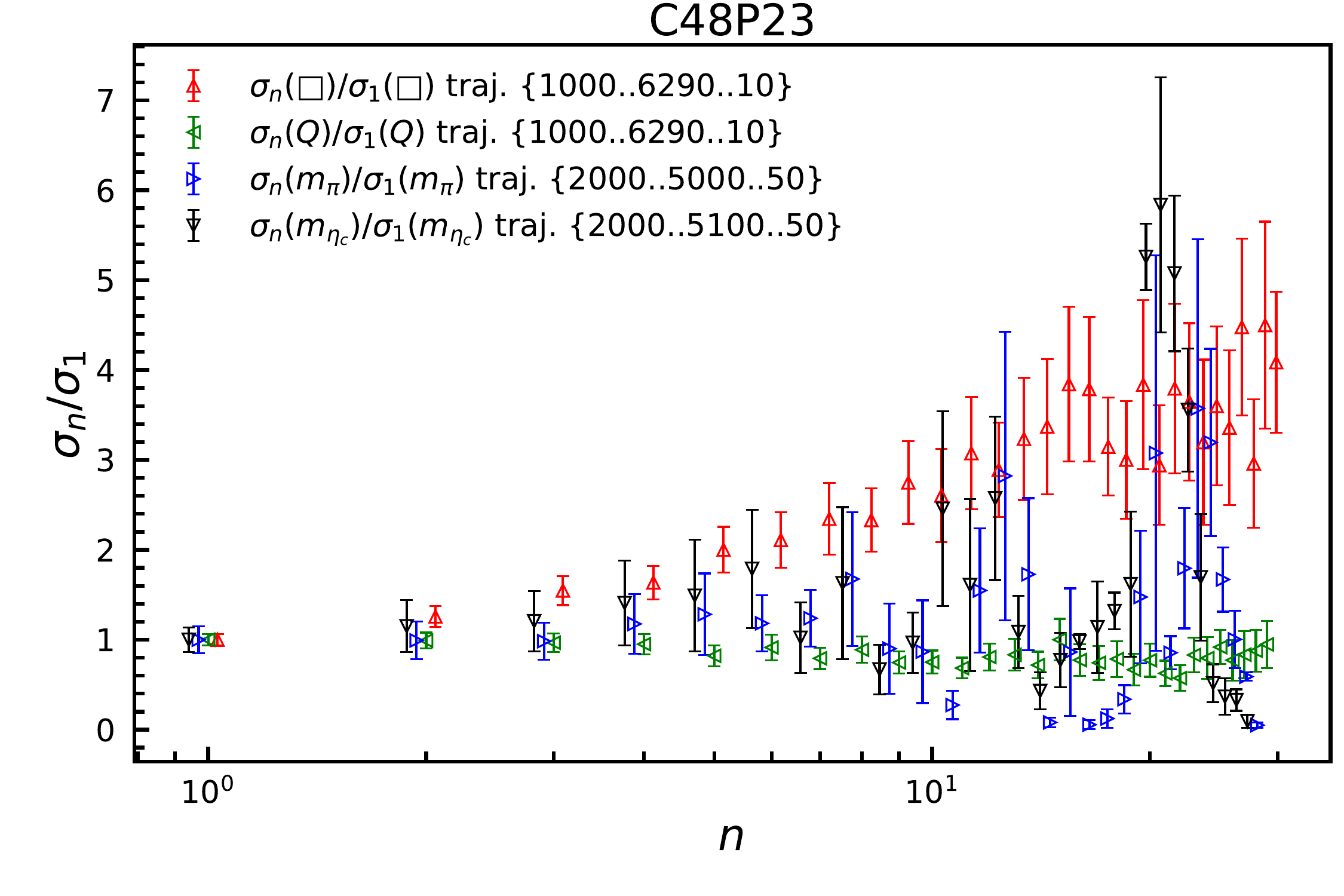} &
       \includegraphics[width=.47\textwidth]{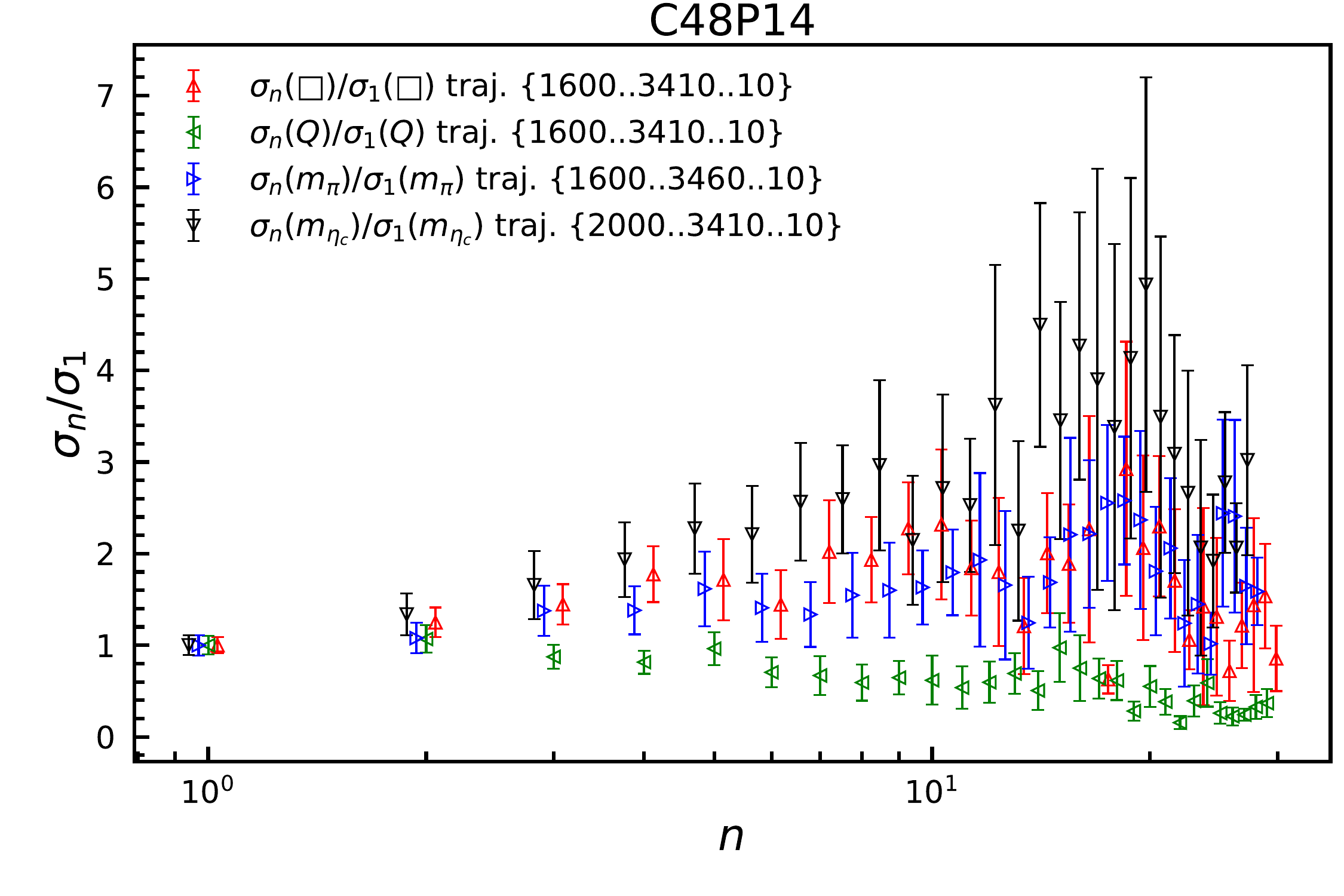} 
   \end{tabular}
    \caption{Part 1 of the autocorrelation analysis. Each panel represents a different ensemble, illustrating the variance of four key measurements -- plaquette value, topological charge, pion mass, and $\eta_c$ mass, as functions of bin size $n$. The numbers $\{n_{\rm min},n_{\rm max},\delta n\}$ displayed in the label of each quantity correspond to the smallest and largest trajectory numbers, as well as the number of trajectories we skipped to obtain an independent configuration used in the figure.}
          \label{fig:autocorr_1}
\end{figure*}

    \item \textbf{Error in Variance Calculation:}
    The standard error of the variance, providing a measure of its uncertainty, is computed as:
    \[
    \text{SE}(\sigma^2_{n}) = \sqrt{\frac{1}{M(M-1)} \sum_{k=1}^{M} \left( s^2_{n,k} - \sigma^2_{n} \right)^2}
    \]
    where $s^2_{n,k}$ is the variance within each bin:
    \[
    s^2_{n,k}(O) = \frac{1}{n-1} \sum_{i=(k-1)n+1}^{kn} (O_i - \bar{O}_{n,k})^2
    \]
\end{enumerate}
 We anticipate the variance of binned measurements will remain stable, assuming that the autocorrelation effects are sufficiently minimized. Here we compute both local and global quantities: the average plaquette value $\square=\left\langle\operatorname{Tr} U_p\right\rangle/3$, and the topological charge,
$$
Q=\int d^4 x q(x), \quad q(x)=\frac{g^2}{32 \pi^2} \epsilon_{\mu \nu \rho \sigma} \operatorname{Tr}\left[F^{\mu \nu}(x) F^{\rho \sigma}(x)\right]
$$ along with the masses of pion and $\eta_c$.
Figs.~\ref{fig:autocorr_1} and \ref{fig:autocorr_2} show the normalized variance $\sigma_n/\sigma_1$ of these $4$ quantities as functions of the bin size $n$ within each ensemble. {The mild $n$ dependence of $\sigma_n/\sigma_1$ suggests that the autocorrelation between configurations is acceptable, even though some of the cases have large statistical uncertainty.} This is particularly evident in the case of $Q$, which does not exhibit any indications of topological charge freezing at the {finest} lattice spacing $a=0.0520(2)(3)$ fm.

\begin{figure*}[pt] 
   \centering
   \begin{tabular}{cc}
       \includegraphics[width=.47\textwidth]{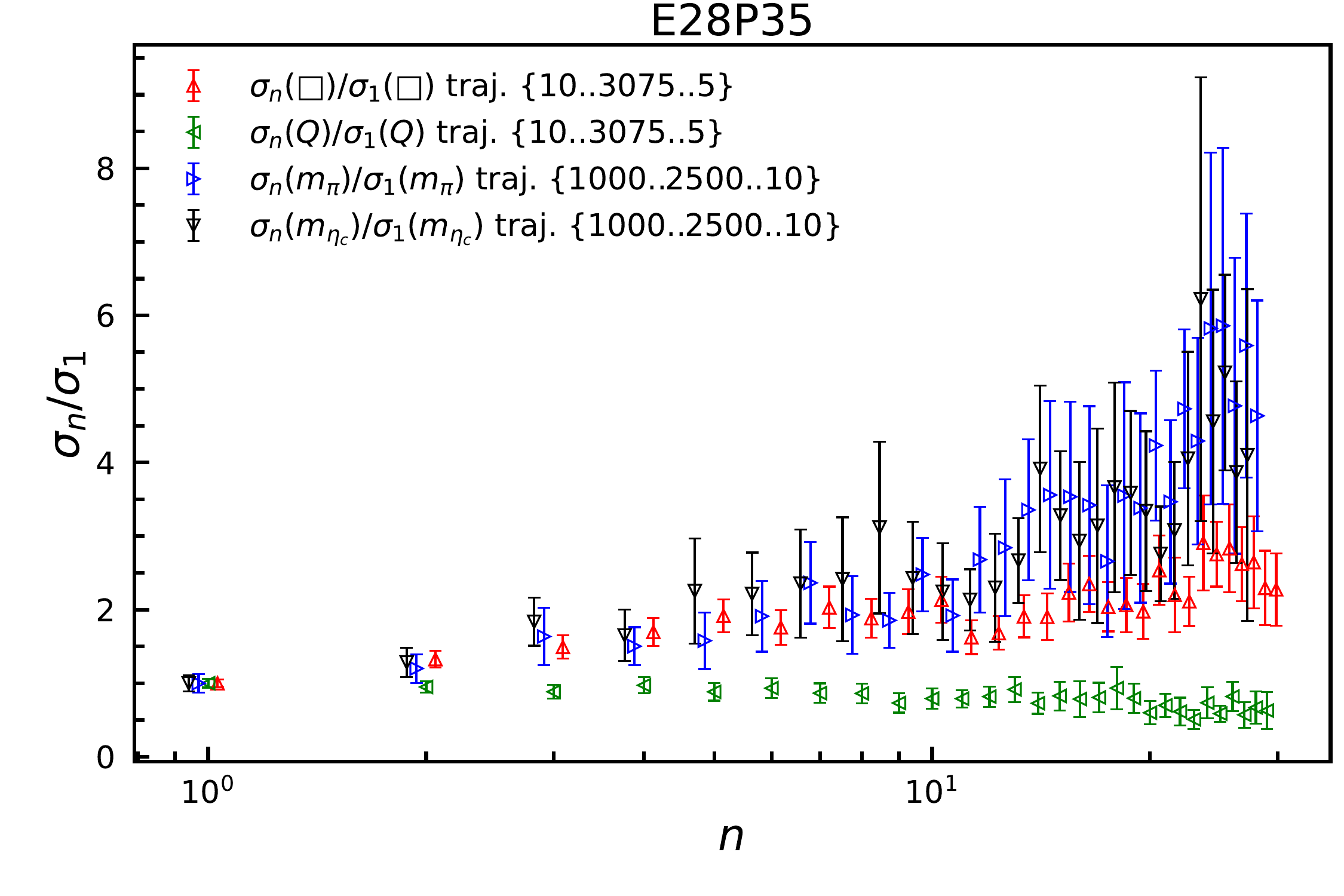}& 
       \includegraphics[width=.47\textwidth]{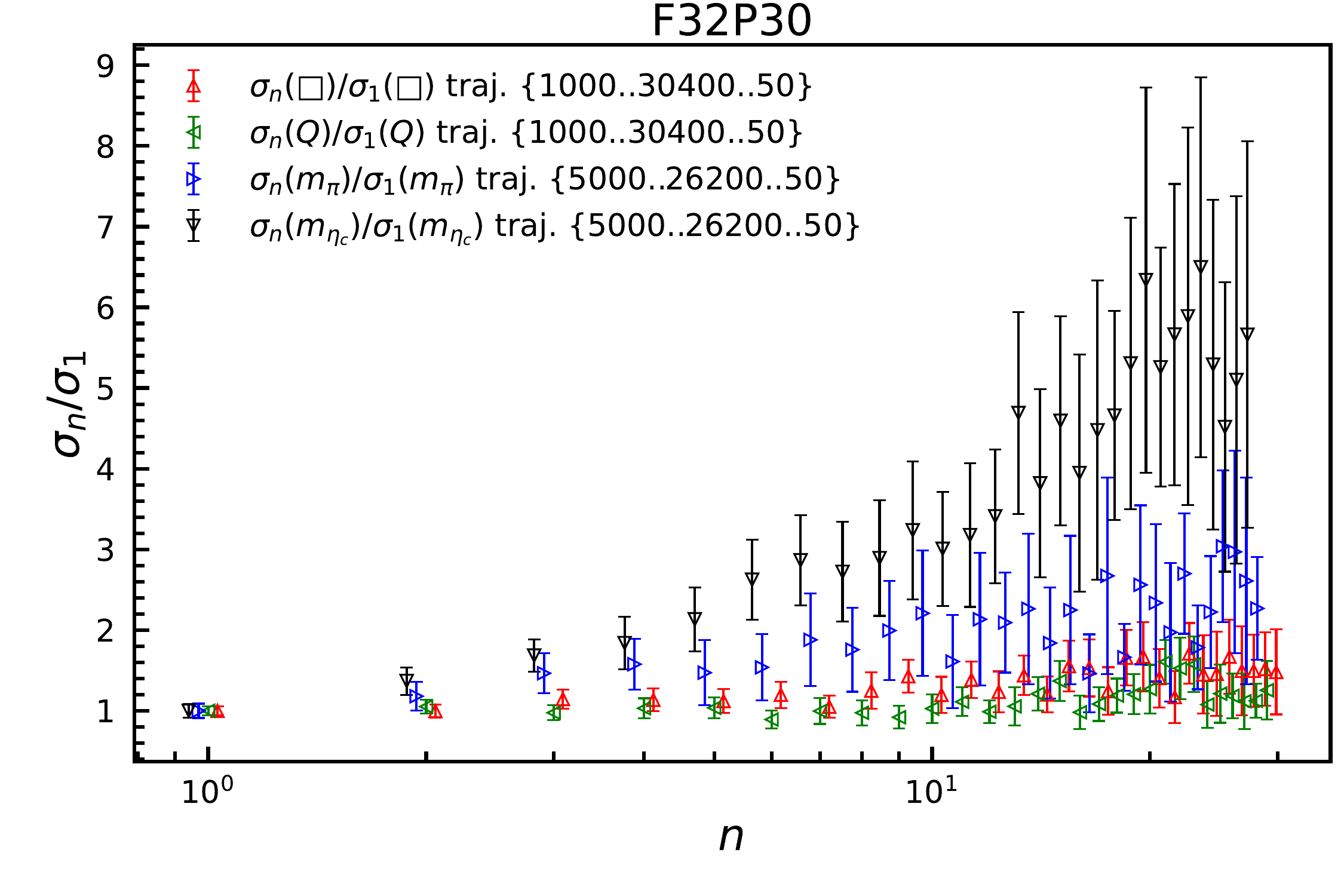} \\
       \includegraphics[width=.47\textwidth]{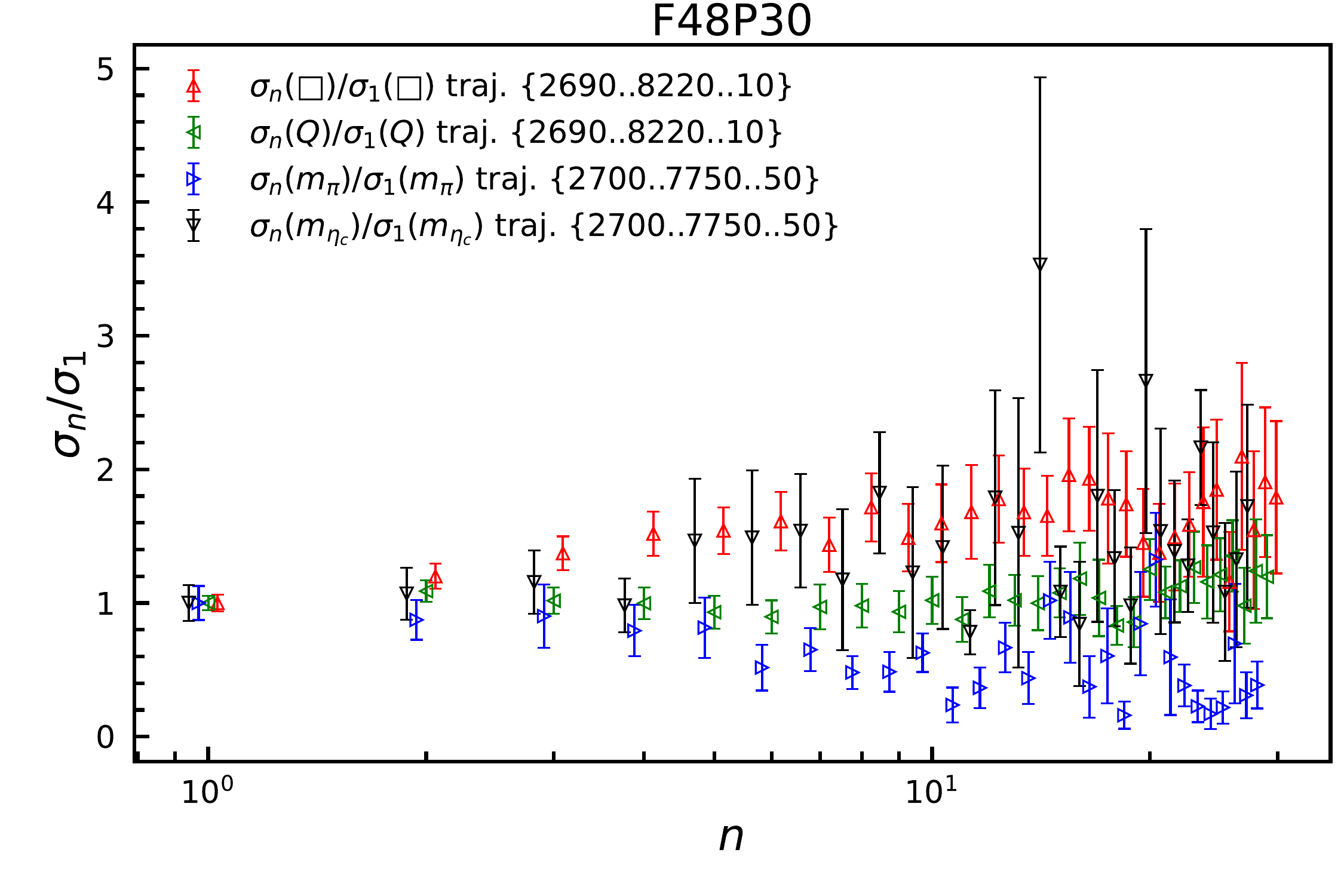} &
       \includegraphics[width=.47\textwidth]{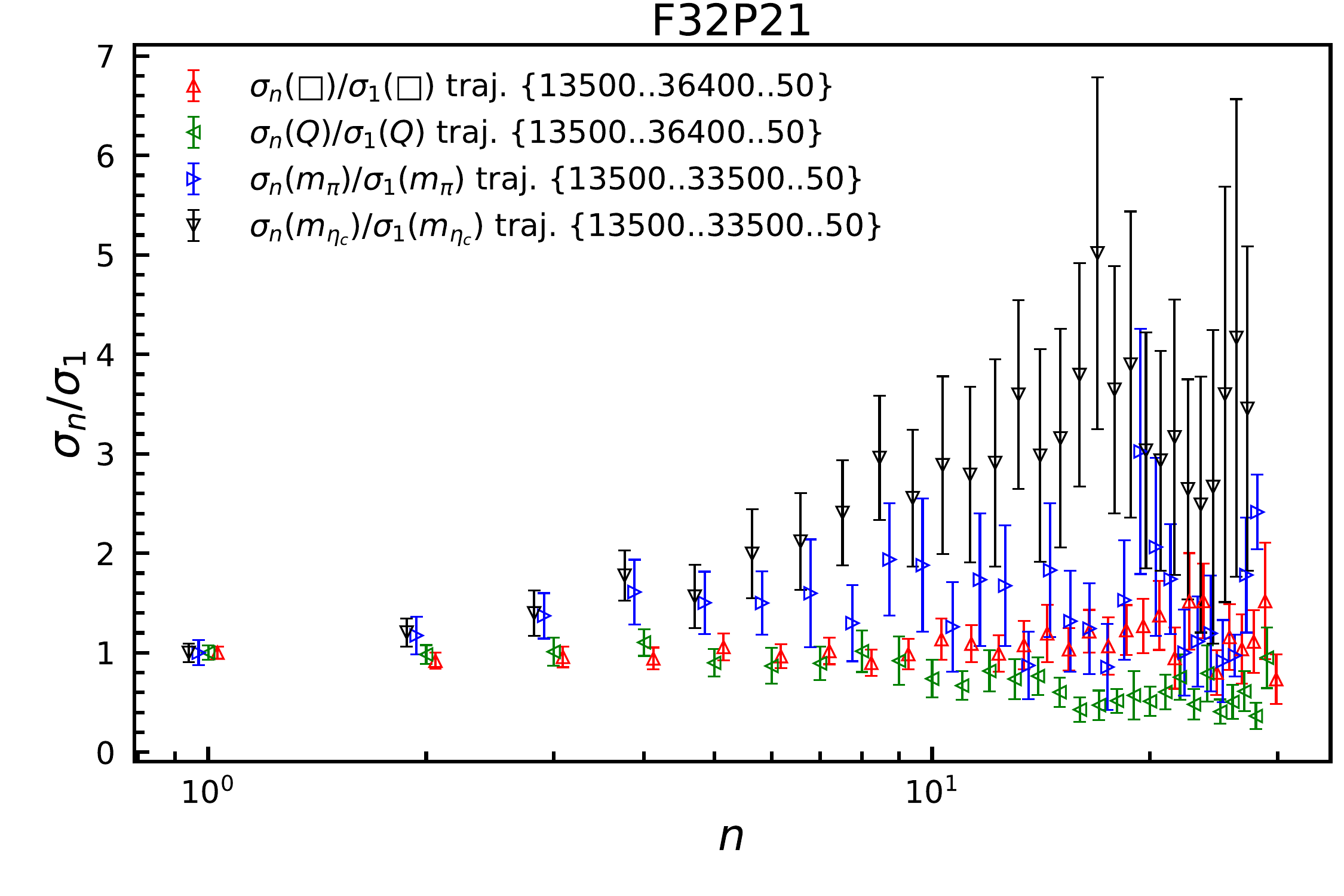} \\
       \includegraphics[width=.47\textwidth]{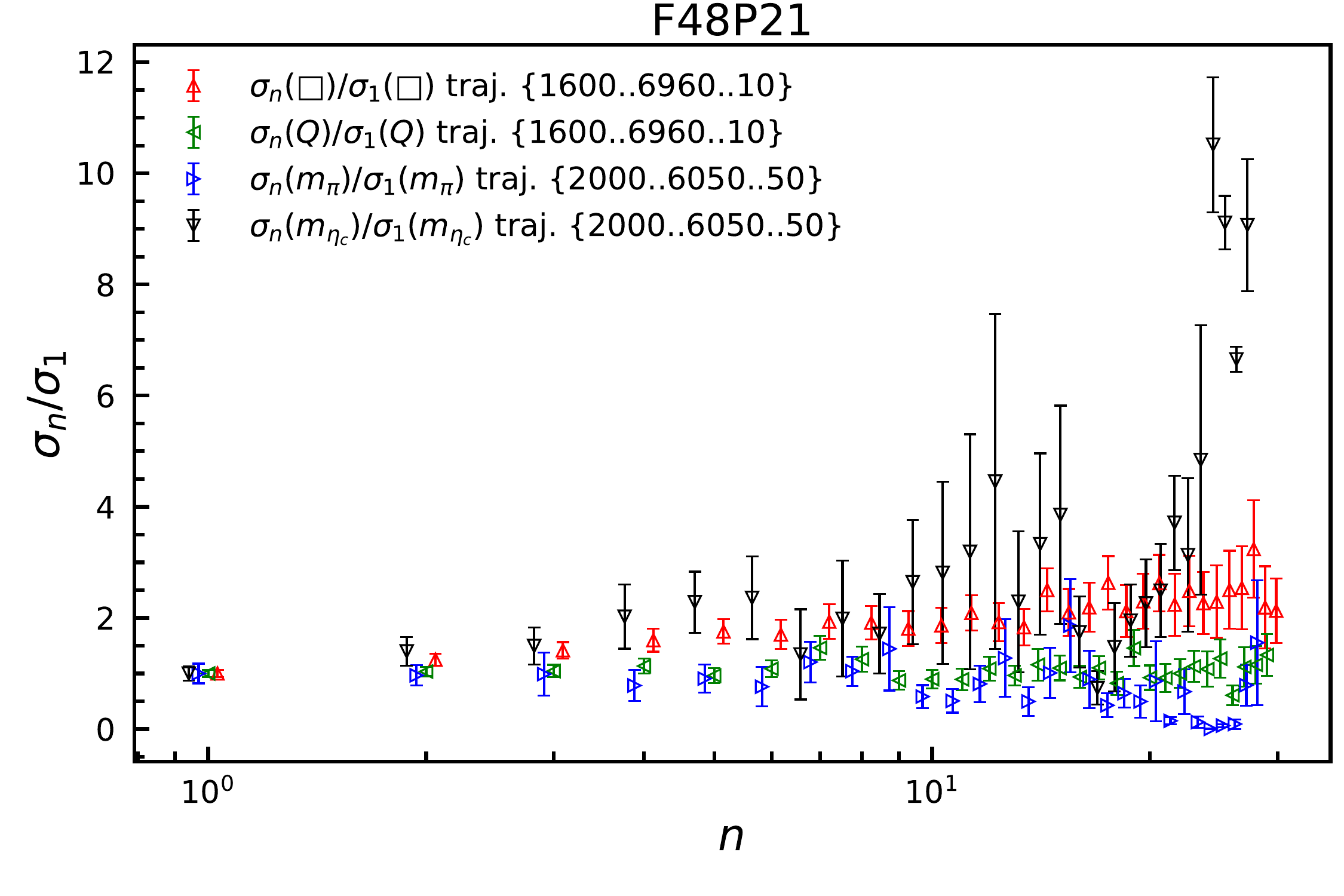} &
       \includegraphics[width=.47\textwidth]{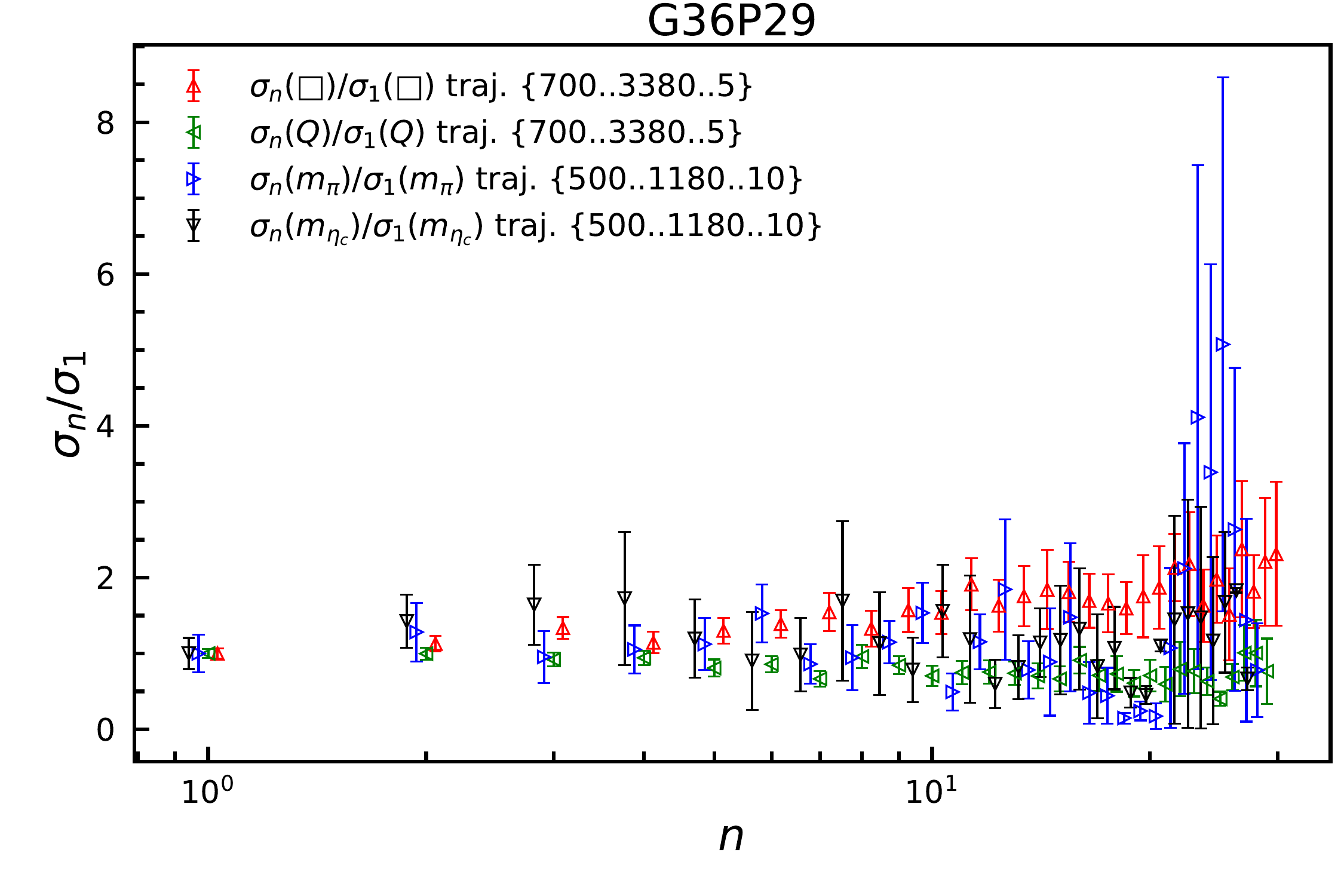} \\
       \includegraphics[width=.47\textwidth]{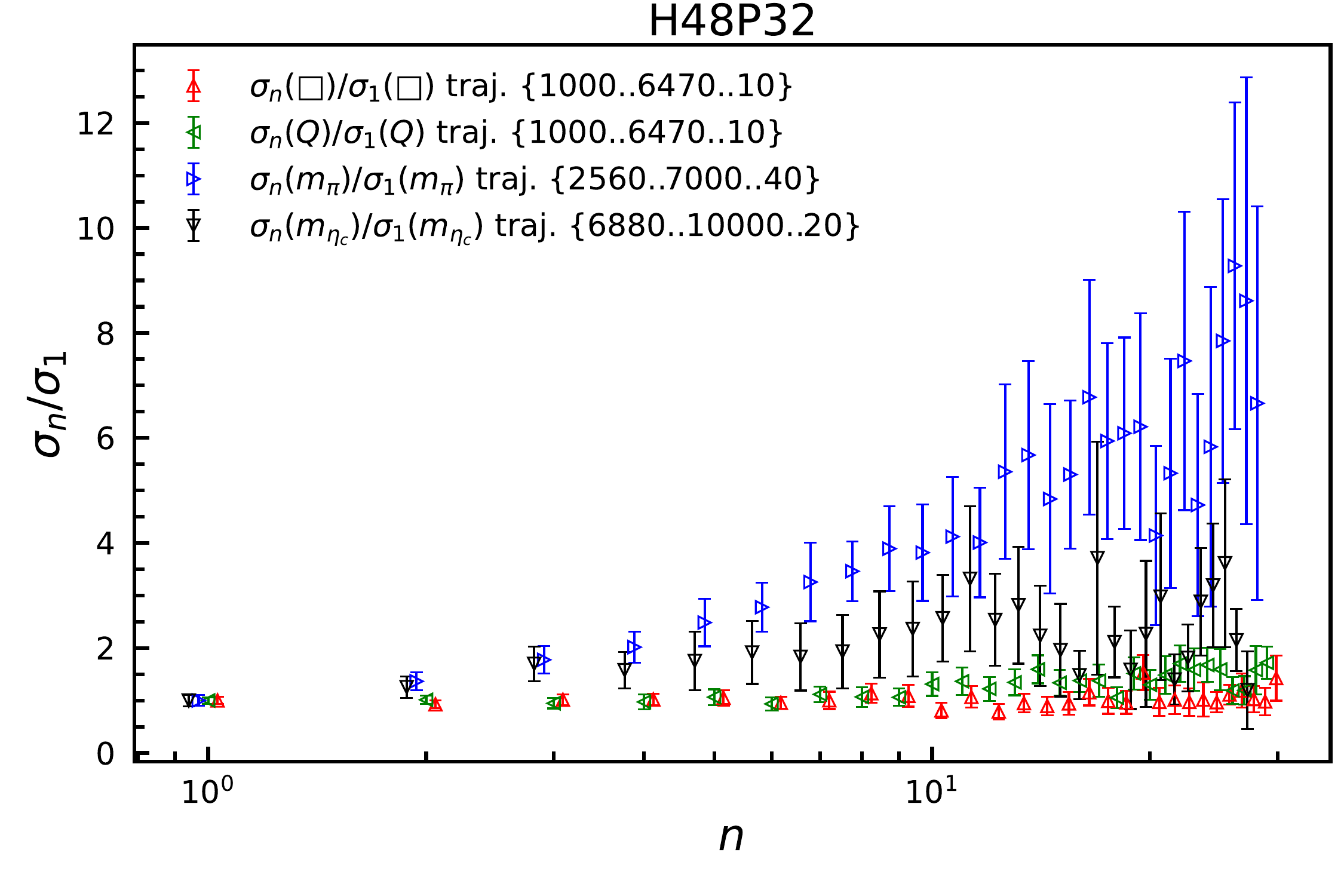} &
   \end{tabular}
   \caption{Part 2 of the autocorrelation analysis.}
          \label{fig:autocorr_2}
\end{figure*}

\clearpage

\subsection{Consistency checks on the extraction of quark and hadron masses}

{In this section, we will examine the consistency of the charm quark mass derived from the PCAC relation of various pseudoscalar mesons, and illustrate the process by which hadron masses are extracted through multiple-state fits.}

\begin{figure}[t]
\includegraphics[width=0.45\textwidth]{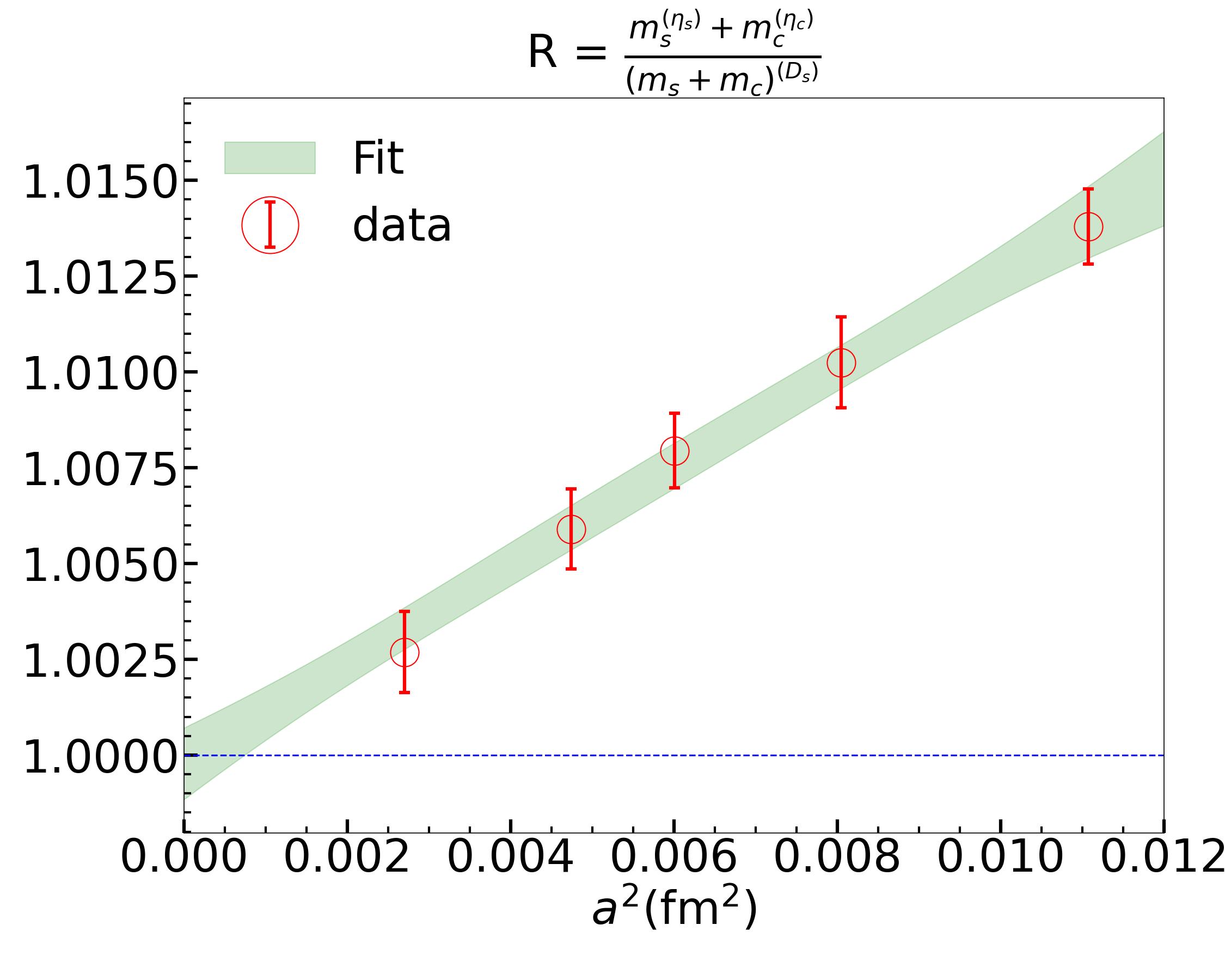}
\caption{The ratio of the PCAC quark mass determined from the quarkonium, and also $D_s$. Two definitions are consistent up to ${\cal O}(a^4)$ correction.}
\label{fig:quark_mass_def}
\end{figure}

\begin{figure}[thb]
\includegraphics[width=0.9\textwidth]{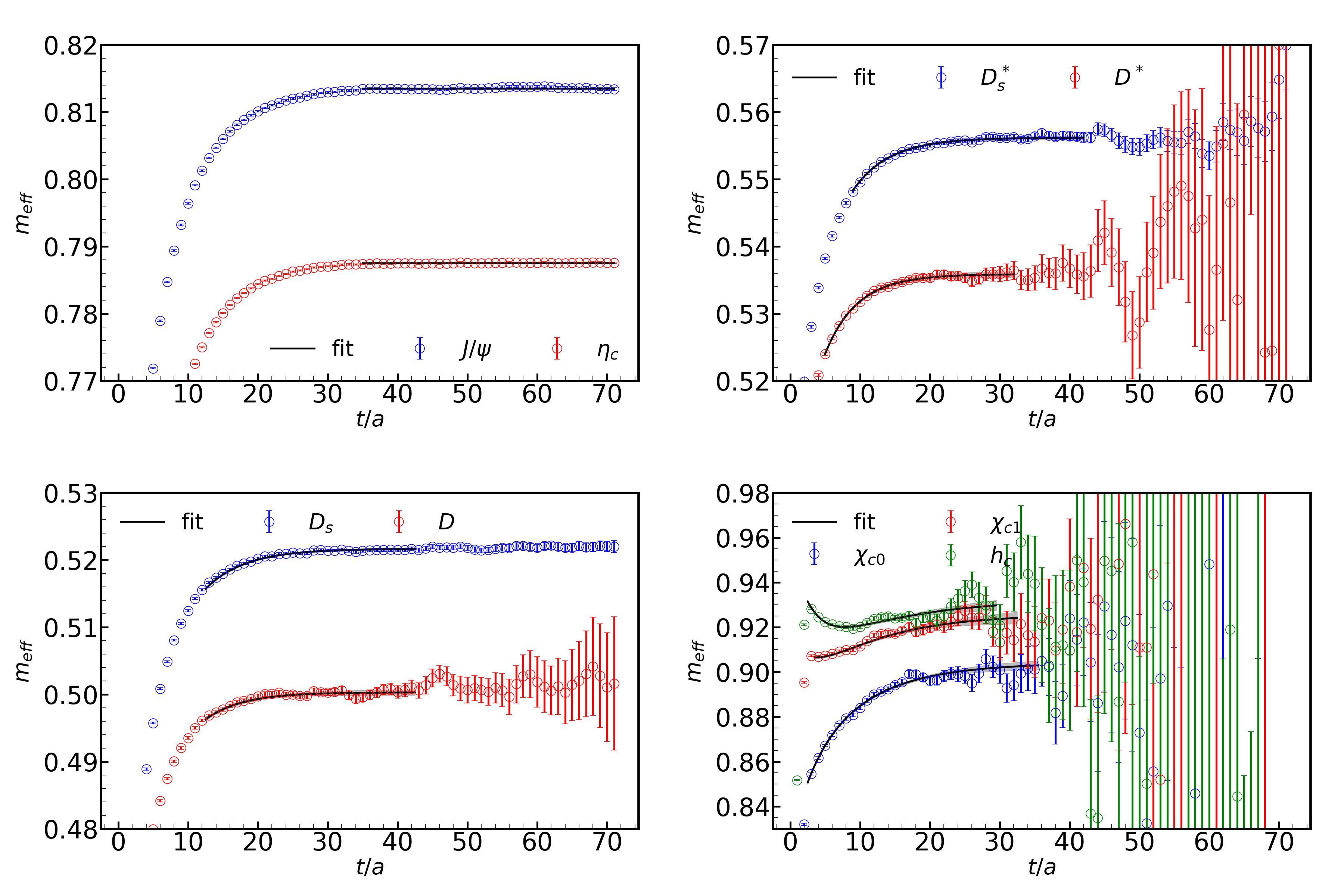}
\caption{Dimensionless effective mass on H48P32. For $J/\psi$ and $\eta_c$ (left top), we used a 1-state fit. For $D$, $D_s$(left bottom) and $D_s^*$, $D^*$(right top) we used 2-state fit. For P-wave charmonium(right bottom), we used a 3-state fit.}
\label{fig:eff_mass}
\end{figure}

In Fig.~\ref{fig:quark_mass_def}, we show the ratio of two determinations of $m_s^{\rm PC}+m_c^{\rm PC}$ from quarkonium ($\eta_s$ and $\eta_c$) and also $D_s$, at five lattice spacings, with the {valence} strange and charm quark masses. 
Two determinations deviate from each other by approximately 1.4\% at the coarsest lattice spacing but are consistent within the statistical uncertainty of 0.1\% after a polynomial extrapolation up to $\mathcal{O}(a^4)$ to the continuum limit. 

When analyzing the effective mass obtained from the wall-wall correlator, we find that its uncertainty is significantly larger than that of the wall-local correlator. This increased uncertainty primarily arises from the statistical fluctuations associated with the non-local correlations between quark and anti-quark at the sink time slices, which are independent across different time slices. Consequently, when the ground state mass and associated weights (related to the decay constant) are extracted from the wall-wall correlator using the exponential fit, they exhibit comparable noise levels to those obtained from the wall-local correlator, particularly for pseudo-scalar states. 

For instance, in the \( J/\psi \) channel on the H48P32 ensemble with the finest lattice spacing, the relative statistical uncertainties for the ground state mass and its weight using the wall-local correlator are 0.12\% and 0.37\%, respectively. In contrast, the corresponding uncertainties for the wall-wall correlator are 0.16\% and 0.45\%. Eventually, the statistical uncertainty for the extracted \( f_{J/\psi} \) from both correlators is 0.24\%, as the correlation between the wall operators in the two correlators partially cancels out the uncertainty. However, the overall statistical uncertainty of the continuum-extracted \( f_{J/\psi} \) increases to 0.66\% due to the greater uncertainty introduced by the continuum extrapolation, particularly from the \( \mathcal{O}(a^4) \) term.

In Fig.~\ref{fig:eff_mass} we show the dimensionless effective mass on ensemble H48P32. The bare quark mass is $am_c$=0.0580 and the bare strange or light quark mass is the unitary one. For $\eta_c$ and $J/\psi$, we used a 1-state fit. For the other particles with increasing statistical uncertainty at large $t$, we used the 2-state or even 3-state fit to enlarge the fit range and then enhance the reliability of the fit.

\begin{figure}
    \centering
    \includegraphics[width=0.9\linewidth]{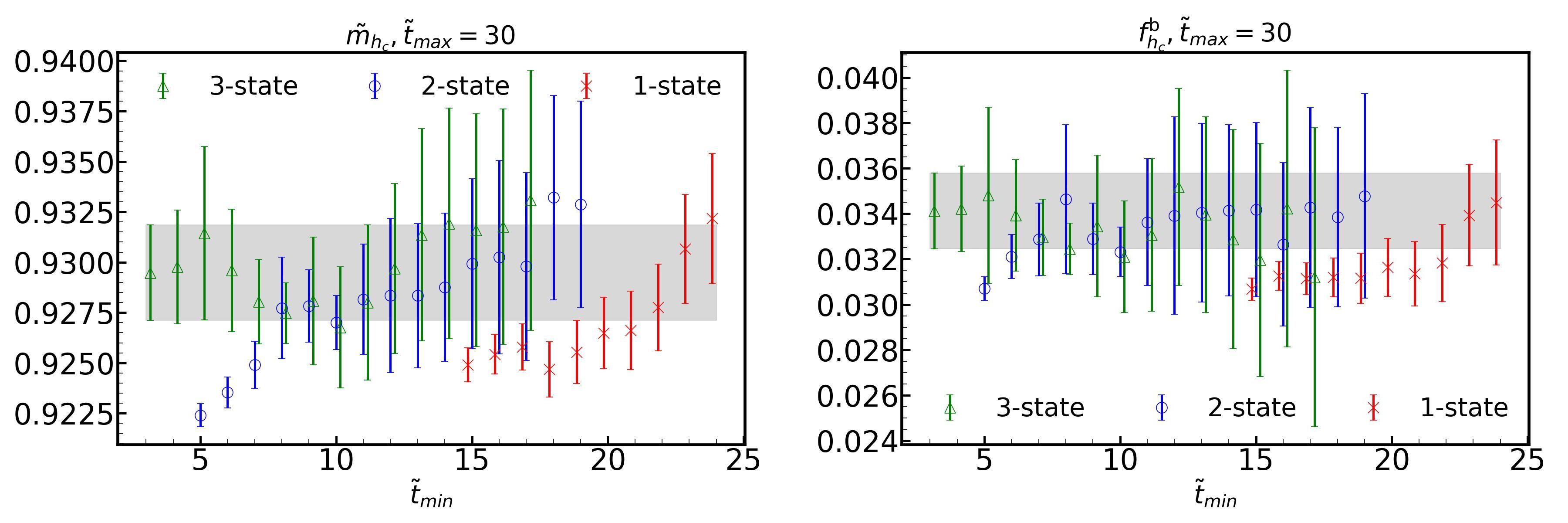}
    \caption{Consistency check of 1, 2, and 3-state fit of $h_c$ on H48P32.} 
\label{fig:fit_state}
\end{figure}

{To show the consistency of 1, 2, and 3-state fit, we take $h_c$ on H48P32 as an example and show the fit results of dimensionless $\tilde{m}_{h_c}$ and $\tilde{f}_{h_c}$ in Fig.~\ref{fig:fit_state}. We fixed the maximum $\tilde{t}_{\rm max}$ = 30, varied $\tilde{t}_{\rm min}$, and gave the results of 1, 2, and 3-state fit respectively. The gray band in the figure corresponds to the 3-state fit value we finally selected with $\tilde{t}_{min}=3$. We can see that this band is consistent with the other cases within the uncertainty.}

\subsection{Renormalization and heavy quark improved normalization}

{In this section, we will cover the normalization and renormalization of quark bilinear operators involving the charm quark, as well as how employing heavy quark improved normalization can help suppress the discretization errors of charmed hadron matrix elements.}

\subsubsection{Heavy quark improved normalization}

{The Wilson term of the Wilson-type fermion, which includes the clover fermion utilized in this study, introduces a momentum-dependent but mass-independent ${\cal O}(ap^2)$ correction, 
\begin{align}
    S^{-1}(p,m)&=\frac{\rm i}{a}\sum_{\mu}\gamma_{\mu}\mathrm{sin}(ap_{\mu})+m+\frac{1}{a}\sum_{\mu}(1-\mathrm{cos}(ap_{\mu}))={{\rm i}p\!\!\!/+m+\frac{1}{2}ap^2+{\cal O}(a^np^{n+1},n\ge2)}.
\end{align}

In the context of the free propagator, this ${\cal O}(a)$ effect can be represented as corrections that are independent of momentum but reliant on mass~\cite{Capitani:2000xi},
\begin{align}
    S^{-1}(p,m)&={{\rm i}p\!\!\!/+m+\frac{1}{2}ap^2+{\cal O}(a^np^{n+1},n\ge2)}={{\rm i}\frac{p\!\!\!/}{1+\frac{\rm i}{2}ap\!\!\!/}+m+{\cal O}(a^np^{n+1},n\ge2)}\nonumber\\
    &=\frac{(1+\frac{1}{2}am)({\rm i}p\!\!\!/+m+\frac{{\rm i}}{2}amp\!\!\!/)}{(1+\frac{1}{2}am)(1+\frac{\rm i}{2}ap\!\!\!/)}+{\cal O}(a^np^{n+1},n\ge2)=\frac{(1+\frac{1}{2}am)({\rm i}(1+\frac{1}{2}am)p\!\!\!/+m)}{1+\frac{1}{2}a({\rm i}(1+\frac{1}{2}am)p\!\!\!/+m)}+{\cal O}(a^np^{n+1},n\ge2)\nonumber\\
    &=\frac{{\rm i}p\!\!\!/+\frac{m}{1+\frac{1}{2}am}}{(\frac{1}{1+\frac{1}{2}am})^2+\frac{a}{2(1+\frac{1}{2}am)}({\rm i}p\!\!\!/+\frac{m}{1+\frac{1}{2}am})}+{\cal O}(a^np^{n+1},n\ge2)\nonumber\\
    &=\big(\frac{(\frac{1}{1+\frac{1}{2}am})^2}{{\rm i}p\!\!\!/+\frac{m}{1+\frac{1}{2}am}}+\frac{a}{2+am}\big)^{-1}+{\cal O}(a^np^{n+1},n\ge2)\label{eq:correction_1}
\end{align}
where $m$ represents the free quark mass unaffected by any $\alpha_s/a$ power divergence of the Wilson-type fermion.

The quark propagator in Eq.~(\ref{eq:correction_1}) encompasses three types of ${\cal O}(a)$ corrections, including the $am$ correction to all orders:

1. The additive constant correction $\frac{a}{2+am}$ corresponds to a delta function $\delta_{x,y}$ in the coordinate space propagator $S(x,y)$ and is irrelevant to the time-ordered on-shell correlation functions.

2. The quark mass rescaling factor $\frac{1}{1+\frac{1}{2}am}$ can be absorbed into the redefinition of the bare quark mass parameter.

3. The quark field rescaling factor $(\frac{1}{1+\frac{1}{2}am})^2$ introduces a multiplicative factor $(1+\frac{1}{2}am)^2=1+ma+{\cal O}(m^2a^2)$ on all quark bilinear operators.

With a non-zero gauge coupling, the bare quark mass of the Wilson-type fermion undergoes additive renormalization. Consequently, the dimensionless quark mass $ma$ in Eq.~(\ref{eq:correction_1}) would be replaced by $\tilde{m}_q^{\rm sub}\equiv\tilde{m}^b-\tilde{m}^{\rm crti}$, where $\tilde{m}^{\rm crti}$ is the quark mass parameter that drives the pion mass towards zero and can be found in Table~\ref{tab:ensemv}. Subsequently, the quark field rescaling factor would be redefined as $1+(1+{\cal O}(\alpha_s))\tilde{m}^{\rm sub}_q+{\cal O}((\tilde{m}^{\rm sub}_q)^2)$. In principle, the gauge interaction can also cause the $m_qa$ correction of the quark bilinear operator $\bar{q}\Gamma q$ with different $\Gamma$ to vary, while the impact would be small.

In this work, we determine this rescaling factor using the vector current normalization. Unlike its continuum counterpart, $Z_V$ under lattice regularization is influenced by both $\alpha_s$ and discretization effects and can be determined from the vector current conservation condition~\cite{Martinelli:1994ty,Zhang:2020rsx}, 
\begin{align}
1=Z_{V}(H)\frac{\langle H|V_4|H\rangle}{\langle H|H\rangle}=Z_{V}(H)\frac{C_{3}(t_f,t)}{C_2(t_f)}|_{0\ll t\ll t_f},\label{eq:ZV_c}
\end{align}
where $C_2(t_f)$ represents the two-point function of an arbitrary hadronic state $H$ with a source-sink separation of $t_f$, and $C_3(t_f,t)$ is the three-point function of H with a source-sink separation of $t_f$ and also an additional vector current inserted at the time slice $t$. Since $C_3$ contains one more quark propagator than $C_2$, and $Z_{V}(H)=1$ should be invariant across different hadronic states in the continuum due to charge conservation, $Z_V(H)$ serves as an ideal quantity for determining the the quark field rescaling factor non-perturbatively.

In the left panel of Fig.~\ref{fig:ZV_quark_mass_dep}, we show that $R_V(\tilde{m}^{\rm sub}_q,a)\equiv Z_V({\rm PS})/Z_V(m_{\pi}\rightarrow 0)$ on the ensembles with $m_{\pi}\sim 300$ MeV and $L\sim$ 2.5 fm
is almost linear on $\tilde{m}^{\rm sub}_q$, with the correction from the $(\tilde{m}^{\rm sub}_q)^{n\ge 2}$ terms around 0.01 for $\tilde{m}^{\rm sub}_q\sim 1$,
\begin{align}
   R_V(\tilde{m}^{b}_q,a)&=1+[1.091(2)+0.518(6) \Lambda_{\chi} a] (\tilde{m}^{b}_q-\tilde{m}^{\rm crti})+[-0.016(3)+0.030(8) \Lambda_{\chi} a] (\tilde{m}^{b}_q-\tilde{m}^{\rm crti})^2\nonumber\\
   &=1+1.091(2)m^{\rm sub}_qa+[0.518(6) \Lambda_{\chi}-0.016(3)m^{\rm sub}_q]a^2+{\cal O}(a^3)
\end{align}
 where $\Lambda_{\chi}$ = 1~GeV and the value of $\tilde{m}^{\rm crti}$ can be found in Table~\ref{tab:ensemv}. It suggests that the correction on the tree level quark mass dependence is majorly an ${\cal O}(a^2)$ effect. Simultaneously, the coefficient of the $(m^{\rm sub}_qa)^{2}$ term is approximately an order of magnitude smaller than the tree-level estimate. This implies that the $m^{\rm sub}_q$ is subject to a sizable $m^{R}_qa$ error compared to the renormalized quark mass $m_q^R$.

It is interesting to observe that the lattice spacing dependence of $R_V$ can be reduced even more in the heavy quark region by representing it as a linear function of $m_{\rm PS}^2a^2$, as illustrated in the right panel of Fig.~\ref{fig:ZV_quark_mass_dep}. Such an observation also suggests that the $m_{q}^2a^2\propto m_{\rm PS}^2a^2$ error is important in the heavy quark region.} 

{In contrast to the clover fermion, the overlap fermion is free of the additive renormalization of the quark mass and then the ${\cal O}(m_qa)$ improvement can be applied automatically using Eq.~(\ref{eq:correction_1})~\cite{Capitani:2000xi}. Thus} $Z_V=Z_A$ only depends on $m_{\rm PS}^2a^2$ with a tiny coefficient $\sim 0.02$~\cite{Wang:2020nbf,He:2022lse} {for the overlap fermion. Thus the quark field rescaling factor is essential here to suppress the discretization error of the heavy meson matrix elements}.

\begin{figure}[t]
\includegraphics[width=0.9\textwidth]{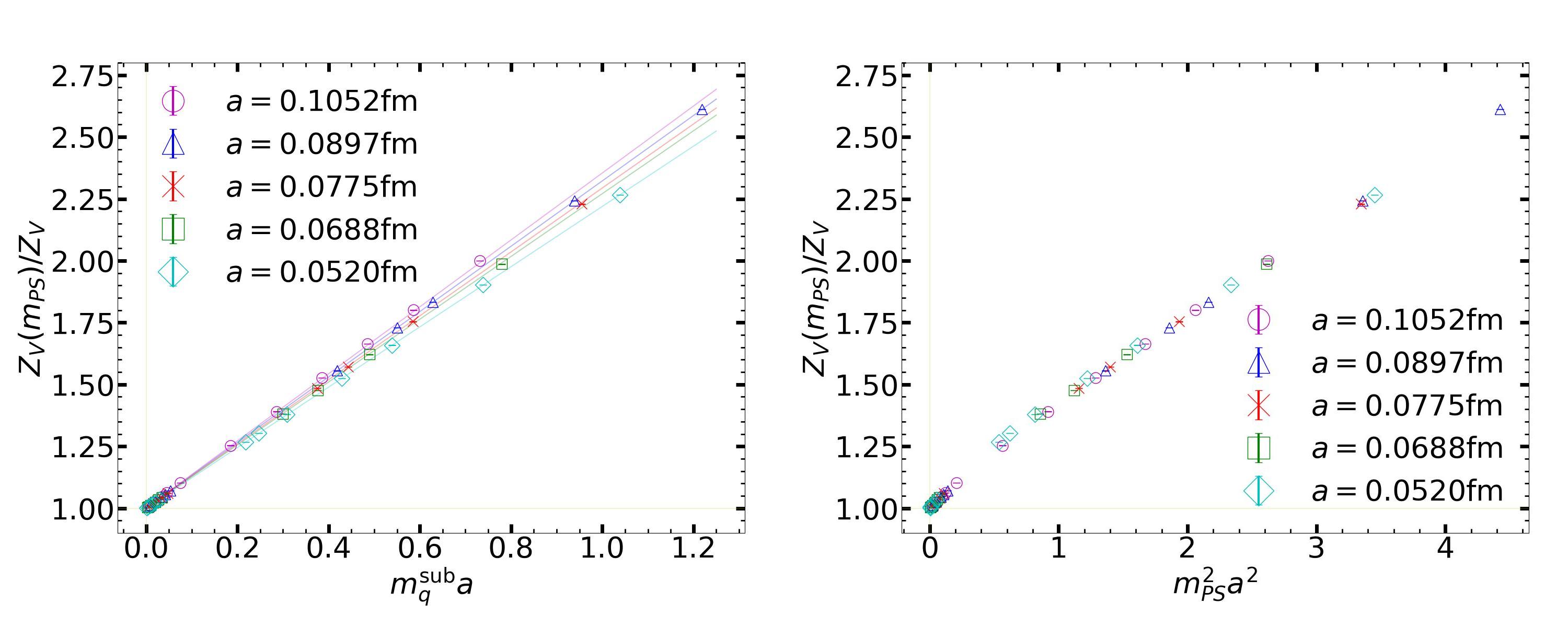}
\caption{The dimensionless {subtracted bare quark mass $m^{\rm sub}_qa$ dependence (left panel) and pseudo-scalar mass $m_{\rm PS}^2a^2$ dependence (right panel)} of $Z_V$, at different lattice spacing.
}
\label{fig:ZV_quark_mass_dep}
\end{figure}

{Assuming that the quark mass dependence of the quark bilinear operator $O_{\Gamma}\equiv \bar{q}\Gamma q$ majorly comes from the massive quark propagator and then insensitive to $\Gamma$, the ratio $Z_{O_{\Gamma}}/Z_V$ can effectively mitigate a substantial portion of the quark mass dependence. This allows for the reuse of this ratio in the chiral limit to determine the renormalization constant $Z^Q_{X_{\Gamma}}\equiv Z^Q_V\frac{Z_{O_{\Gamma}}}{Z_V}$ of $X_{\Gamma}\equiv \bar{Q}\Gamma Q$ with heavy quark $Q$, where $Z^Q_V\equiv Z_V({\rm PS})$ is extracted from Eq.~(\ref{eq:ZV_c}) using the non-singlet pseudo-scalar state with interpolation field $\bar{Q}\gamma_5Q$, and the ratio $\frac{Z_{O_{\Gamma}}}{Z_V}$ is obtained in the chiral limit. 

For the heavy-light quark bilinear operator $ Y_{\Gamma}\equiv\bar{Q}\Gamma q$, we define this renormalization constant as $Z_Y^Q=\frac{Z^Q_V}{Z_V}Z_{O_{\Gamma}}=\sqrt{Z_VZ_V^Q}\frac{Z_{O_{\Gamma}}}{Z_V}$, assuming that the quark mass dependence majorly comes from the quark propagator and then can be approximated by $\sqrt{\frac{Z^Q_V}{Z_V}}$ for each quark. It is slightly different from the linear combination $\frac{Z_V+Z_V^Q}{2}$~\cite{Kuberski:2024pms} used by the ALPHA collaboration for the heavy-light quark operator.

After the renormalization with the above renormalization constants, the remaining quark mass dependence of $Z^Q_{X_{\Gamma},Y_{\Gamma}}$ can be estimated by incorporating an extra $a\alpha_s$ term in the continuum extrapolation of the renormalized hadron matrix elements, and turns out to be consistent with zero within 2$\sigma$. The above renormalization procedure is referred to as the heavy quark improved normalization in the subsequent analysis.}

\subsubsection{Numerical test}

For the clover fermion,
the ratio between $Z_A$ and $Z_V$ can be extracted from the corresponding amputated vertex functions $\Lambda_{A,P}$ in the off-shell quark state with momentum $p$ under the Landau gauge~\cite{Martinelli:1994ty,Hu:2023jet},
\begin{align}
\frac{Z_A}{Z_V}&=\frac{\frac{1}{48}\text{Tr}[\Lambda_{V}(p)\gamma_\mu]}{\frac{1}{48}\text{Tr}[\Lambda^\mu_{A}(p)\gamma_\mu\gamma_5]},
\end{align}
where $\Lambda_{\cal O}(p)=S^{-1}(p)G_{\cal O}(p)S^{-1}(p)$ can be obtained from the quark propagators $S(p)=\sum_xe^{-ip\cdot x}\langle \psi(x)\bar{\psi}(0)\rangle$ and also $G_\mathcal{O}(p)=\sum_{x,y}e^{-ip\cdot (x-y)}\langle
\psi(x)\mathcal{O}(0)\bar{\psi}(y)\rangle$. $Z_A/Z_V$ can deviate from 1 due to additive chiral symmetry breaking, and it is essential to include this effect to obtain correct $f_{\pi,K}$ after the continuum extrapolation~\cite{Hu:2023jet}.

The renormalization of the PCAC quark mass requires the ratio of the renormalization constants $Z_A/Z_P$ can also be extracted from the corresponding amputated vertex functions,
\begin{align}
\frac{Z_A}{Z_P(\mu)}&=\frac{\frac{1}{12}\text{Tr}[\Lambda_{P}(p)\gamma_5]}{\frac{1}{48}\text{Tr}[\Lambda^\mu_{A}(p)\gamma_\mu\gamma_5]}|_{p^2=\mu^2},
\end{align}
subtracting the $1/m_q$ mass pole of the Goldstone meson~\cite{Liu:2013yxz} and extrapolating to the chiral limit. 

The meson decay constants also require the renormalization constants $Z_{S,T}$.
More detailed discussion on the ratio 
\begin{align}
\frac{Z_T}{Z_V}=\frac{\frac{1}{48}\text{Tr}[\Lambda_{V}(p)\gamma_\mu]}{\frac{1}{72}\text{Tr}[\Lambda^{\mu\nu}_{T}(p)\gamma_{\nu}\gamma_\mu]},\ \frac{Z_S}{Z_V}=\frac{\frac{1}{12}\text{Tr}[\Lambda_{V}(p)}{\frac{1}{72}\text{Tr}[\Lambda^{\mu\nu}_{T}(p)\gamma_{\nu}\gamma_\mu]},
\end{align}
can be found in our previous work~\cite{Hu:2023jet}.

\begin{figure}[t]
\includegraphics[width=0.45\textwidth]{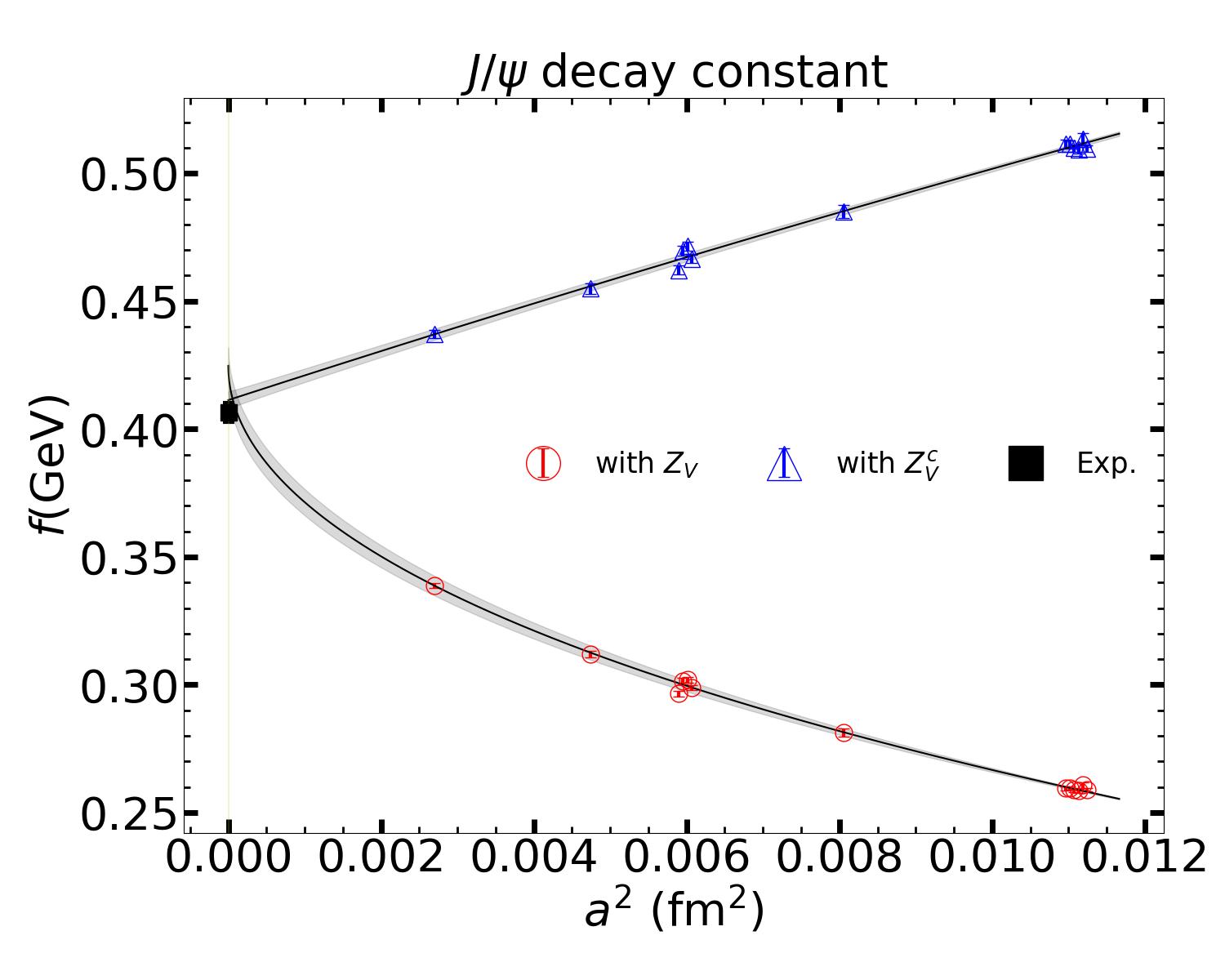}
\caption{Comparison of $f_{J/\psi}$ using heavy quark improved and original normalization constants $Z_V$, at different lattice spacing and continuum. The experimental value is also shown as black boxes for comparison. }
\label{fig:jpsi_decay_constant}
\end{figure}

Fig.~\ref{fig:jpsi_decay_constant} displays the values of $f_{J/\psi}$ at various lattice spacings using either $Z_V^c\equiv Z_V(\eta_c)$ (blue triangles) or $Z_V\equiv Z_V(\pi)_{m_{\pi}\rightarrow 0}$  (red circles). It is evident that employing the heavy quark improved normalization $Z_V^c$ leads to significantly smaller discretization errors and also better agreement with the experimental value 406.5(3.7)(0.5) MeV~\cite{ParticleDataGroup:2022pth,HPQCD2020} after the continuum extrapolation, compared to the case with the chiral extrapolated normalization $Z_V$. {The latter uses the following ansatz for the joint fit,
\begin{align}
    &X(m_{\pi},m_{\eta_s},a)=X(m_{\pi}^{\rm phys},m_{\eta_s}^{\rm phys},0) +d^X_1 (m_{\pi}^2-(m_{\pi}^{\rm phys})^2)+d^X_2 (m_{\eta_s}^2-(m_{\eta_s}^{\rm phys})^2) + \tilde{d}^X_3 a+ \tilde{d}^X_4 a^2,\label{eq:x_dep_a}
\end{align}
to incorporate the ${\cal O}(m_qa)$ effect in the matrix element, resulting in larger uncertainty post continuum extrapolation. A similar functional form is also adopted for the continuum extrapolation of other matrix elements using $Z_V$ in the subsequent discussion.}

\begin{figure}[t]
\includegraphics[width=0.45\textwidth]{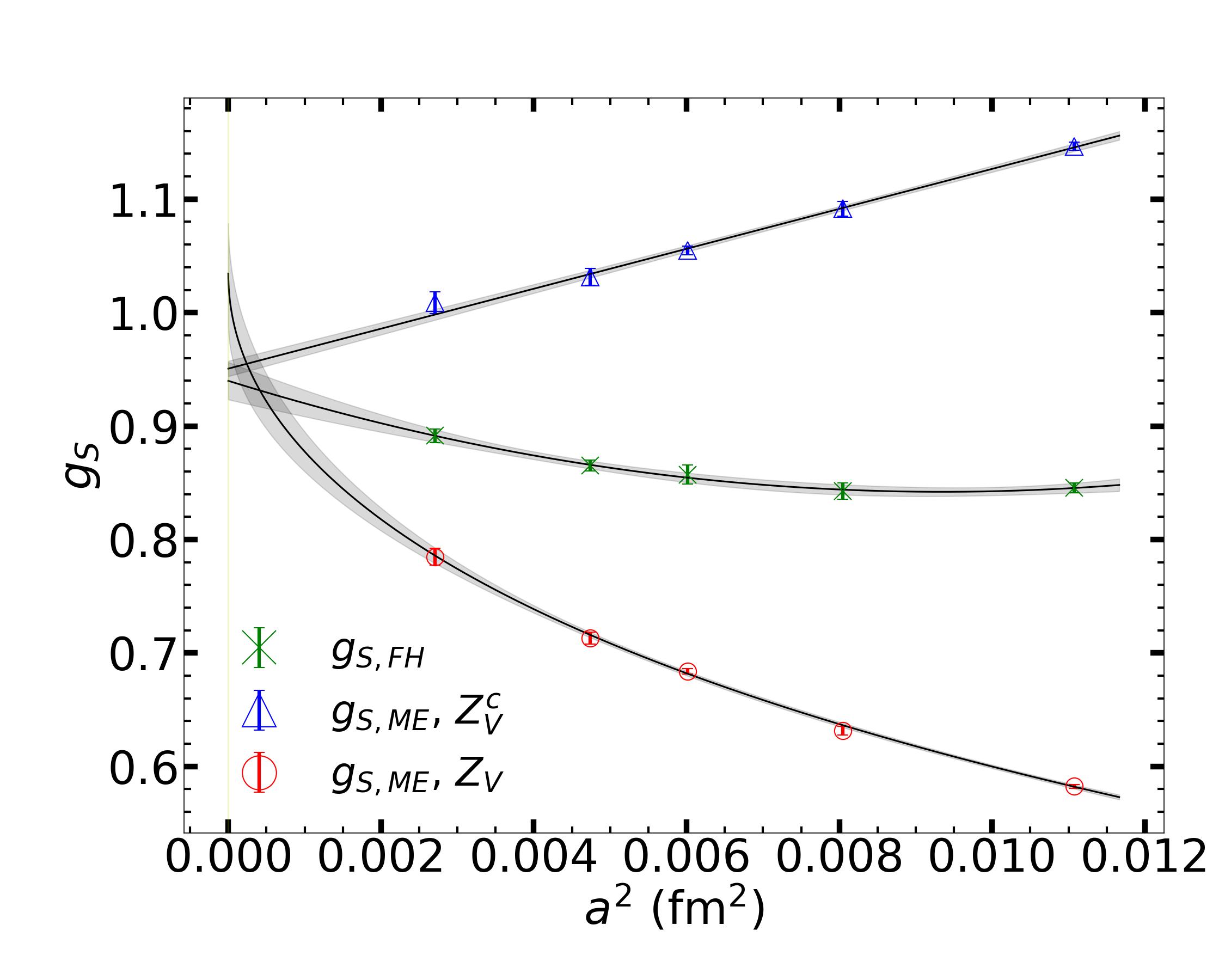}
\caption{Comparison of $\bar{c}c$ scalar matrix element in $\eta_c$ using original and heavy quark improved normalization constants $Z_V$, and also that from the Feynman-Hellman theorem, at different lattice spacing and continuum.}
\label{fig:gs}
\end{figure}

The improvement in suppressing discretization errors for heavy flavor matrix elements can be extended to other currents as well. For instance, considering $g_{S,{\rm ME}}(H)\equiv \frac{\langle H|S|H\rangle}{\langle H|H\rangle}$ as illustrated in Fig.~\ref{fig:gs}, one can use the original $Z_S$ (red circles) for the $X_I\equiv \bar{c}c$ matrix element in the $\eta_c$, akin to the treatment for $S\equiv \bar{l}l$ in the pion, or opt for the heavy quark improved $Z_{X_I}^{c}\equiv Z_V^c\frac{Z_S}{Z_V}$ (blue triangles) to mitigate discretization errors. Additionally, we present $g_{S,{\rm FH}}\equiv \frac{\partial m_{\eta_c}}{\partial m^R_c}=\frac{Z_S}{Z_A}\frac{\partial m_{\eta_c}}{\partial m^{\rm PCAC}_c}$ (green crosses) for comparison, which should align with $g_{S,{\rm ME}}$ in the continuum. Upon continuum extrapolation, all these quantities converge, while $g_{S,{\rm ME}}$ renormalized by $Z_S$ exhibits a larger discretization error.

The improvement can be extended to the flavor-changed current, like $Y_{\Gamma}\equiv \bar{c}\Gamma l$, by utilizing the improved renormalization constant $Z_{Y_{\gamma}}^c\equiv \sqrt{Z_VZ_V^c}\frac{Z_{O_{\gamma}}}{Z_V}$. As shown in Fig.~\ref{fig:dsv_decay_constant}, for the decay constant of $D_s^*$, we can see that the use of the heavy quark improved renormalization constant can reduce the discretization error significantly. Similar improvement could be used for the decay constant of the other open-charm meson, such as $f_D$, as discussed in the following section.

\begin{figure}[thb]
\includegraphics[width=0.45\textwidth]{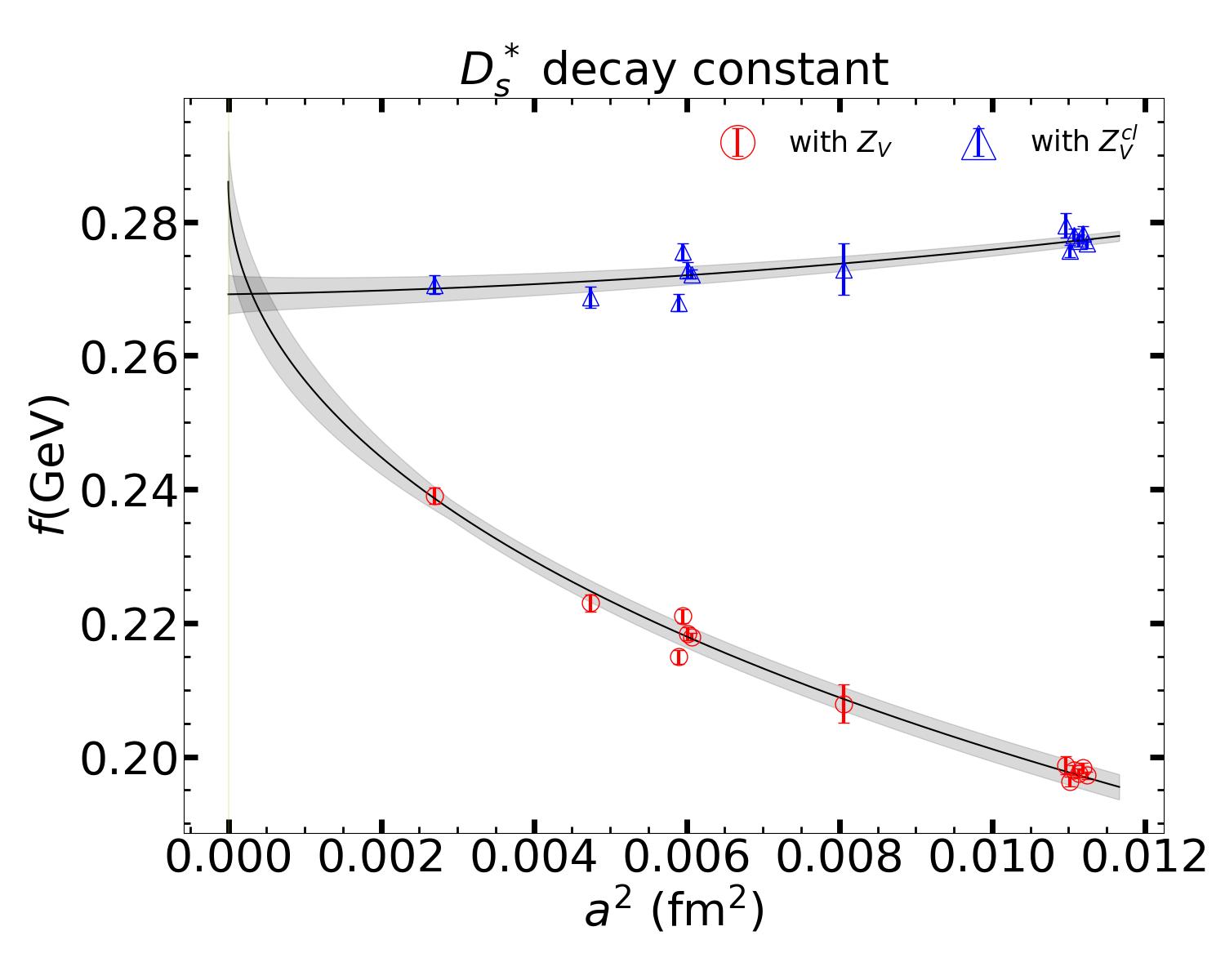}
\caption{Comparison of $f_{D_s^*}$ using heavy quark improved normalization constant $Z_{V}^{cl}\equiv\sqrt{Z_VZ_V^c}$ and original normalization constants $Z_V$ at different lattice spacing and continuum. The discretization error for the former is smaller.}
\label{fig:dsv_decay_constant}
\end{figure}

\begin{figure}[thb]
\includegraphics[width=0.45\textwidth]{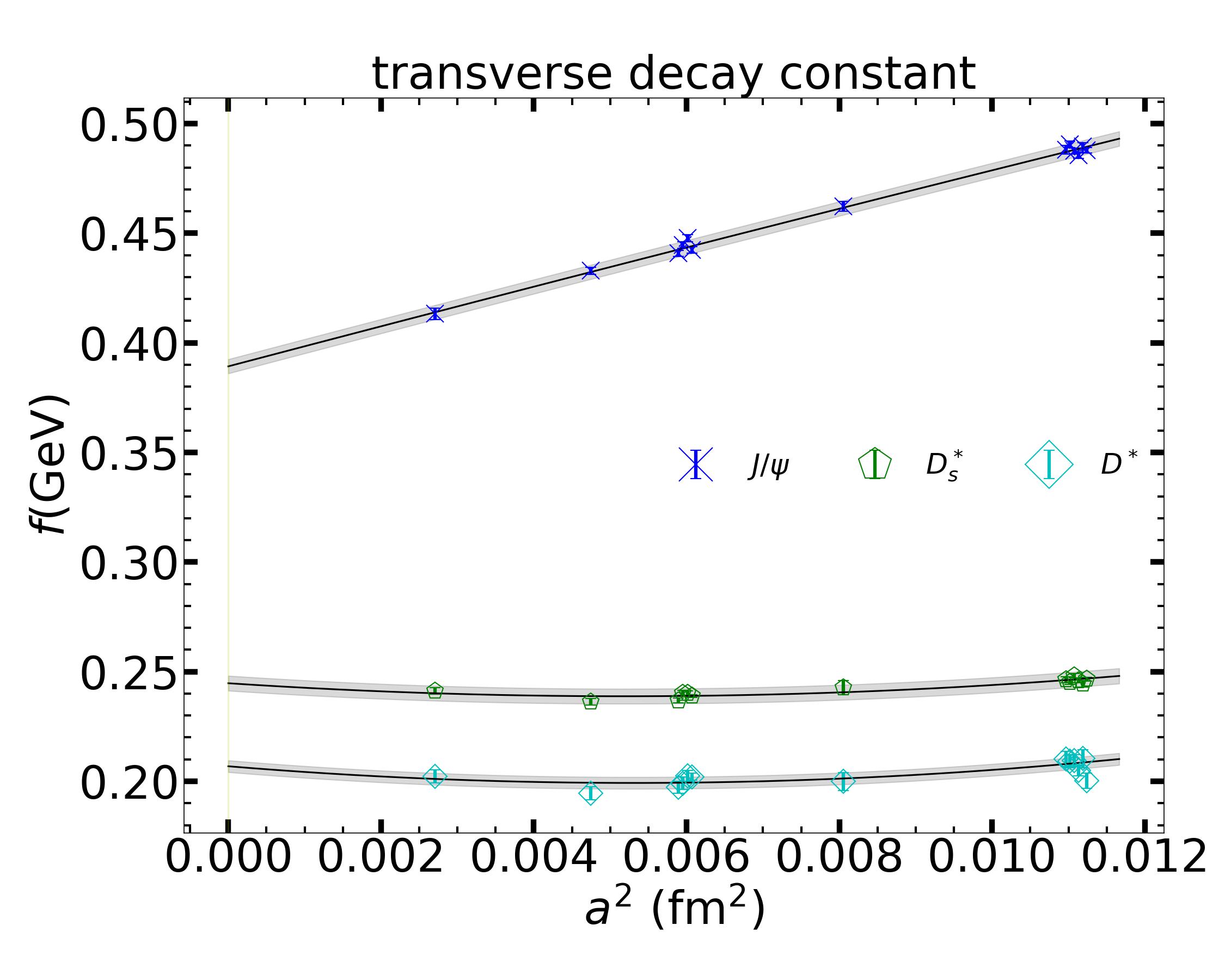}
\includegraphics[width=0.45\textwidth]{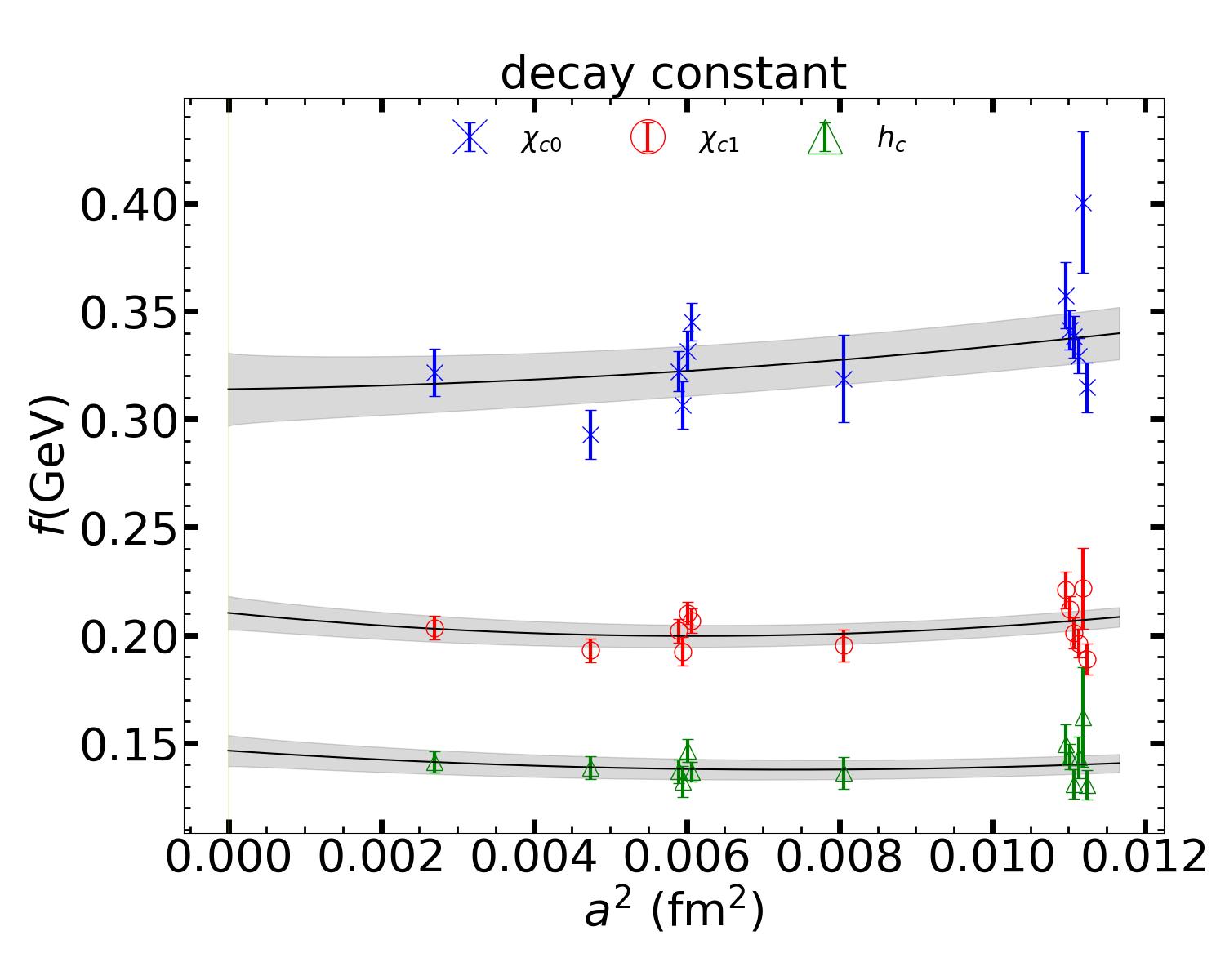}
\caption{Lattice spacing dependence for the transverse decay constants (left panel) and the P-wave charmonium decay constants (right panel), with the heavy quark improvement.}
\label{fig:decay_constant_lat_dep2}
\end{figure}

{In addition, we show the lattice spacing dependence of the transverse decay constant of the charmed vector mesons and also those of the P-wave charmonium in Fig.~\ref{fig:decay_constant_lat_dep2}. We can see that the lattice spacing dependence of those cases is also majorly linear on $a^2$, with the heavy quark improved normalization.} 

{As discussed above, the ${\cal O}(am^{\rm sub}_q)$ term is necessary to describe the quark mass dependence dependence of $Z_V({\rm PS};a)$. Thus the discretization error of the hadron matrix elements with the charm quark is ${\cal O}(a)$ and can enlarge the uncertainty of the continuum extrapolation significantly}. Since using $Z_V^c$ can significantly suppress the discretization error {and provide the consistent result}, we will only use $Z_V^c$ for all the final prediction when the charm quark is involved.

At the end of this section, we apply the joint fit of the renormalization constants using the following form to obtain the values at the chiral limit of the sea quark,
\begin{align}
\frac{Z_{X\neq V}}{Z_V}(a,m_{\pi})=\frac{Z_X}{Z_V}(a,0)\times(1+c^X_1m_{\pi}^2),
\end{align}
where we assume the light quark dependence at different lattice spacing to be similar which is supported by the $\chi^2/\mathrm{d.o.f}\sim 1$ we obtained. For $Z_V$ and $Z_V^c$ which is much more precise due to the high precision pseudoscalar correlation functions, we also include the strange sea and finite volume effects in the fit,
\begin{align}
&Z_V(a,m_{\pi},m_{\eta_s},1/L)=Z_V(a,0,0,0)\nonumber\\&\quad\quad\quad\quad (1+c_1m_{\pi}^2+c_2m_{\eta_s}^2+c_3e^{-m_{\pi}L}),
\end{align}
and re-scale the uncertainty by $\sqrt{\chi^2/\mathrm{d.o.f}}$. All the chiral-extrapolated renormalization constants are collected in Table~\ref{tab:rc} and the numbers on each ensemble without chiral extrapolation can be found in Table~\ref{tab:Z}. It can be observed that $Z_V^c$ is a factor of 2 larger than $Z_V$ at the coarsest lattice spacing, while becomes much closer at the finest one.

\begin{table*}[ht!]
\caption{$Z_V$,$Z_V^c$ and ratios of different renormalization constants at different lattice spacing}
\begin{tabular}{c c|c c c c c c c}
   &  &  $Z_V$ & $Z_V^c$ & $Z_A/Z_V$ & $Z_S/Z_V$& $Z_P/Z_V$& $Z_T/Z_V$\\
   \hline
    \multirow{5}{*}{$a$ (fm)} &0.10524(05)(62) & 0.8021(10) &1.5790(21) &1.0718(07)&1.2019(33) &0.9209(45)&1.0811(09)\\
    &0.08973(20)(53) & 0.8221(12) &1.4107(22) &1.0516(14)& 1.0958(63)&0.8583(70)&1.0963(18) \\
    &0.07753(03)(45)& 0.8387(10) &1.3104(17) &1.0527(08)& 1.0492(33)&0.8399(36)&1.1017(10)\\
    &0.06887(12)(41) &0.8502(12) &1.2445(19) &1.0418(09)& 1.0053(71)&0.8133(42)&1.1097(15)\\
    &0.05199(08)(31) &0.8727(12) &1.1334(17) &1.0344(10)& 0.9175(86)&0.7801(50)&1.1236(16) \\
    \hline
   \multirow{3}{*}{Para} & $c_1$ &$-$0.0207(18) &$-$0.0119(08) &0.0346(68) &$-$0.008(34)&0.036(44) &0.021(12)\\
   & $c_2$ &$-$0.0044(11) &$-$0.0126(06)\\
   & $c_3$ &$-$0.0013(05) &$-$0.0070(03)\\
\end{tabular}
\label{tab:rc}
\end{table*}

\begin{table*}[ht!]                   
\caption{The renormalization constants on each ensemble.}  
\begin{tabular}{c c | c c c c c c c}              
$a$ (fm)  & Symbol & $Z_V$ & $Z_V^s$ (optional) & $Z_V^c$ & $Z_A/Z_V$ & $Z_P/Z_V$ &$Z_S/Z_V$ & $Z_T/Z_V$ \\
\hline 
0.10524(05)(62) & C24P34 &0.79676(32)& 0.85034(10) &1.57130(20) & 1.07677(72) &0.921(12)& 1.205(05)&1.0856(20)\\ 
& C24P29 & 0.79814(23) &0.85184(06) & 1.57353(18) & 1.07244(70) &0.922(17)& 1.194(06)&1.0825(10)\\
& C32P29 & 0.79810(13) & 0.85167(04)& 1.57163(14) & 1.07648(63) &0.923(07)&1.199(08) &1.0826(10)\\
& C32P23 &0.79957(13) & 0.85350(04) & 1.57644(12) & 1.07375(40) &0.929(08)& 1.197(06)&1.0818(10)\\
& C48P23 &0.79954(05) & 0.85339(02)& 1.57326(08)& 1.07317(70) &0.923(09)& 1.208(10)&1.0819(10)\\
& C48P14 &0.79957(06) & 0.85359(02) & 1.57415(08) & 1.07320(68) &0.915(11)&1.204(04) &1.0831(10)\\
\hline
 0.08973(20)(53) & E28P35 &0.81768(04)&0.85877(02) &1.40361(05)&1.06029(53) &0.862(06)& 1.095(05)&1.0991(10) \\
\hline
 0.07753(03)(45) & F32P30 &0.83548(12)&0.86900(03) & 1.30566(07)& 1.05549(54)&0.841(10) &1.042(09) &1.1036(10)\\
& F48P30 &0.83511(04)& 0.86880(02)&1.30451(06)&1.05687(66) & 0.841(04)& 1.053(09)&1.1052(10)\\
& F32P21 &0.83579(09) & 0.87031(03)&1.30782(08)&1.05416(50) &0.845(05) & 1.042(06)&1.1008(10)\\
& F48P21 & 0.83567(05)& 0.86880(02)&1.30673(04)&1.05434(88) &0.840(07)&1.051(04) &1.1033(10)\\
\hline
0.06887(12)(41)& G36P29 &0.84636(09)& 0.87473(05)&1.23990(13)&1.04500(22) & 0.814(03)& 1.006(07)&1.1118(10) \\
\hline
0.05199(08)(31)& H48P32 &0.86855(04)&0.88780(01) & 1.12882(11)
& 1.03802(28)&0.783(04)&0.916(09) &1.1260(10)\\
\end{tabular}  
\label{tab:Z}
\end{table*}

\subsection{{Systematic uncertainty and error budget}}\label{w0}

{In this section, we will address our approach to handling systematic uncertainties, as well as the contributions from various sources to both the statistical and systematic uncertainties in our final predictions regarding charm physics.}

\begin{figure}[thb]
    \centering
    \includegraphics[width=0.45\linewidth]{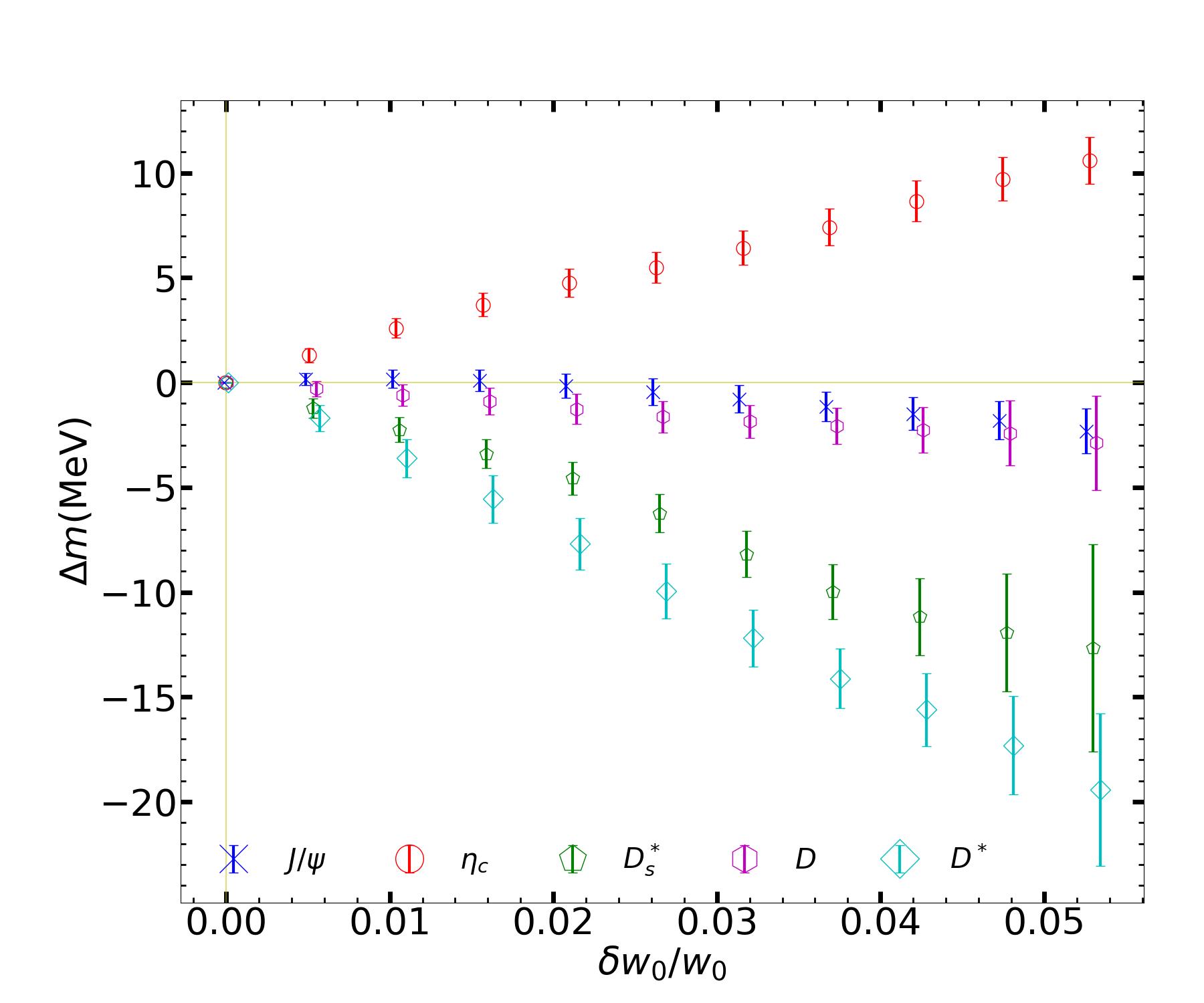}
    \caption{Meson mass changes with the center value of $w_0$.}
    \label{fig:aw0}
\end{figure}
For the systematic error from the uncertainty of the scale parameter $w_0=0.1736(9)$~fm~\cite{FlavourLatticeAveragingGroupFLAG:2021npn}, we vary the central value of $w_0$ by 1 sigma and repeat the whole analysis, and then redo the entire analysis and consider the difference
in the central value of the result as a systematic uncertainty from the lattice spacing. 

{Specifically, using \( w_0 = 0.1736 \) fm, we can determine the lattice spacing for \( \beta = 6.200 \) to be \( 0.10524(05) \) fm. Additionally, on the physical pion mass ensemble C48P14 with \( \beta = 6.200 \), we find the dimensionless bare valence strange and charm quark mass parameters to be $-$0.2335(1) and 0.4205(7), respectively, given \( m_{\eta_s} = 689.89(49) \) MeV and \( m_{D_s} = 1966.7(1.5) \) MeV. Through a joint fit that incorporates similar determinations across all ensembles to mitigate the effects of unphysical light and strange sea quark masses, as well as discretization errors using Eq.~(\ref{eq:x_dep}), we predict \( m_{D_s^*} = 2.1130(22) \) GeV.}

{Next, we increase the central value of \( 0.1736 \) fm by \( 1 \sigma \) (i.e., \( 0.0009 \) fm, or 0.5\%). This adjustment changes the dimensionless bare valence strange and charm quark mass parameters on the C48P14 ensemble into $-$0.2329(1) and 0.4286(7), respectively. The resulting modifications to these bare valence quark masses largely offset the impact of the change in \( w_0 \), keeping our predictions for the joint fits at \( m_{D_s^*} = 2.1116(22) \) GeV. We consider the change in central values, \( 0.0014(10) \) GeV, as the systematic uncertainty attributable to \( w_0 \), which constitutes only 0.07\% of \( m_{D_s^*} \) and is significantly smaller than the relative uncertainty associated with \( w_0 \).}

{To demonstrate this effect more clearly, we sequentially changed the central value of $w_0$ by 1 sigma until 10 sigma ($\sim$ 5\% away from the original central value), and calculated the changes in the mass of various mesons when the central value of $w_0$ changed, as shown in Fig.~\ref{fig:aw0}. We can see that the systematic error caused by the error of $w_0$ is majorly canceled, especially for $J/\psi$. In the case of $m_{D^*}$, which exhibits the highest sensitivity among the cases depicted in Fig.~\ref{fig:aw0}, the change is 1\% {when $w_0$ is changed by 5\%. It means that the impact of the $w_0$ uncertainty is also suppressed by a factor of 5}.

If we repeat the joint fit using the following functional form that includes an additional finite volume term, 
\begin{align}
    &X(m_{\pi},m_{\eta_s},a,1/L)=X(m_{\pi}^{\rm phys},m_{\eta_s}^{\rm phys},0,0) +d^X_1 (m_{\pi}^2-(m_{\pi}^{\rm phys})^2)+d^X_2 (m_{\eta_s}^2-(m_{\eta_s}^{\rm phys})^2) + d^X_3 a^2+ d^X_4 a^4+d^X_Le^{-m_\pi L},
    \label{eq:x_dep2_sm}
\end{align}
The systematic uncertainties arising from the finite volume correction, derived from the value of $d^X_Le^{-m_\pi L}$ for the C48P23 ensemble with the largest $m_{\pi}L$, is $\sim$0.2\% ($-$0.413(75)~MeV) for $f_{D^*}$ which is particularly sensitive to finite volume effects among the quantities we studied in this work. However, this impact is much smaller than the statistical uncertainty associated with our final prediction of $f_{D^*}$.  Following a similar approach, the systematic uncertainty for other quantities remains at 0.1\% or even lower, also significantly smaller than the statistical uncertainties, as summarized in Table~\ref{tab:error1}-\ref{tab:error3} of this supplemental material.}

\begin{table*}[ht!]
\centering
\caption{The $a\alpha_s$ term coefficient of different quantities.}
\begin{tabular}{c|ccccccccc}
& $m_{J/\psi}$&$m_{\eta_c}$&$m_{\chi_{c_0}}$&$m_{\chi_{c_1}}$&$m_{h_c}$&$m_D$&$m_{D^*}$&&$m_{D_s^*}$ \\
\hline
$d^X_{\alpha_s}$ & 0.024(18)&$-$0.043(24)&$-$0.005(35)&0.041(45)&0.080(42)&0.064(62)&0.067(84)&&0.138(41)\\
\hline
&$f_{J/\psi}$&$f_{\eta_c}$&$f_{\chi_{c_0}}$&$f_{\chi_{c_1}}$&$f_{h_c}$&$f_D$&$f_{D^*}$&$f_{D_s}$&$f_{D_s^*}$\\
\hline
$d^X_{\alpha_s}$&$-$0.075(53)&$-$0.188(96)&0.023(48)&0.031(39)&0.018(35)&0.13(15)&$-$0.003(96)&0.064(76)&0.081(58)\\
\hline
&$f_{J/\psi}^T$&$m_c(2\rm GeV)$&&&&&$f_{D^*}^T$&&$f_{D_s^*}^T$\\
\hline
$d^X_{\alpha_s}$&$-$0.12(13)&$-$0.003(60)&&&&&0.126(59)&&0.12(12)
\end{tabular}
\label{tab:a}
\end{table*}

{
Simultaneously, we also perform the joint fit by including an additional  $a\alpha_s$ term with the prior $d^X_{\alpha_s}\in[-1,1]$, 
\begin{align}
    &X(m_{\pi},m_{\eta_s},a)=X(m_{\pi}^{\rm phys},m_{\eta_s}^{\rm phys},0) +d^X_1 (m_{\pi}^2-(m_{\pi}^{\rm phys})^2)+d^X_2 (m_{\eta_s}^2-(m_{\eta_s}^{\rm phys})^2) + d^X_3 a^2+ d^X_4 a^4+d^X_{\alpha_s}a\Lambda_\chi^2\log u_0
    \label{eq:x_dep3_sm}
\end{align}
where $\Lambda_\chi = 1~\rm GeV$. The relation $\mathrm{log}u_0\propto\alpha_s$ is used to approximate the renormalized $\alpha_s$ without using the bare coupling $\beta$ which suffers from kinds of ambiguity. As shown in table~\ref{tab:a}, except for the $a\alpha_s$ term of $m_{D_s^*}$, which is 3$\sigma$ away from zero, that of all the other cases are consistent with zero within 2$\sigma$. 

The difference between the central value of $X(m_{\pi}^{\rm phys},m_{\eta_s}^{\rm phys},0)$ with and without $a\alpha_s$ term term is considered a systematic uncertainty. As shown in the summary Table~\ref{tab:error1}-\ref{tab:error3} of this supplemental materials,  the impact on the $X(m_{\pi}^{\rm phys},m_{\eta_s}^{\rm phys},0)$ is consistently much smaller than the statistical uncertainty, thanks to the heavy quark improved normalization propsoed in the previous section.
}

{
Eventually, we collect the statistical and systematic uncertainties into those from different origins, in Table~\ref{tab:error1}-\ref{tab:error3}.

\begin{table*}[ht!]
\caption{Error budget of meson masses. The primary sources of statistical and systematic uncertainties are highlighted in bold font.} 
\begin{tabular}{c|c c c c c c c c}
     & $m_{\eta_c}$ &$m_{J/\psi}$ & $m_{\chi_{c_0}}$ &$m_{\chi_{c_1}}$ &$m_{h_c}$ &$m_D$ & $m_{D^*}$ &$m_{D_s^*}$\\
     \hline
     Central value &2.9745&3.0898&3.423&3.511&3.522&1.8637&2.0115&2.1130 \\
     \hline
     \hline
     Statistical (total) &0.0018 &0.0016 &0.013 & 0.016 & 0.013&0.0018 &0.0025 &0.0022\\
    \hline
    2-point function & 0.0007 & 0.0006 & \textbf{0.013} & \textbf{0.016}& \textbf{0.013}& \textbf{0.0016} &\textbf{0.0023} &\textbf{0.0020}\\
    Lattice spacing & \textbf{0.0017} & \textbf{0.0014} & 0.002 & 0.002 &0.002 &0.0007 &0.0008 &0.0008\\
    Sea $\pi$ mass & 0.0005 & 0.0005 & 0.001 & 0.001 & 0.001 & 0.0005 & 0.0005 & 0.0005\\
    Sea $\eta_s$ mass& 0.0002 & 0.0002 & $<$0.001 & $<$0.001 &$<$0.001 &0.0001 & 0.0001 & 0.0002\\
    \hline
    \hline
    Systematic (total)& 0.0031 & 0.0027 & 0.003 & 0.006 & 0.011 & 0.0015 &0.0028 &0.0022\\
    \hline
    $D_s$ mass& \textbf{0.0028} & \textbf{0.0027} &\textbf{0.003} & 0.003 & 0.003 &\textbf{0.0015} &0.0016 &\textbf{0.0016}\\
     Fit ansatz & $<$0.0001 & $<$0.0001 &$<$0.001 & 0.001 & 0.001 &0.0002 &0.0014& 0.0005 \\
     Scale setting &0.0013 & $<$0.0001 & 0.001 & \textbf{0.004} & \textbf{0.009}&0.0002 &\textbf{0.0018} &0.0014\\
     Finite volume&$<$0.0001&$<$0.0001&$<$0.001&$<$0.001&$<$0.001&0.0001&0.0002&0.0001 \\
     $a\alpha_s$ term&0.0001&$<$0.0001&0.001&0.004&0.005&0.0002&0.0002&0.0001 \\
\end{tabular}
\label{tab:error1}
\end{table*}

Table~\ref{tab:error1} presents the meson mass cases. The statistical uncertainty is defined as the statistical fluctuations due to the limited number of configurations in the ensembles we used. It encompasses uncertainties from the original 2-point function, the determination of lattice spacing using gradient flow (excluding the systematic uncertainty associated with the scale parameter $w_0$ we used from the literature~\cite{FlavourLatticeAveragingGroupFLAG:2021npn}), as well as the statistical uncertainties of the dimensionless pion and $\eta_s$ masses in the sea. It is important to note that the statistical uncertainty of the original 2-point function includes contributions from the continuum extrapolation with both $a^2$ and $a^4$ terms in the global fit. The largest statistical uncertainties in our meson mass predictions arise from the 2-point function, except for the cases of $\eta_c$ and $J/\psi$, where the uncertainty from the determination of lattice spacing is the predominant factor.

\begin{table*}[ht!]
\centering
\caption{Error budget of pseudo-scalar and longitudinally polarized vector mesons, which includes an additional statistical uncertainty arising from the normalization/renormalization constant \( Z_{V,A} \).}
\begin{tabular}{c|c c c c c c}
      &$f_{\eta_c}$ &$f_{J/\psi}$ & $f_D$ & $f_{D^*}$  & $f_{D_s}$ & $f_{D_s^*}$ \\
     \hline
Central value &0.3944 &0.4114 &0.2108 &0.2292 &0.2470 &0.2691 \\
\hline
\hline
Statistical (total)&0.0029 &0.0034 &0.0020 &0.0026  & 0.0021 &0.0030\\
\hline
2-point function &0.0017 &\textbf{0.0027} &\textbf{0.0014} &\textbf{0.0021} &\textbf{0.0016} &\textbf{0.0027} \\
Lattice spacing & 0.0013 &0.0011 &0.0011 &0.0012 &0.0009 &0.0008 \\
Sea $\pi$ mass & 0.0001 & 0.0001 & 0.0001 & 0.0001 & 0.0001 & 0.0001\\
Sea $\eta_s$ mass & $<$0.0001 & $<$0.0001 & $<$0.0001 & $<$0.0001 & $<$0.0001 & $<$0.0001 \\
Renormalization &\textbf{0.0019} &0.0016 & 0.0009 &0.0008 &0.0010 &0.0009 \\
\hline
\hline
 Systematic (total) &0.0018 &0.0022 &0.0011 &0.0017 &0.0004 &0.0003\\
 \hline
 $D_s$ mass &0.0003 &0.0003 &0.0002 &0.0002 &0.0002 &0.0002\\
Fit ansatz & 0.0003 &\textbf{0.0018} & 0.0003&0.0008 &0.0002 &\textbf{0.0002} \\
  Scale setting & \textbf{0.0018} &0.0012 &\textbf{0.0010} &\textbf{0.0014} &\textbf{0.0002} &0.0001\\
  Finite volume&$<$0.0001 &$<$0.0001 &0.0001 &0.0004 &$<$0.0001&0.0001\\
  $a\alpha_s$ term&0.0003&0.0002&0.0004&0.0001&0.0002&0.0001 \\
\end{tabular}
\label{tab:error2}
\end{table*}

The systematic uncertainties considered in this work include those arising from the experimental $D_s$ mass, the fit ansatz with additive or multiplicative discretization errors, the uncertainty associated with the lattice average of the scale parameter $w_0$ for the $N_f$=2+1 case~\cite{FlavourLatticeAveragingGroupFLAG:2021npn}, and the finite volume effects estimated from the corrections applied to the ensemble with the largest $m_{\pi}L$. We also account for the impact of including an additional $a\alpha_s$ term. Notably, the systematic uncertainties from those of $m_{D_s}$ and $w_0$ are comparable to or even exceed the statistical uncertainties.

\begin{table*}[ht!]
\centering
\caption{Error budget of charm quark mass and the other decay constants, which further includes an additional systematic uncertainty arising from the perturbative matching.}
\begin{tabular}{c|c c c c c c c}
    & $m_c(m_c)$  & $f_{J/\psi}^T$ & $f_{\chi_{c_0}}$ & $f_{\chi_{c_1}}$ & $f_{h_c}$  & $f_{D^*}^T$ & $f_{D_s^*}^T$\\
     \hline
Central value &1.2933  & 0.3893& 0.314 &0.2103 &0.1465  &0.2068 &0.2447 \\
\hline
\hline
Statistical (total)&0.0072  &0.0032 &0.017 &0.0079 &0.0073 &0.0027 &0.0034\\
\hline
2-point function &0.0019 &\textbf{0.0025} &\textbf{0.015} &\textbf{0.0069} &\textbf{0.0070} &\textbf{0.0019} &\textbf{0.0030}\\
Lattice spacing &0.0018  &0.0011 &0.001 &0.0013 &0.0012 &0.0010 &0.0008\\
Sea $\pi$ mass &$<$0.001  &0.0001 &0.001 &0.0006 &0.0003 &0.0001 &0.0001\\
Sea $\eta_s$ mass &$<$0.001  &$<$0.0001 &$<$0.001 &0.0001 &$<$0.0001 &$<$0.0001 &$<$0.0001\\
Renormalization &\textbf{0.0067} &0.0014 &0.007 &0.0013 &0.0008 &0.0010 &0.0011 \\
\hline
\hline
 Systematic (total) &0.0095  &0.0024 &0.004 &0.0034 &0.0031 &0.0014 &0.0006\\
 \hline
 $D_s$ mass &0.0014 &0.0003 &$<$0.001 &0.0003 &0.0001 &0.0002 &0.0002\\
Fit ansatz &0.0003  &\textbf{0.0020} &0.001 &0.0008 &\textbf{0.0027} &\textbf{0.0012} &0.0001 \\
  Scale setting &0.0040 &0.0011 &0.003 &\textbf{0.0033} &0.0014 &0.0006 &0.0002\\
  Matching &\textbf{0.0085} &0.0005 &\textbf{0.003} & $-$ & 0.0002 &0.0003 &0.0003\\
   Finite volume&0.0001&$<$0.0001&$<$0.001&$<$0.0001&$<$0.0001&0.0003&0.0001 \\
  $a\alpha_s$ term&0.0001&0.0003&$<$0.001&0.0003&$<$0.0001&0.0002&\textbf{0.0004} \\
\end{tabular}
\label{tab:error3}
\end{table*}

For the decay constants of pseudo-scalar and longitudinally polarized vector mesons, we include an additional statistical uncertainty arising from the normalization/renormalization constant \( Z_{V,A} \), as indicated in Table~\ref{tab:error2}. The systematic uncertainty from \( m_{D_s} \) is significantly reduced in this context, while the uncertainties associated with both renormalization constants and the fit ansatz become more prominent. It is worth noting that the statistical uncertainty from the original 2-point function remains the dominant factor, except for \( f_{\eta_c} \), where it can only be suppressed when data from finer lattice spacing becomes available.

For the charm quark mass and the other decay constants, we also include the systematic uncertainty from the perturbative matching, which is fully correlated across all lattice spacings. However, this effect is not dominant except for the charm quark mass, whose uncertainty primarily arises from the renormalization and matching processes.}

\clearpage
\end{widetext}

\end{document}